\documentclass{aa} 

\usepackage{graphicx}
\usepackage{txfonts}
\usepackage[usenames]{color} 
\usepackage{tabularx} 
\usepackage{footnote}
\makesavenoteenv{table*}
\usepackage{rotating}
\usepackage{float}
\usepackage{longtable} 
\usepackage{lscape} 

\usepackage{array}
\newcolumntype{H}{>{\setbox0=\hbox\bgroup}c<{\egroup}@{}}
\usepackage{multirow}
\usepackage{scalerel}
\usepackage{tikz}
\usetikzlibrary{svg.path}
\definecolor{orcidlogocol}{HTML}{A6CE39}
\tikzset{
  orcidlogo/.pic={
    \fill[orcidlogocol] svg{M256,128c0,70.7-57.3,128-128,128C57.3,256,0,198.7,0,128C0,57.3,57.3,0,128,0C198.7,0,256,57.3,256,128z};
    \fill[white] svg{M86.3,186.2H70.9V79.1h15.4v48.4V186.2z}
                 svg{M108.9,79.1h41.6c39.6,0,57,28.3,57,53.6c0,27.5-21.5,53.6-56.8,53.6h-41.8V79.1z M124.3,172.4h24.5c34.9,0,42.9-26.5,42.9-39.7c0-21.5-13.7-39.7-43.7-39.7h-23.7V172.4z}
                 svg{M88.7,56.8c0,5.5-4.5,10.1-10.1,10.1c-5.6,0-10.1-4.6-10.1-10.1c0-5.6,4.5-10.1,10.1-10.1C84.2,46.7,88.7,51.3,88.7,56.8z};
  }
}

\newcommand\orcid[1]{\href{https://orcid.org/#1}{\mbox{\scalerel*{
\begin{tikzpicture}[yscale=-1,transform shape]
\pic{orcidlogo};
\end{tikzpicture}
}{|}}} \href{#1}{#1}}
\usepackage[breaklinks=true,colorlinks, linkcolor=blue,urlcolor=blue,citecolor=blue]{hyperref}

\usepackage{threeparttablex} 
 
\usepackage[switch]{lineno}
\usepackage[normalem]{ulem}
\usepackage{xcolor}

\usepackage{booktabs}

\usepackage[switch]{lineno}

\begin{document}

\newcommand\redsout{\bgroup\markoverwith{\textcolor{red}{\rule[0.5ex]{2pt}{1.0pt}}}\ULon}

   \title{Solar energetic particles injected inside and outside a magnetic cloud}

   \subtitle{The widespread solar energetic particle event on 2022 January 20}

   \author{L.~Rodríguez-García
          \inst{1,2}
          \and
          R.~Gómez-Herrero\inst{2} \and  N.~Dresing\inst{3}
          \and L.~A.~Balmaceda\inst{4,5} \and E.~Palmerio\inst{6}\and A.~Kouloumvakos\inst{7} 
          \and I.~C.~Jebaraj\inst{3} \and F.~Espinosa~Lara\inst{2}
           \and M.~Roco\inst{2}  \and C.~Palmroos\inst{3}   \and A.~Warmuth\inst{8} \and G.~Nicolaou\inst{9}\and G.~M.~Mason\inst{7}\and
           J. Guo\inst{10} \and T.~Laitinen \inst{11}  \and I.~Cernuda\inst{2} \and T.~Nieves-Chinchilla\inst{4} \and
           A.~Fedeli\inst{3}\and C.~O.~Lee \inst{12}\and C.~M.~S.~Cohen\inst{13}
           \and C.~J.~ Owen\inst{9} \and  G.~ C.~Ho\inst{14} \and O.~Malandraki\inst{15}
          \and R.~Vainio\inst{3}\and  J.~Rodríguez-Pacheco\inst{2}
          }
   \institute{European Space Agency (ESA), European Space Astronomy Centre (ESAC), Camino Bajo del Castillo s/n, 28692 Villanueva de
la Cañada, Madrid, Spain \\
              \email{laura.rodriguezgarcia@esa.int}
         \and Universidad de Alcalá, Space Research Group (SRG-UAH), Plaza de San Diego s/n, 28801 Alcalá de Henares, Madrid, Spain\and Department of Physics and Astronomy, University of Turku, FI-20014 Turku, Finland\and Heliophysics Science Division, NASA Goddard Space Flight Center, Greenbelt, MD 20771, USA 
             \and Physics and Astronomy Department, George Mason University, 4400 University Drive, Fairfax, VA 22030, USA \and Predictive Science Inc., San Diego, CA 92121, USA \and The Johns Hopkins University Applied Physics Laboratory, 11101 Johns Hopkins Road, Laurel, MD 20723, USA\and Leibniz-Institut für Astrophysik Potsdam (AIP), D-14482  Potsdam, Germany \and Mullard Space Science Laboratory, University College London, Dorking RH5 6NT, UK \and Deep Space Exploration Laboratory/School of Earth and Space Sciences, University of Science and Technology of China, Hefei 230026, China \and Jeremiah Horrocks Institute, University of Central Lancashire, Preston PR1 2HE, UK \and Space Sciences Laboratory, University of California, Berkeley, CA 94720, USA \and California Institute of Technology, Pasadena, CA 91125, USA  \and Southwest Research Institute, San Antonio, TX 78238, USA \and National Observatory of Athens/IAASARS, I.\ Metaxa \& Vas.\ Pavlou, GR-15236 Penteli, Greece}

   \date{Received 6 September 2024; Accepted 5 December 2024}

 
  \abstract
  {On 2022 January 20, the Energetic Particle Detector (EPD) on board Solar Orbiter measured a solar energetic particle (SEP) event showing unusual first arriving particles from the anti-Sun direction. Near-Earth spacecraft separated by 17$^{\circ}$ in longitude to the west of Solar Orbiter measured classic anti-sunward-directed fluxes. STEREO-A and MAVEN, separated by 18$^{\circ}$ to the east and by 143$^{\circ}$ to the west of Solar Orbiter, respectively, also observed the event, suggesting that particles spread over at least 160$^{\circ}$ in the heliosphere.   }
   {The aim of the present study is to investigate how SEPs are accelerated and transported towards Solar Orbiter and near-Earth spacecraft, as well as to examine the influence of a magnetic cloud (MC) present in the heliosphere at the time of the event onset on the propagation of energetic particles.  
} 
   {We analysed remote-sensing data, including flare, coronal mass ejection (CME), and radio emission to identify the parent solar source of the event. We investigated  energetic particles, solar wind plasma, and magnetic field data from multiple spacecraft.   } 
   {Solar Orbiter was embedded in a MC erupting on 16 January from the same active region as that related to the SEP event on 20 January. The SEP event is related to a M5.5 flare and a fast CME-driven shock of $\sim$1433 km s$^{-1}$, which accelerated and injected particles within and outside the MC. Taken together, the hard SEP spectra, the presence of a Type II radio burst, and the co-temporal Type III radio burst being observed from 80 MHz that appears to emanate from the Type II burst, suggest that the shock is likely the main accelerator of the particles.     }
   {Our detailed analysis of the SEP event strongly suggests that the energetic particles are mainly accelerated by a CME-driven shock and are injected into and outside of a previous MC present in the heliosphere at the time of the particle onset. The sunward-propagating SEPs measured by Solar Orbiter are produced by the injection of particles along the longer (western) leg of the MC still connected to the Sun at the time of the release of the particles. The determined electron propagation path length inside the MC is around 30\% longer than the estimated length of the loop leg of the MC itself (based on the graduated cylindrical shell model), which is consistent with the low number of field line rotations.   }

   \titlerunning{The widespread SEP event on 2022 January 20: The role of the ICME} \authorrunning{L.\ Rodríguez-García et al.}
   \maketitle
%
\begin{figure*}[htb]
\centering
  \resizebox{1.0\hsize}{!}{\includegraphics{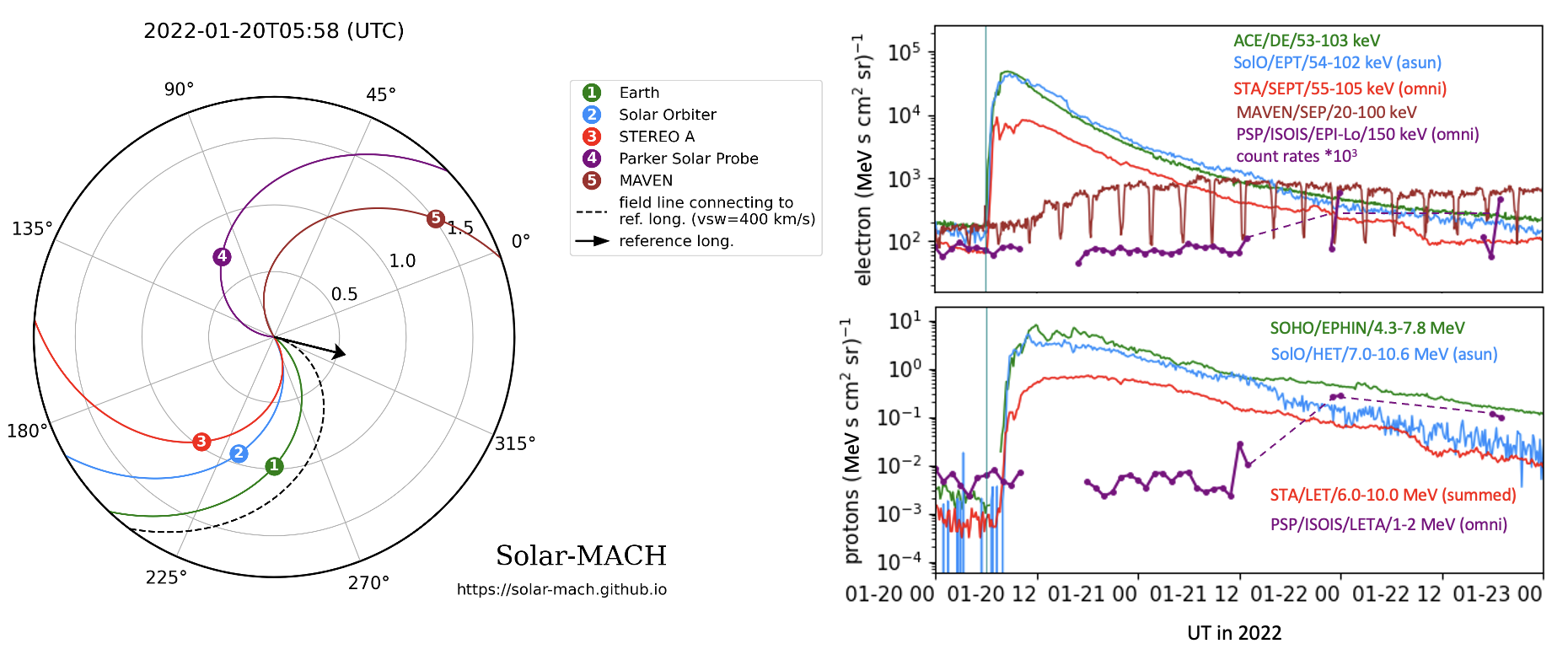}}
     \caption{Longitudinal spacecraft constellation and magnetic connectivity  at 05:58 UT on 2022 January 20 (left) along with multi-spacecraft SEP measurements (right). \textit{Left}: Spacecraft configuration using the \textit{Solar-MACH} tool \citep{Gieseler2022}, available online at \url{https://doi.org/10.5281/zenodo.7100482}. \textit{Right:} The upper panel shows near-relativistic electron intensities and the lower  panel $\sim$5~MeV  proton intensities observed by the spacecraft indicated with the same colour code shown on the left panel. The blue vertical line indicates the time of the soft-X ray peak of the flare ($\sim$05:58 UT) associated with the SEP event. }
     \label{fig:solar_mach}
\end{figure*}

\section{Introduction}
\label{sec:Introduc}

Solar energetic particle (SEP) events are sporadic enhancements of particle intensities associated with transient
solar activity. In the inner heliosphere, these intensity enhancements are usually measured in situ at energy ranges spanning many orders of magnitude, from a few keV to hundreds of MeV or even above 1 GeV. For the most energetic events, near-relativistic and relativistic electrons and protons are observed. The mechanisms proposed to explain the origin of large SEP events include: (1) acceleration during magnetic reconnection processes associated with solar jets \citep{Krucker2011} and flares \citep{2007Kahler}; (2) acceleration at shocks driven by fast coronal mass ejections (CMEs) \citep[e.g.][]{2002Simnett, Desai2016, Kouloumvakos2019, Jebaraj24}; and/or (3) acceleration during magnetic restructuring in the aftermath of CMEs and in the current sheets formed in the wake of CMEs \citep[e.g.][]{Kahler1992, 2004MaiaPick, 2005Klein}.  

SEP events are often classified into two categories, impulsive and gradual \citep{Cane1986, Reames1999}, on account of their observed properties, such as timescale, spectrum, composition, and charge state, and the associated radio bursts. During most gradual events, SEPs are detected over a very wide range of heliolongitudes. These widespread SEP events have been extensively researched \citep[e.g.][]{Reames1996, Lario2006,Lario2013,Lario2016, Wibberenz2006, Dresing2012,Dresing2014,2023Dresing, Papaioannou2014, Richardson2014, Gomez-Herrero2015, Paassilta2018,guo2018September, guo2023first, Xie2019, 2021Rodriguez-Garcia, 2022KouloumvakosNov29,2023Dresing,2024Khoo} thanks to constellations of spacecraft widely distributed throughout the heliosphere, such as Helios \citep{Porsche1981}, \textit{Ulysses} \citep{Wenzel1992}, the SOlar and Heliographic Observatory \citep[SOHO;][]{Domingo1995SOHO}, the Solar TErrestrial RElations Observatory \citep[STEREO;][]{Kaiser2008STEREO}, MErcury Surface Space ENvironment GEochemistry and Ranging \citep[MESSENGER;][]{Solomon2007MESSENGER}, and more recently Solar Orbiter \citep[][]{Muller2020,Zouganelis2020}, Parker Solar Probe \citep[PSP;][]{Fox2016}, BepiColombo \citep[][]{2021Benkhoff}, Mars Atmosphere and Volatile EvolutioN \citep[MAVEN;][]{Jakosky2015}, and even Mars Science Laboratory \citep[MSL;][]{grotzinger2012mars} on the surface of Mars.

CMEs are large eruptions of magnetised plasma that are ejected from the Sun into the heliosphere as a result of the release of a huge quantity of energy stored in the coronal magnetic field. Remote-sensing observations of CMEs close to the Sun provide evidence for the existence of magnetic flux-rope (MFR) structures within CMEs \citep{vourlidas_flux_2014}. These consist of confined plasma within a helically organised magnetic structure. In interplanetary (IP) space, evidence of MFRs is found in structures known as magnetic clouds \citep[MCs;][]{Burlaga1981}. In the best examples, the in situ MFR signatures show a monotonic rotation of the magnetic field direction through a large angle, along with a low plasma temperature and low plasma $\beta$.  

SEPs can be injected inside IP CMEs (hereafter ICMEs) either due to impulsive acceleration at flares occurring at the footpoints of the parent ICME and/or when a new CME-driven shock intercepts one or both legs of other ICMEs \citep{Richardson1991, Masson2012, 2016DresingINJE, Palmerio2021, 2023Wimmer}. \cite{2011Kahler_a} found some electron events inside ICMEs in which the active regions (ARs) responsible for the accelerated particles were different from the CME source region, suggesting an interconnection with adjacent loops. Independently of the source, SEPs propagating inside ICMEs provide a valuable tool for studying their magnetic topology \citep[e.g.][]{RichardsonCane1996, Larson1997, 2001Malandraki, 2011Kahler_a, 2011Kahler_b, 2016DresingINJE, 2017Gomez_Herrero}.  In particular, near-relativistic electrons may be used as probes of the magnetic structure inside the MC, as they only require a few minutes to travel 1 au from their source.

On 2022 January 20, a widespread SEP event was observed by different spacecraft located in the inner heliosphere, namely the near-Earth probes Solar Orbiter, STEREO-A, and MAVEN, and was seen to span a longitudinal range of $\sim$160$^{\circ}$ in the ecliptic plane (assuming that PSP did not observe the SEP event). The SEP origin was associated with an M-class flare and a wide and fast CME erupting near the west limb from Earth's perspective. Figure~\ref{fig:solar_mach} (left) illustrates the spacecraft locations in the heliographic equatorial plane along with nominal Parker field lines connecting each spacecraft with the Sun in the centre of the plot, using measured solar wind speeds when available. The black arrow marks the longitude of the associated flare (W76), and the dashed black spiral depicts the nominal magnetic field line connecting to this location. Near-Earth spacecraft (1, green) show the best nominal connection to the flare site. Solar Orbiter (2, blue) ---separated 17$^{\circ}$ eastwards from Earth--- and STEREO-A (3, red) ---separated 18$^{\circ}$ eastwards from Solar Orbiter--- were also well connected to the flaring region. PSP (4, purple) ---separated 147$^{\circ}$ eastwards from Earth--- and MAVEN$^{\circ}$ (5, brown) ---separated 126$^{\circ}$ westwards from Earth--- had the larger longitudinal separation between the solar source and the footpoint of the respective nominal field lines. 

The top panel of Fig.~\ref{fig:solar_mach} (right) shows near-relativistic ($\sim$20--150~keV)  electron intensities and the bottom panel shows $\sim$5~MeV proton intensities measured by the different spacecraft, with the readouts colour-coded according to the key in the left plot. The legend shows the telescopes' pointing directions (when available), which were used to compute the intensities of the first arriving particles for each spacecraft. As expected given their closest magnetic connection, near-Earth spacecraft observed the highest intensities, measuring standard anti-sunward-directed particles. However, Solar Orbiter, close to Earth's location, measured unusual sunward-directed fluxes for the first arriving particles. 

The detection of predominantly sunward-propagating beams is relatively uncommon. This sunward particle flux might be related to a source located beyond the observer (i.e. a connection to an IP shock) or to a particular interplanetary magnetic field (IMF) configuration (i.e. folded magnetic lines or a close structure, such as an IP flux rope). For example, over the STEREO mission until 2017, only six SEP events were found with dominant sunward particle fluxes \citep{2017Gomez_Herrero}. Over the Solar Orbiter mission, from a survey of approximately 300 solar energetic electron (SEE) events observed from November 2020 until December 2022 by EPD, as listed by Warmuth et al. in prep., only this SEP event on 2022 January 20 presents clear sunward electron fluxes.  We note however that the survey of Warmuth et al. in prep.  is based on an unambiguous association between flare and SEE events, which might not favour events injected into IP clouds. When including energetic protons, the SEP event on 2022 February 15 analysed by \cite{2024Wein} also shows clear sunward fluxes, but these are not related to a solar origin. 

In an attempt to shed some light on which physical mechanisms are behind the unusual
sunward-directed particles observed by Solar Orbiter, the present
study has been designed to meet two main objectives: (1) to identify the solar source of this widespread SEP event, and (2) to investigate the acceleration and propagation conditions that could affect the observed SEP properties at the different but closely spaced observers, in particular at the location of Solar Orbiter. The paper is structured as follows.  The instrumentation used in this study is introduced in Sect.~\ref{sec:Instrumentation}.  A summary of the SEP event observations and analysis is provided in Sect.~\ref{sec:particle_observation_analysis}. We include the remote-sensing observations and data analysis of the SEP parent solar source in Sect.~\ref{sec:solar_parent_activity}.  A detailed analysis of the ICME present in the heliosphere at the time of the particle release is shown in Sect.~\ref{sec:Magnetic_cloud_16}. Section~\ref{sec:interplanetary_propagation} traces the interplanetary propagation of the particles within the ICME. In Sect.~\ref{sec:summary_discussion}, we summarise and discuss the main findings of the present study and in Sect. \ref{sec:Conclusions} we outline our main conclusions. 

\begin{table*}
\caption{Magnetic connectivity between spacecraft and the Sun at 05:58 UT on 2022 January 20.} 

\label{table:ADAPT-WSA connectivity}      
\centering          
\begin{tabular}{l cccc|cccccHH}     
\toprule
(1) & (2) & (3) & (4) & (5) & (6) & (7) & (8) & (9) & (10)& (11) & (12)\\
  & & & & & \multicolumn{2}{c}{Parker}&\multicolumn{3}{c}{PFSS$^{(a)}$}&&\\ 
  Spacecraft & r & Lon.$^{(b)}$ & Lat.$^{(b)}$ & V$_{obs}$ & Lon.$^{(b)}$ & Lat.$^{(b)}$ &Lon.$^{(b)}$ & Lat.$^{(b)}$ & Polarity  & Lon.$^{(b)}$ & Lat.$^{(b)}$ \\
    &  (au) & ($^{\circ}$) & ($^{\circ}$) & (km s$^{-1}$)  & ($^{\circ}$) & ($^{\circ}$) & ($^{\circ}$) & ($^{\circ}$)& (O, M) & ($^{\circ}$) & ($^{\circ}$) \\
  \hline
  Flare & --- & 325 & 8 & --- & --- & --- & --- & --- & ---& --- & --- \\
  L1 & 0.98 & 249.1 & $-$5.0 & 480 & 299.9&-12.0&317.2& $-$17.2 & (-1, -1) & 287.1 & -4.1  \\
  Solar Orbiter  & 0.92 & 232.2 & $-$1.4 & 510  &277.4&-8.4& 315.6 & $-$16.5 & (+1, -1) & 280.2 & -0.4 \\
  STEREO-A  & 0.96 & 214.4 & $-$1.2 & 357 &281.8 &-8.2&316.0 & $-$16.6 & (-1, -1) & 270.7 & -0.3 \\
\bottomrule
\end{tabular}
\tablefoot{Columns 1--4 present the respective spacecraft and its location in Carrington coordinates (the first row provides the flare location). Column 5 lists the measured solar wind speed (one--hour--averaged at the SEP onset), 6--7 and 8--9 respectively provide the backmapped magnetic footpoints of the observer at 2.5 R$_\sun$ and at the solar surface using the PFSS model. Column 10 gives the observed (O) and modelled (M) polarity.
\tablefoottext{a}{PFSS footpoints at 1~R$_\sun$;}
\tablefoottext{b}{Longitude and latitude values are given in the Carrington coordinate system.}

}
\end{table*}

\section{Instrumentation} 
\label{sec:Instrumentation}

The study of the wide spread of particles and the relation with the parent solar source requires the analysis of both remote-sensing and in situ data from a wide range of instrumentation on board different spacecraft. We used data from Solar Orbiter, PSP, MAVEN, STEREO, SOHO, Wind  \citep{1997Ogilvie}, the Advanced Composition Explorer \citep[ACE;][]{Stone1998ACE}, the Solar Dynamics Observatory \citep[SDO;][]{Pesnell2012}, the Geostationary Operational Environmental Satellites (GOES), and the \textit{Fermi} spacecraft.  

Remote-sensing observations of CMEs and related solar activity phenomena were provided by the Atmospheric Imaging Assembly \citep[AIA;][]{Lemen2012} on board SDO, the C2 and C3 coronagraphs of the Large Angle and Spectrometric COronagraph \citep[LASCO;][]{Brueckner1995} instrument on board SOHO, and the Sun Earth Connection Coronal and Heliospheric Investigation \citep[SECCHI;][]{Howard2008SECCHI} instrument suite on board STEREO-A. In particular, we used imaging data from the COR1 and COR2 coronagraphs and the Extreme Ultraviolet Imager \citep[EUVI;][]{Wuelser2004}, which are part of the SECCHI suite. We used STEREO-Heliospheric Imager \citep[HI;][]{2009Eyles} data to track the evolution of the CME in the heliosphere. Radio observations were provided by the Radio and Plasma Wave Investigation \citep[SWAVES;][]{Bougeret2008_S/WAVES} instrument on board the STEREO mission, the YAMAGAWA solar radio spectrograph \citep{2017Iwai}, and the e-Callisto network, in particular data from the Astronomical Society of South Australia \citep[ASSA;][]{2009Benz}. The solar flare is primarily studied with X-ray observations provided by the Spectrometer/Telescope for Imaging X-rays \citep[STIX;][]{Krucker2020} on board Solar Orbiter, the Gamma-ray Burst Monitor \citep[GBM;][]{2009Meegan} on board the \textit{Fermi} spacecraft, and the soft X-ray Sensor \citep[XRS;][]{Garcia1994} on board GOES\footnote{\url{https://satdat.ngdc.noaa.gov/sem/goes/data/avg/}\label{goes_data}}. 

The properties of energetic particles near 1 au were measured by the SupraThermal Electrons and Protons (STEP) instrument, the Electron Proton Telescope (EPT), the High Energy Telescope (HET), and Suprathermal Ion Spectrograph (SIS) of the Energetic Particle Detector \citep[EPD;][]{Rodriguez-Pacheco2020} instrument suite on board Solar Orbiter. We also used the Solar Electron and Proton Telescope \citep[SEPT,][]{Mueller2008SEPT}, the Low-Energy Telescope  \citep[LET,][]{Mewaldt2008LET}, and the High-Energy Telescope \citep[HET,][]{vonRosenvinge2008HET}, and the Suprathermal Ion Telescope \citep[SIT,][]{1998Mason} on board STEREO \citep[all of them part of the IMPACT instrument suite,][]{Luhmann2008IMPACT}; the Electron Proton and Alpha Monitor \citep[EPAM,][]{Gold1998EPAM}, and the Ultra-Low Energy Isotope spectrometer \citep[ULEIS][]{2008Mason} on board ACE; the Electron Proton Helium INstrument (EPHIN), part of the Comprehensive Suprathermal and Energetic Particle Analyzer \citep[COSTEP,][]{Mueller-Mellin1995COSTEP} and the Energetic Relativistic Nuclei and Electron Instrument \citep[ERNE,][]{Torsti1995ERNE} on board SOHO; and the 3D Plasma and Energetic Particle Investigation \citep[3DP;][]{1995Lin3DPWind} on board Wind. SEP observations within 1~au were provided by the Integrated Science Investigation of the Sun \citep[IS$\odot$IS;][]{McComas2016} suite on board PSP. Low-energy  electrons and ions are covered by the Energetic Particle Instrument-Low \citep[EPI-Lo;][]{Hill2017}, while high-energy particles are measured by the Energetic Particle Instrument-High \citep[EPI-Hi;][]{Wiedenbeck2017}. SEP data beyond 1~au were measured by the Solar Energetic Particle \citep[SEP;][]{Larson2015} instrument on board MAVEN.

Solar wind plasma and magnetic field observations were provided by the Magnetometer \citep[MAG;][]{2020Horbury} and the Solar Wind Analyzer \citep[SWA;
][]{2020Owen} on board Solar Orbiter. We used the Electron Analyser System (EAS),  part of the SWA instrument, to measure the pitch-angle distribution of the suprathermal electrons.  We also used the Plasma and Suprathermal Ion Composition \citep[PLASTIC;][]{Galvin2008} investigation and the Magnetic Field Experiment \citep[MAG;][]{Acuna2008} on board STEREO; and the Magnetic Fields Investigation \citep[MFI;][]{1995Lepping} as well as the Solar Wind Experiment \citep[SWE;][]{1995Ogilvie} on board Wind. Magnetograms from the Global Oscillations Network Group \citep[GONG;][]{Harvey1996GONG} are available from the National Solar Observatory website\footnote{\url{https://gong.nso.edu/data/magmap/index.html}}.

\begin{figure}[htb]
\centering
  \resizebox{1.0\hsize}{!}{\includegraphics{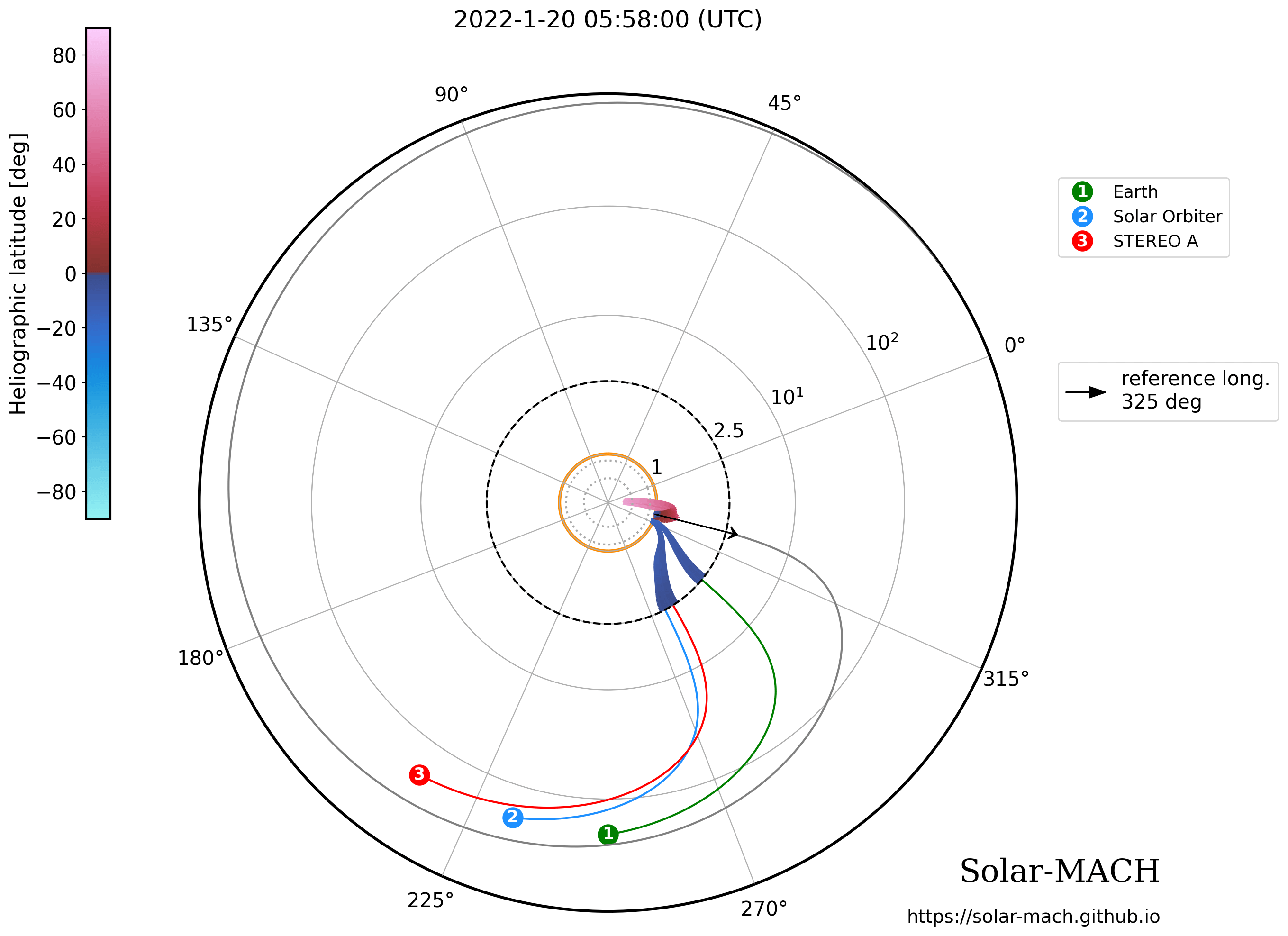}}
     \caption{Semi-logarithmic representation of the spacecraft constellation in the Carrington coordinate system at 05:58~UT on 2022 January 20. The orange circle at the centre indicates the Sun and the black arrow corresponds to the flare location. Colour-coded solid circles mark the various spacecraft of the constellation, and the lines connected to them represent the nominal Parker spiral solutions calculated using their heliocentric distances and the observed solar wind speeds. The potential field source surface (at 2.5~\(R_\odot\) in this case), which is the outer boundary of the potential-field model, is shown with the dashed circle. Below the source surface the magnetic field lines are extrapolated using a PFSS model, where the colour of the lines corresponds to heliospheric latitude. The reddish closed lines around the flare location are also given by the PFSS model. Below the source surface the plot is linear and above it is logarithmic in distance.  }
     \label{fig:connectivity}
\end{figure}

\section{Solar energetic particle event on 2022 January 20: In situ observations and analysis near 1 au}
\label{sec:particle_observation_analysis}

We summarise here the particle observations and analysis of the SEP event on 2022 January 20.  As shown in Fig.~\ref{fig:solar_mach} right, Solar Orbiter, STEREO-A, near-Earth spacecraft, and MAVEN (only electrons) observed the SEP event. The periodic decrease observed in the MAVEN electron data is due to its elliptical orbit. The dips occur when going into and out of the periapsis. PSP, regardless of the data gaps during the observing time, shows a late increase observed in the middle of January 21. However, the enhancement at PSP may be related to an eruption that originated in the vicinity of AR 12934 (close to the southeastern limb as seen from Earth) around $\sim$08:30 UT on January 21.  Thus, based on the available observations, it can be argued that the SEPs accelerated by this solar event resulted in a particle spread over at least 160$^{\circ}$ around the Sun, from STEREO-A to MAVEN. 

The right panel of Fig.~\ref{fig:solar_mach} also shows how the event features, such as intensity-time profiles, onset times, and peak intensities vary across the different observers.  MAVEN observed very gradually growing electron fluxes with a small increase, compatible with its large connection angle to the source region. We focus in this study on the near-1 au observations of the SEP event, namely Solar Orbiter, near-Earth spacecraft, and STEREO-A, which are well connected to the solar source. Therefore PSP and MAVEN data are not included in the detailed study shown below. 

This SEP event is named as SEP-C25-0019 by \cite{2024Dresing} who compiled a list\footnote{\url{https://zenodo.org/records/11280649}} of 45 multi-spacecraft SEP events observed during solar cycle 25. Based on $\sim$1 MeV electron and $\geq$25 MeV proton peak intensities, the 2022 January 20 is respectively in position 22 and 20 of the list.  
 
\subsection{Magnetic connectivity}
\label{sec:connectivity}

A fundamental parameter for interpreting the SEP event profiles at different locations is the longitudinal separation between the solar source and the footpoint of the IMF lines connecting to the respective observer. The location and magnetic connectivity of the different spacecraft around the estimated time of the soft-X ray peak of the flare ($\sim$05:58 UT) is shown in Fig.~\ref{fig:connectivity} and detailed in Table \ref{table:ADAPT-WSA connectivity}, where Cols. 2--4 present the spacecraft locations at the time of the soft-X ray peak of the flare.  

 We note that the determination of the magnetic connectivity presented here is based on the assumption of nominal IP magnetic field lines following the shape of a Parker spiral, from which magnetic field lines are tracked downwards to the photosphere using the Potential Field Source Surface \citep[PFSS;][]{Schatten1969,Altschuler1969,WangSheeley1992} model. As explained later this assumption is likely not valid for Solar Orbiter during the SEP event.  Figure~\ref{fig:connectivity} shows the instantaneous connectivity derived with the PFSS coronal field solution. The corresponding footpoint connectivity is listed in Cols.~6--7 of Table~\ref{table:ADAPT-WSA connectivity} and the observed solar wind speed that is used to calculate the Parker spiral is shown in Col.~5.  Columns 8--9 of Table~\ref{table:ADAPT-WSA connectivity} shows the magnetic connection points from the various spacecraft to the photosphere. Based on Cols. ~8--9, the connectivity of near-Earth, Solar Orbiter, and STEREO-A to the solar surface is very close, $\sim$316$^{\circ}$ longitude and ${\sim}$$-$16$^{\circ}$ latitude. Then, the difference with the solar flare region is of $\sim$9$^{\circ}$ in longitude and $\sim$24$^{\circ}$ in latitude for the three aforementioned spacecraft. Column 10 shows the magnetic field polarity observed (O) and modelled (M), indicating a good agreement between the Parker spiral--PFSS model and observations except for Solar Orbiter. This discrepancy might be related to the IP structure present at Solar Orbiter at the time of the SEP event, which is not considered in the whole back-mapping process described above. We note that the observed magnetic polarity is derived from the magnetic field vector observed in situ by the corresponding spacecraft, being positive (+1) for outward IMF and negative (-1) for inward IMF.

\begin{figure}[htb]
\centering
  \resizebox{0.8\hsize}{!}{\includegraphics{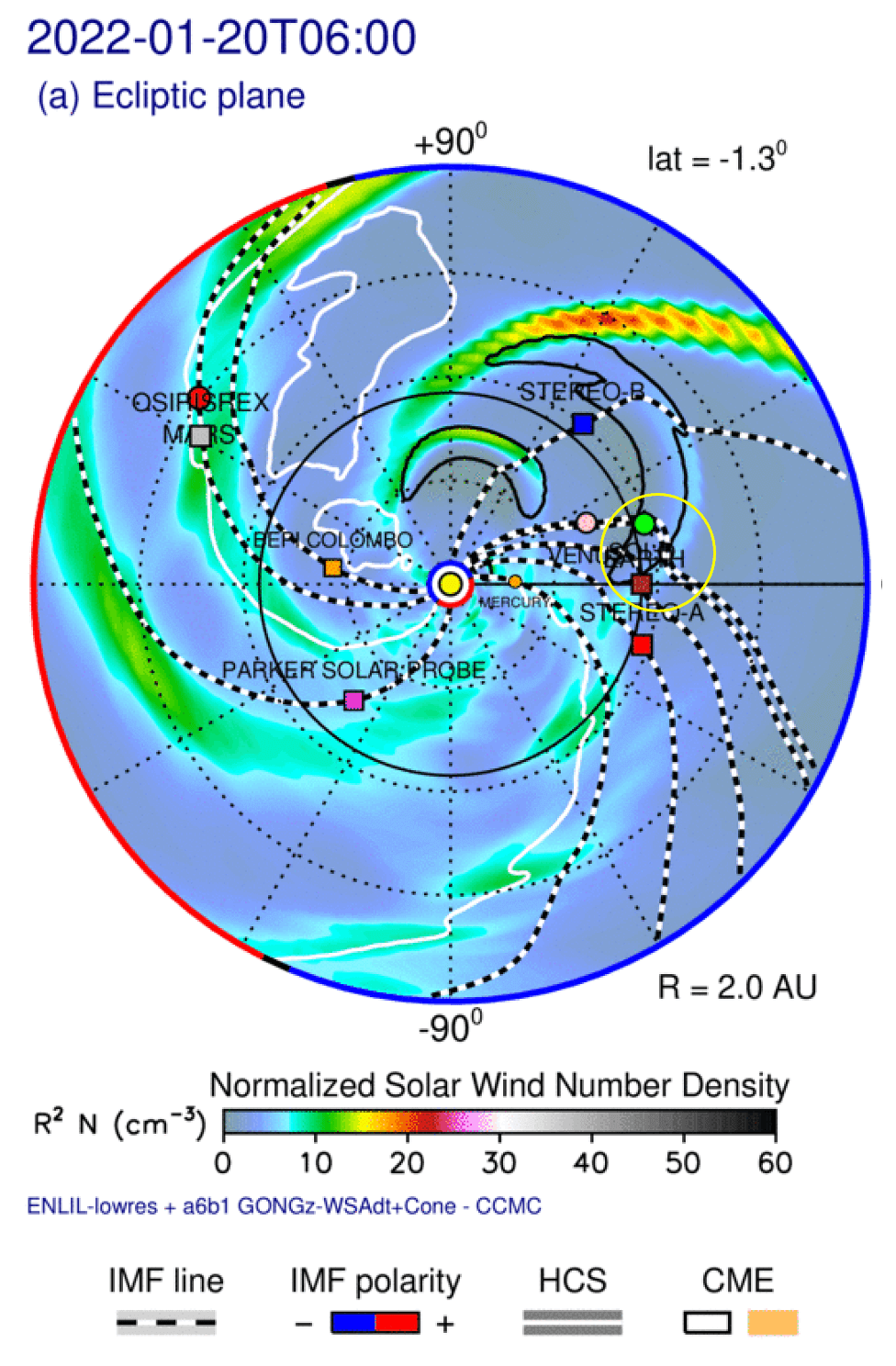}}
     \caption{Snapshot of the radially scaled solar wind density from the ENLIL simulation in the ecliptic plane at 06:00 UT on 2022 January 20. The black and white dashed lines represent the IMF lines and the black contours track the ICMEs. The white lines correspond to the HCS, which separates the regions with opposite magnetic polarity, shown in blue (negative) or red (positive) on the outer edge of the simulation region. The yellow circle indicates the flank arrival of an ICME to Solar Orbiter (details given in the text). Credit: Community Coordinated Modeling Center (CCMC). }
     \label{fig:enlil}
\end{figure}


To further investigate the impact of previous CMEs in the magnetic connectivity of the different observers we performed a detailed simulation of the IP conditions during the event. Figure~\ref{fig:enlil} shows a snapshot of the solar wind density in the WSA-ENLIL+Cone model \citep[hereafter ENLIL model;][]{Odstrcil2004} simulation around the SEP onset time on 2022 January 20 at 06:00~UT. We describe in detail the input parameters for the ENLIL model and the link to the online simulation in Appendix~\ref{append:ENLIL_model}.
The black contours track the ejecta of the ICME. They are manifested in the simulation as coherent and outward propagating high-density regions. The black and white dashed lines represent the IMF lines connecting the Sun with the various observer positions. The simulation shows several stream interaction regions present near Solar Orbiter, Earth, STEREO-A, and Mars at the time of the onset of the particle event that might modify the magnetic connectivity and SEP propagation conditions. However the connectivity given by the ENLIL model is similar to the one given by the nominal Parker Spiral. According to ENLIL, there is a wide ICME just leaving behind the Earth environment but reaching Solar Orbiter through its eastern flank during the SEP event on January 20 (indicated with a yellow circle in Fig.~\ref{fig:enlil}). This ICME, which is studied in detail in Sect.~\ref{sec:Magnetic_cloud_16}, was ejected on 2022 January 16 from the same AR as the one related to the SEP event on January 20.
Results of the ENLIL simulation are also presented in the six bottom panels of Figs. \ref{fig:Earth_solo_part_plasma} and \ref{fig:STA_particles_plasma} that show the in-situ plasma and magnetic field data (discussed in the following sections) over-plotted with the pink line showing the result of the ENLIL simulation from mid 18 January to 22 January. The pink dashed lines represent the ENLIL simulation results without including the CMEs. As discussed in the following sections, ENLIL follows the general trend of the measured solar wind speed at the locations of Solar Orbiter and near-Earth locations.

\subsection{Solar energetic particle observations and interplanetary context}
\label{sec:in situ}

The heliospheric conditions in which particles propagate at the time of release may affect the SEP timing and intensity profiles \citep[e.g.][]{Laitinen2013, Dalla2020, Lario2022}. We used both multi-point solar wind and IMF observations as well as the results of the ENLIL model presented above to provide a comprehensive understanding of the interplanetary structures and their possible influence on the propagation of the SEPs. 
In this section, we discuss multi-spacecraft SEP observations as well as in situ plasma and  magnetic field. To classify the different in situ signatures within an ICME, we considered the following criteria. We defined the ICME start with the IP shock, followed by the sheath region and by the magnetic obstacle (MO). Within the MO, the core of the structure---that is, the MC---is restricted to periods where the following features are shown: (1) an increase in the magnetic field strength, (2) a monotonic magnetic field rotation (flux rope) resulting in large net rotation of at least one of the magnetic field components, (3) low proton temperature, and (4) plasma $\beta$ below 1 \citep{Burlaga1981}.

\begin{figure*}[htb]
\centering
  \resizebox{1.0\hsize}{!}{\includegraphics{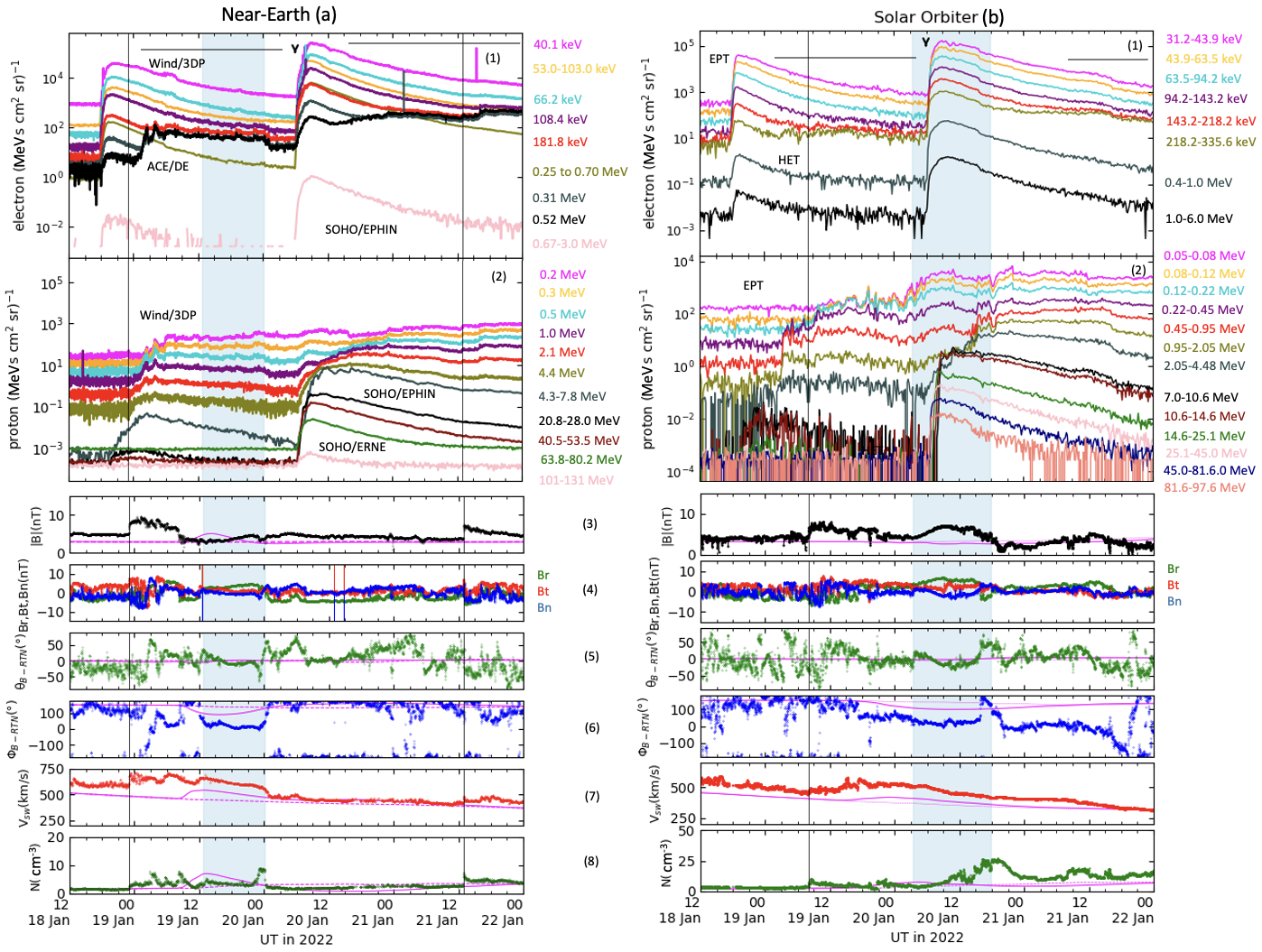}}
     \caption{In situ SEP time profiles and plasma and magnetic field observations by near-Earth spacecraft (a), and Solar Orbiter (b). Top: Energetic electron (1) and proton (2) temporal profiles.  The arrow in the top panels indicates the flare peak time. Bottom: In situ plasma and magnetic field observations. The panels present, from top to bottom, the magnetic field magnitude (3), the magnetic field components (4), the magnetic field latitudinal (5) and azimuthal (6) angles, $\theta$\textsubscript{B-RTN} and $\phi$\textsubscript{B-RTN}, the solar wind speed (7), and the proton density (8), where RTN stands for radial-tangential-normal coordinates \cite[e.g.][]{Hapgood1992}. Solid vertical lines indicate IP shocks, blue shaded areas indicate MC. The horizontal line on the top panel indicates a period of proton contamination. The pink lines represent the ENLIL simulation results. }
     \label{fig:Earth_solo_part_plasma}
\end{figure*}

\subsubsection{Solar energetic particle observations and interplanetary context: Earth}
\label{subsec:Interplanetary_context_Earth}
Figure~\ref{fig:Earth_solo_part_plasma}a shows the particle, plasma, and magnetic field observations by near-Earth spacecraft. The first two panels show the SEP event on January 20 using the Sun-directed telescopes of the different instruments. The onset of the solar flare is indicated with an arrow head at the top of the panel 1, which shows a fast rise of energetic electrons that reach energies of at least 0.7~MeV. The proton intensity-time profile (panel 2) also shows a fast intensity increase for ion energies above 4 MeV up to 100 MeV, showing a gradual increase for the lower ion energies. Electrons and protons arrived to the spacecraft from the Sun at pitch-angle 180 (inwards polarity), as discussed in Sect.~\ref{sec:Earth_anisot}. Near-Earth spacecraft also observed a prior SEP event on January 18 followed by a shock-driving ICME (indicated by the first vertical line and blue shading) arriving  just before the onset of the January 20 SEP event, as described below. A second IP shock indicated by a second vertical line at 12:56 UT on January 21 locally accelerated low-energy protons ($\lesssim$4 MeV). There are periods when the protons of $\sim$500 keV (blue line in panel 2) contaminate the electron channels ($\sim$200--500 keV), as observed before and during the passage of the ICME and before and after the second IP shock (second vertical line). The possible contaminated periods are indicated with a horizontal line in panel 1.  
 
The solar wind speed at the onset of the particle event is ${\sim}480$~km~s\textsuperscript{-1}, as shown in panel 7. At the time of the SEP event, an ICME (blue shading) had recently left the near-Earth environment and did not appear to affect the particle propagation at this location. This ICME is likely the same structure arriving later at Solar Orbiter, as discussed in Sect.~\ref{subsec:Interplanetary context_Solar Orbiter}. The ICME starts with the arrival of an IP shock (first vertical line) at 22:57 UT on January 18. At this time we observe a simultaneous increase in the magnetic field magnitude (panel 3) and solar wind speed (panel 7), followed by a region of increased magnetic field and large fluctuations in the orientation of the magnetic field, which corresponds to the sheath region. After the sheath, we observe a region of coherent magnetic field rotation indicated with the blue shaded area, starting at 12:40 UT on January 19 lasting until 00:11 UT on January 20. The general trend of the solar wind speed is well simulated by ENLIL, which predicts the IP shock arrival within the model time uncertainties \citep{Wold2018}. 

 \begin{figure}[htb]
\centering
  \resizebox{1.0\hsize}{!}{\includegraphics{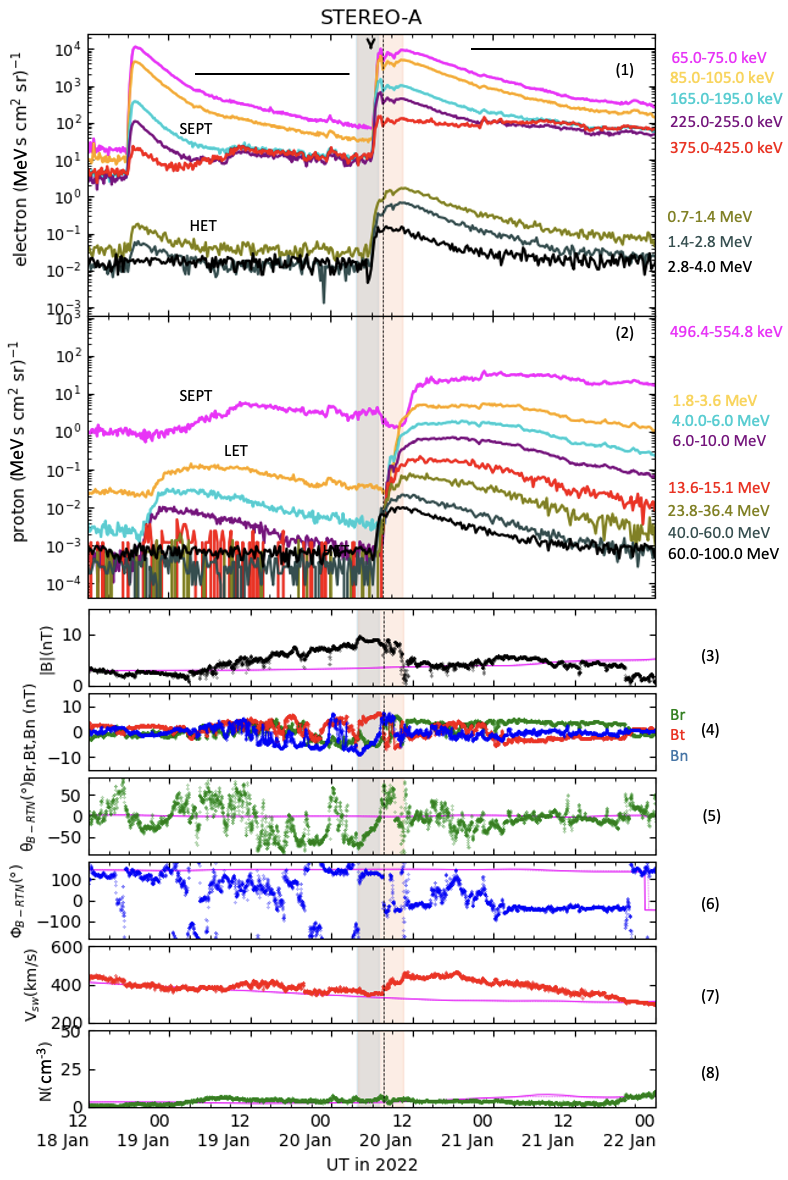}}
     \caption{In situ SEP time profiles, and plasma and magnetic field observations by  STEREO-A. Panels and IP structures are as in Fig. \ref{fig:Earth_solo_part_plasma}. The salmon-shaded area indicate an SIR. The vertical dashed line indicates the SI within the SIR. The grey shaded area shows a flux-rope embedded in an SIR. }
     \label{fig:STA_particles_plasma}
\end{figure}

\subsubsection{Solar energetic particle observations and interplanetary context: Solar Orbiter}
\label{subsec:Interplanetary context_Solar Orbiter}

Panels 1 and 2 of Fig.~\ref{fig:Earth_solo_part_plasma}b show the SEP event observed by Solar Orbiter. The data correspond to the particles measured by the anti-sunward looking telescopes, which measured the earliest onsets and highest intensities, as shown in Figs. \ref{fig:anisotro_elec} and \ref{fig:anistrop_protons} 2 and discussed in Sect.~\ref{sec:solar_orbiter_aniso}. The electron event (panel 1) is observed to reach energies above $\sim$1~MeV, showing a fast intensity increase in both EPT and HET measurements. 
The energetic ion observations (panel 2) by Solar Orbiter/EPD/HET show a clear energy dispersion, reaching energies up to ${\sim}80$~MeV, used for the velocity dispersion analysis (VDA) discussed in Sect.~\ref{sec:Particle_timing_solo}.
The intervening MC present at the time of the SEP onset (presented below) likely played a major role in the observed electron and proton anisotropies, as described in Sect.~\ref{sec:solar_orbiter_aniso}.
Solar Orbiter also measured the previous SEP event on January 18, as can be seen in panel 1 of Fig.~\ref{fig:Earth_solo_part_plasma}b. 
The lower energy EPT protons ($\lesssim$7 MeV) showed an increase prior to the SEP event on January 20, probably related to the arrival of an ICME, as discussed below. This increase might also affect the onset time determination and therefore these channels were not included in the VDA analysis, as discussed in Sect.~\ref{sec:Particle_timing_solo}. The ion contamination is also present in the decay phase of both events (January 18 and 20), indicated with the horizontal lines in panel 1, clearly visible in the high energy electrons ($\gtrsim$218 keV).

The solar wind speed at the time of the SEP event onset is $\sim$510~km~s\textsuperscript{-1}, as shown in panel 7, measured by the SWA instrument, which is fairly reproduced by ENLIL (pink line). As shown by the magnetic field and plasma data in panels 3--8, using MAG and SWA, the SEP event onset takes place during the passage of an ICME at Solar Orbiter. An IP shock is impacting the spacecraft at 08:02~UT on January 19, which is indicated by the vertical line in Fig. \ref{fig:Earth_solo_part_plasma}b.  An MC arrives at 03:28~UT on 20 January, just before the particle onset, being observed until 17:52~UT (blue shaded area). The energetic electrons and the higher energetic ion particles ($\gtrsim$7 MeV) are propagating inside the ICME, which, however, seems to have a modulation effect on the flux of ions ${\lesssim}5$~MeV.

\subsubsection{Solar energetic particle observations and interplanetary context: STEREO-A}

\label{subsec:Interplanetary_context_STEREO-A}

Observations of the SEP event at STEREO-A are shown in Fig.~\ref{fig:STA_particles_plasma}. STEREO-A observed the earliest increases in the north telescope and the highest intensities in the south telescope, as presented in Sect.~\ref{sec:STEREO-A_aniso}. However, after the solar superior conjunction of the STEREO spacecraft (from January to August 2015) until the approach to the Earth in August 2023, the STEREO-A spacecraft was rolled 180 degrees about the spacecraft--Sun line in order to allow the high-gain antenna to remain pointing at Earth. Consequently, the nominal pointing directions of the SEP suite of instruments were different from what was originally intended, and therefore we used omnidirectional fluxes in the plot. 

With a fast intensity increase, a clear electron and proton event is observed up to ${\sim}3$~MeV energies (panel 1) and ${\sim}60$~MeV (panel 2), respectively, where clear velocity dispersion is also observed. The prior SEP event that occurred on 2022 January 18 is also measured by STEREO-A, whose background might affect the determination of the onset times, discussed in the VDA in Sect.~\ref{sec:VDA_STEREO-A}. Just before the January 20 SEP event, the low-energy ions ($\sim$500 keV) show a small increase that coincides with the arrival of an MO (grey shaded area) and a stream interaction region (SIR, shaded in salmon colour) as detailed below. Both structures are present at the time of the flare peak time (arrow in top panel). The ion contamination is also present in the decay phase of both events (January 18 and 20), indicated with the horizontal lines in panel 1, clearly visible in the high energy electrons ($\gtrsim$165 keV). 

As shown by the magnetic field and plasma data in panels 3--8 in Fig. \ref{fig:STA_particles_plasma}, the SEP event onset at STEREO-A takes place during the passage of an MO from 03:45 UT to 07:00 UT on January 20 (grey shaded area) embedded in an SIR (salmon area). From 03:54 UT to 10:38 UT on January 20, the speed rises from ${\sim}400$ to ${\sim}500$~km~s\textsuperscript{-1}; sudden changes of the magnetic field polarity close to the stream interface (SI; dashed vertical line), and drops in the magnetic field strength together with temperature increases (not shown), which suggests that local reconnections are occurring. The SI is indicated with the dashed line, which coincides with the maximum total pressure (not shown). The solar wind speed at the time of the SEP event onset is $\sim$357~km~s\textsuperscript{-1}, as shown in panel 7, fairly simulated by ENLIL. 

\begin{figure*}[htb]
\centering
  \resizebox{1.0\hsize}{!}{\includegraphics{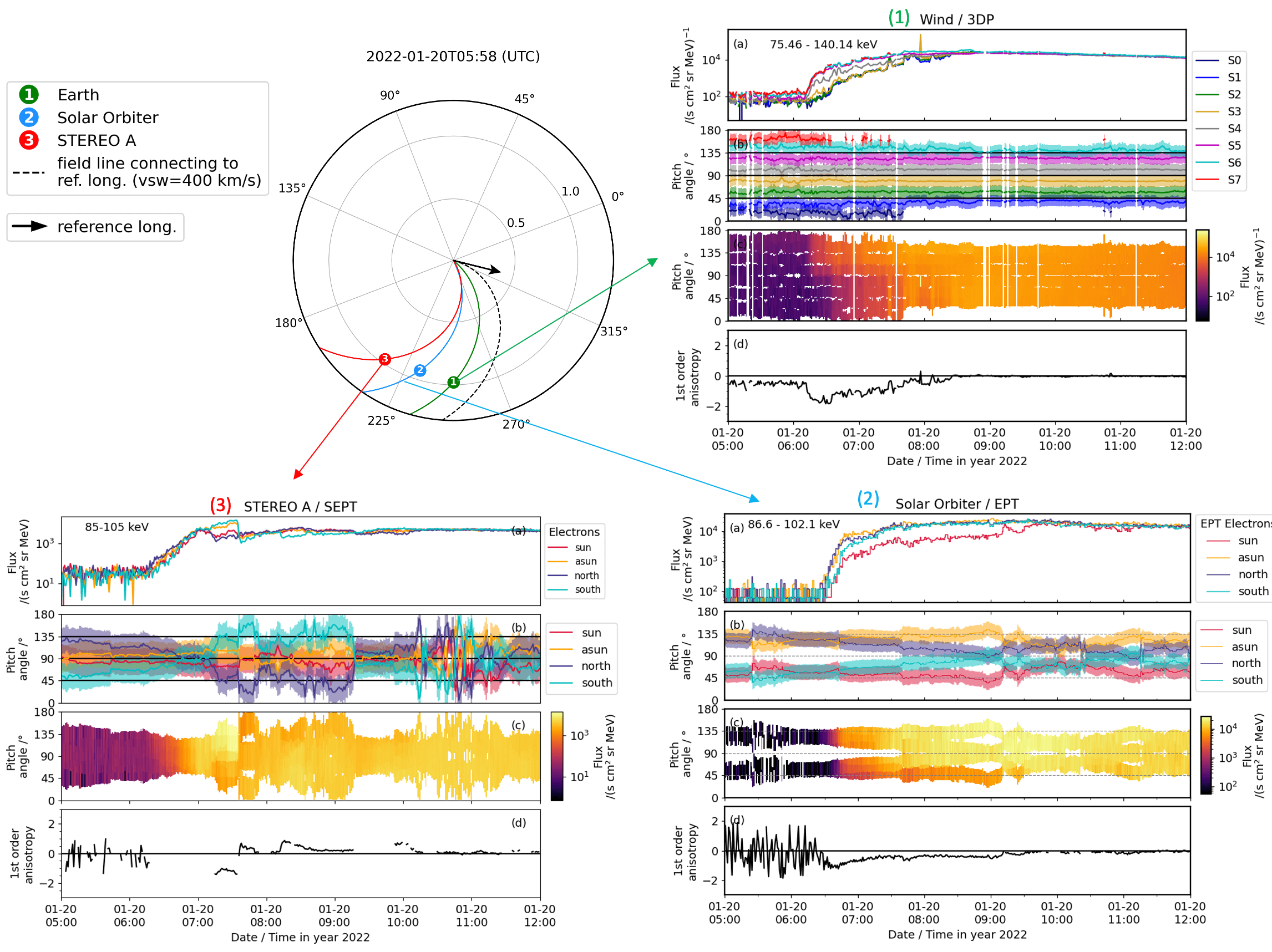}}
     \caption{Pitch-angle distribution of electrons measured by Wind/3DP/75-140 keV (\textit{1}), Solar Orbiter/EPT/86-102 keV (\textit{2}), and STEREO-A/SEPT/85-125 KeV (\textit{3}). Panel descriptions:   \textit{(a)}: Intensities observed by the each of the eight pitch-angle bins of the Wind/3DP (1); and of the centre of the four telescopes of Solar Orbiter/EPT (2) and STEREO/SEPT (3) (Sun in red, anti-Sun in orange, north in blue, and south in green). \textit{(b)}: Pitch-angle coverage of each field of view; \textit{(c)}: Pitch-angle distribution with colour-coded intensities. \textit{(d)}: first-order anisotropy values, in the range [-3, 3] \citep[e.g.][]{Dresing2014}.  The top left panel shows the longitudinal spacecraft constellation and nominal connectivity at 05:58 UT on 2022 January 20.}   
     \label{fig:anisotro_elec}
\end{figure*}

\begin{figure*}[htb]
\centering
  \resizebox{1.0\hsize}{!}{\includegraphics{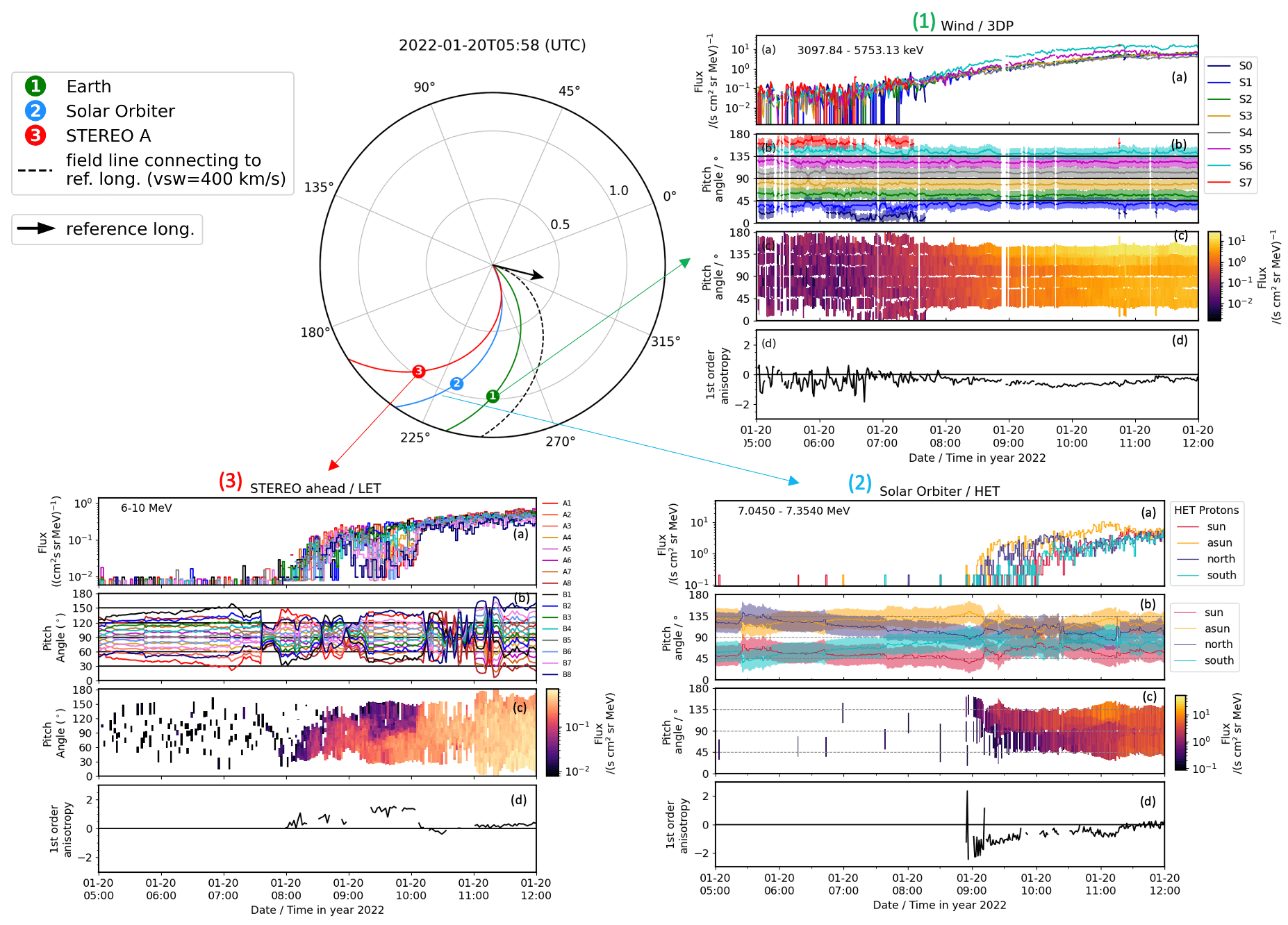}}
     \caption{Pitch-angle distribution of protons measured by Wind/3DP/3MeV  (\textit{1}), Solar Orbiter/HET/7 MeV (\textit{2}), and STEREO-A/LET/6 MeV (\textit{3}). Panel descriptions are the same as those of Fig. \ref{fig:anisotro_elec}. Panel 3 shows the 16 sectors of STEREO-A/LET, eight front-side (reddish colours) and eight back-side sectors (bluish colours).} 
     \label{fig:anistrop_protons}
\end{figure*}

\subsection{Solar energetic particle pitch-angle distributions and first arriving particles near 1 au}
\label{sec:anistropies}
In this section we study the pitch-angle distribution (PAD) of the three spacecraft, namely Solar Orbiter, STEREO-A, and Wind, which all have energetic particle anisotropy information. We used the four apertures of the three-axis stabilised Solar Orbiter and STEREO-A spacecraft, namely EPD/EPT, EPD/HET, STEREO-A/SEPT, and STEREO-A/LET \citep[e.g.][]{Dresing2014, Gomez-Herrero2021}. The coverage of pitch angles by the four apertures of EPD and STEREO-A depends on the orientation of the magnetic field with respect to the telescopes. However, Wind is a spin-stabilised spacecraft, which allows the 3DP instrument to scan different regions of the sky and thus infer a more complete estimate of the 3D particle distribution.  

\subsubsection{Solar energetic particle pitch-angle distributions: Earth}
\label{sec:Earth_anisot}
Particle intensities measured by Wind/3DP are stored into eight pitch-angle bins (sector 0 to sector 7). The panel 1 of Fig.~\ref{fig:anisotro_elec} shows the electron PAD observed by Wind/3DP at ${\sim}75$--140~keV, which shows clear anisotropies during the onset of the SEP event for about two hours until ${\sim}$07:30 UT.  Panel a shows the intensities measured by the eight sectors of the instrument.  During the onset of the SEP event the sectors measuring particles coming from the Sun presented higher intensities, covering pitch-angles from ${\sim}$135$^{\circ}$ to ${\sim}$180$^{\circ}$, as shown in panel b, which shows the pitch-angles of each of the centre of the sectors. Panel c shows the colour-coded PAD intensity. The plot shows a discontinuity at pitch-angle ${\sim}$90$^{\circ}$, where the two hemispheres of pitch-angle seem to be separated. This pitch-angle discontinuity is discussed further in Appendix~\ref{app:pitch-angle space}. The first-order anisotropy  $A$ \citep[e.g.][]{Dresing2014} shown in panel d is negative, corresponding to particles propagating from the Sun, as the magnetic field polarity is negative during the period (Br negative, as shown in panel 4 of Fig.~\ref{fig:Earth_solo_part_plasma}a). We note that large values of A (i.e.\ |A|$\gtrsim$2) indicate highly anisotropic 
flows of particles, whereas small values (i.e.\ |A|$\lesssim$0.2) indicate nearly isotropic
flows \citep[][]{Dresing2014}. 

The panel 1 of Fig. \ref{fig:anistrop_protons} shows that the early phase of the ${\sim}3.1$--5.7~MeV proton event is anisotropic (A$\approx$-1) for more than twelve hours (whole interval not shown), with higher fluxes in the sunward-looking telescope that corresponds to pitch angles near 180$^{\circ}$, consistent with the inward magnetic polarity, showing much longer lasting anisotropies than for electrons.

\subsubsection{Solar energetic particle pitch-angle distributions: Solar Orbiter}
\label{sec:solar_orbiter_aniso}

The four apertures of EPD/EPT and EPD/HET cover four viewing directions that are oriented along the nominal Parker spiral to the Sun and away from the Sun, to the north and to the south with some inclination \citep[Fig.~4 in][]{Rodriguez-Pacheco2020}. The panel 2 of Fig.~\ref{fig:anisotro_elec} shows the electron PAD observed by Solar Orbiter/EPT at ${\sim}87-102$~keV, which shows clear anisotropies during the onset of the SEP event at ${\sim}$06:30 until ${\sim}$09:00 UT. Panel a shows the intensities measured by the sun (red), asun (orange), north (blue), and south (green) telescopes.  During the onset of the SEP event the asun and north telescopes measured slightly higher intensities, covering pitch-angles from ${\sim}$100$^{\circ}$ to ${\sim}$140$^{\circ}$, as shown in panel b, which shows the pitch-angles of the centre of the telescopes.  Panel c shows the colour-coded intensity PAD. The plot shows a discontinuity at pitch-angle ${\sim}$60$^{\circ}$, better seen in Fig. \ref{fig:Pa_space_solo}. However, we note that the coverage around pitch-angle 50$^{\circ}$--100$^{\circ}$ is not ideal during the early phase of the event (from ${\sim}$06:30 to ${\sim}$08:00 UT). We present in more detail this discontinuity in Appendix~\ref{app:pitch-angle space}, including pitch-angle data from the STEP instrument. The anisotropy index shown in panel d is negative, corresponding to particles propagating towards the Sun, as the local magnetic vector is pointing outwards during the period (shown in panel 4 of Fig.~\ref{fig:Earth_solo_part_plasma}b). While the maximum anisotropy value at Solar Orbiter is lower than at Wind, the duration of significant electron anisotropies is about three hours, and therefore slightly longer compared to Wind observations.

The panel 2 of Fig.~\ref{fig:anistrop_protons} shows the ${\sim}$7~MeV proton intensities observed in the four telescopes of Solar Orbiter/HET. HET shows a one-and-a-half-hour anisotropic period starting shortly after 08:30 UT on January 20 (panel d). The pitch-angle coverage is similar during this period (panels b and c). The asun telescope was measuring the higher intensities, covering pitch-angle ${\sim}$140$^{\circ}$ (panel c). This means that particles propagated towards the Sun, as discussed above. 

\begin{table*}
\caption{Timing of the main solar phenomena and inferred SEP injection times t$_{\mbox{inj}}$. All times shifted to 1 au on 2022 January 20.}             
\label{table:timing}
\small
\centering          
\begin{tabular}{l l l l }     
\hline{Time (UT)}  &   Observer / Instr. & Feature & Comment \\      
At 1 au  & & & \\
 \hline
05:51 & SDO/AIA& Early phase of the eruption & Observed by 131{\AA} \& 94{\AA}\\ 

05:52 & SDO/AIA& EUV wave formation & Observed until 06:09, 373 km s$^{-1}$\\

05:52$\pm8$ min  & STEREO-A/SEPT&t$_{\mbox{inj}}$ using electron onsets & VDA (L=2.3$\pm0.5$ au)\\ 

05:54    &       \textit{Fermi}/GBM & 1st nonthermal HXR peak (>300 keV) & Several peaks until 06:00 UT \\

05:55$\pm1$ min         & SDO/AIA & EUV wave intersects western leg ICME on Jan. 16& Details given in Sect. \ref{sec:CME_January_16} \\

05:55   & SDO/AIA & Shock wave at 1.27~$R_{\odot}$& Well-formed bubble over the west limb \\

05:55   & Ground b./ ASSA\&YAM. & TIIa onset & From 250 to 100 MHz, 340 km s$^{-1}$\\

05:55   & Ground b./ ASSA\&YAM. & TIIb onset& Observed during 10 min to 16 MHz, 1400 km s$^{-1}$\\

05:55:40  & Ground b./ ASSA\&YAM. &     TIIIs & Until 05:58 UT. Partly emanating from HBs\\

05:56$\pm4$ min  & SolO/EPT\&HET &t$_{\mbox{inj}}$ using both proton and electron onsets & VDA (L=2.6$\pm0.1$ au)\\

05:56:30  & Ground b./ ASSA\&YAM.         &     HBs & Until 06:00 within 80 to 50 MHz \\

05:57    &       \textit{Fermi}/GBM & Max. nonthermal HXR peak (>300 keV) & Co-temporal with TIIb and TIIIs  \\

05:58$\pm1$ min    &     SOHO/LAS. \& STA/COR & 1st 3D shock intersection with SolO & \(\theta_\mathrm{Bn} = 84^{\circ}\)at 1.47 ~$R_{\odot}$ \\

06:00$\pm1$ min    &     SOHO/LAS. \& STA/COR & 1st 3D shock intersection w/ near-Earth \& STA  & \(\theta_\mathrm{Bn} = 82^{\circ}\)at 1.38 ~$R_{\odot}$ \& \(\theta_\mathrm{Bn} = 71^{\circ}\)at 1.39 ~$R_{\odot}$ \\

06:02$\pm4$ min  & SOHO/ERNE \& Wind/3DP&t$_{\mbox{inj}}$ using both proton and electron onsets & VDA (L=1.4$\pm0.1$ au)\\  

\hline
\end{tabular}
\end{table*} 
\subsubsection{Solar energetic particle pitch-angle distributions: STEREO-A}
\label{sec:STEREO-A_aniso}

SEPT apertures on board STEREO-A have a similar configuration to Solar Orbiter/EPT. However, since the spacecraft was put upside-down after the superior solar conjunction in 2015 until August 2023 as discussed above, the sun and asun telescopes pointed perpendicular to the nominal Parker Spiral within the ecliptic plane. The sun telescope pointed in the [$-$R, $-$T] quadrant, whereas the asun aperture pointed in the [+R, +T] quadrant. The north and south telescopes pointed opposite to the nominal configuration. 

The panel 3 of Fig.~\ref{fig:anisotro_elec} shows the electron PAD observed by STEREO-A/SEPT at ${\sim}85-105$~keV, showing a data gap in the anisotropy panel during the onset of the SEP event, as shown in panel d. Due to the peculiar configuration of STEREO-A and the orientation of the magnetic field vector during this period, the pitch-angle coverage is not appropriate to detect field-aligned particles, as seen in panel b. The anisotropy can therefore not be determined. Coinciding with an increase in the pitch-angle coverage, we observed some electron anisotropy after the onset, from ${\sim}$07:20 UT until ${\sim}$07:50 UT. During this time the asun and south telescopes measured slightly higher intensities, covering pitch-angles from ${\sim}$90$^{\circ}$ to ${\sim}$180$^{\circ}$. The anisotropy index shown in panel d turns from negative to positive at ${\sim}$07:40 UT, when Br changed from negative to positive (panel 4 in Fig.~\ref{fig:STA_particles_plasma}).

The panel 3 of Fig.~\ref{fig:anistrop_protons} shows the 6--10~MeV proton intensities observed in the 16 sectors of STEREO-A/LET, eight front-side (reddish colours)  and eight back-side sectors (bluish colours). LET measured a one-and-a-half-hour anisotropic period starting shortly after 08:00 UT on 20 January, where most of the particles are observed in the sunward-facing sectors.  
The pitch-angle coverage is stable during this period, shown in panel b, which shows the pitch-angles of the sector centres.  As for the electrons, the coverage is not ideal during the onset but sufficient to see the period where the beam has a discontinuity at pitch-angle 90$^{\circ}$ (panel c).

\subsection{Solar energetic particle timing}
\label{sec:particle_timing}
We analysed the timing of the SEP event by using the so-called velocity dispersion analysis (VDA) method, which is based on the assumption that first-observed SEPs of each energy have been injected simultaneously and propagate scatter-free and without adiabatic cooling which may cause energy changes. We note that we refer here to injection of particles as when an already accelerated source of particles becomes magnetically connected to the observer. We include details about the VDA method in Appendix \ref{sect:app_VDA}. For this event, we focused our timing analysis in the three spacecraft located near 1 au, namely Solar Orbiter, near-Earth probes, and STEREO-A, which are all well-connected to the source and show clear energy dispersion to perform VDA. The results presented below are included in Table \ref{table:timing}, which shows the timing of the inferred SEP injection times and of the solar phenomena (discussed in the following sections).  

\subsubsection{Solar energetic particle timing: Near Earth}
\label{sec:Particle_timing_Earth}
\begin{figure}[htb]
\centering
  \resizebox{1.0\hsize}{!}{\includegraphics{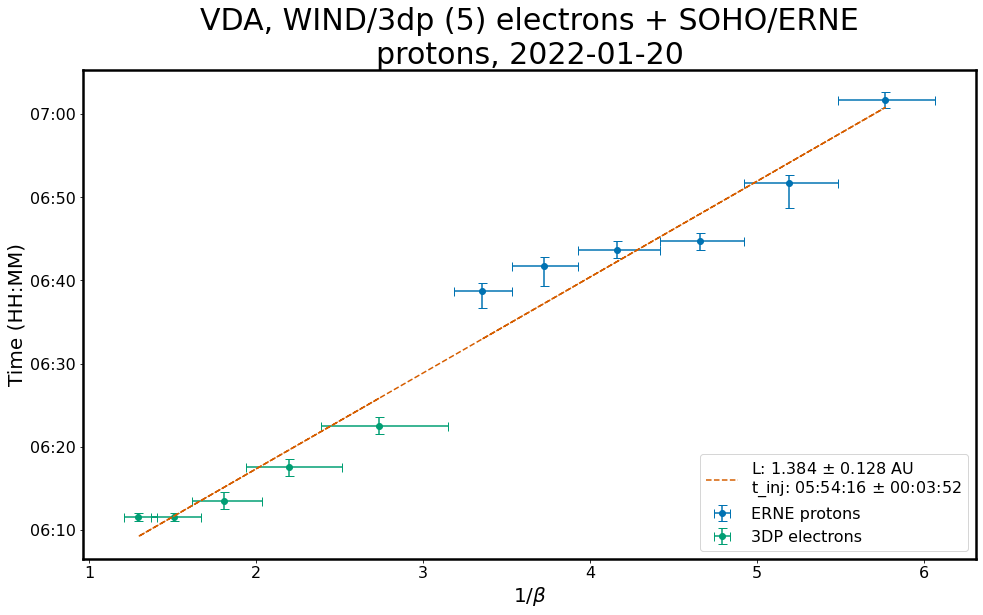}}
     \caption{Velocity dispersion analysis of the onset of the SEP event at near-Earth spacecraft. The horizontal and vertical axes correspond to the reciprocal of the particle velocities ($1/\beta$=$c/v$) and onset times, respectively. The green and blue data points respectively identify the onsets of the 3DP electron and ERNE proton at the corresponding velocities (energies), with the respective errors indicated. The dashed line is the linear regression to fit all points. The legend gives the effective path length (L) and the estimated release time (t\_inj) discussed in the main text. } 
     \label{fig:VDA_L1}
\end{figure}
To estimate the path length and infer the injection time of the particles for the near-Earth spacecraft (Wind and SOHO), we used a modified Poisson-CUSUM method that employs statistical bootstrapping \citep[e.g.][]{Huttunenheikinmaa05, Palmroos2022}. The method and the background windows used for the fitting are explained in detail in Appendix \ref{app:CUSUM_method}. To estimate the onset times we used sector 5 of the Wind/3DP instrument, which covers pitch-angles of anti-sunward propagating particles, as this sector observed the first arriving particles. The electron channels selected for the fitting are 27.84--401.3 keV, where the increase of the peak intensity over the background was at least $\times$4400. For protons we used the ERNE energy channels between 13 and 50 MeV, where velocity dispersion was observed in the onset times and the peak-to-background intensity ratios were $\times$860--3890.

The VDA results are presented in Fig.~\ref{fig:VDA_L1}. Two sets of data points represent the onset times as observed by SOHO/ERNE (protons, blue) and Wind/3DP (electrons, green). The horizontal error bars represent the width of the energy channels, and the vertical error bars represent the 95\% confidence interval of the onset times as provided by the Poisson-CUSUM-bootstrap hybrid method. A first-order polynomial is fitted to the data points with orthogonal distance regression (ODR) algorithm, and it is shown as the orange line over the points. The slope of this line (L = 1.4 $\pm$ 0.1~au) represents the effective path length travelled by the particles, which is close to the nominal Parker spiral length for near Earth ($\sim$1.08~au) using the measured solar wind speed. The intersection with the vertical axis represents the time of the particle injection $t_{inj}$ = 05:54 UT $\pm$ 4 min, or 06:02 UT $\pm$ 4 min shifted $\sim$8.2 min to compare with electromagnetic observations from 1 au. To compare, the VDA performed on ERNE (13--64 MeV) protons and the lowest electron energy channel (0.25--0.7 MeV) of EPHIN (not shown) yielded a path length of $L = 1.3 \pm 0.3$~au and an injection time of 06:00 UT $\pm$ 9 min or 06:08 UT $\pm$ 9 min shifted $\sim$8.2 min, which is consistent with results given by SOHO/ERNE + Wind/3DP.

\subsubsection{Solar energetic particle timing: Solar Orbiter}
\label{sec:Particle_timing_solo}

In the case of Solar Orbiter, we used the anti-sunward measurements of the EPT electrons (33.37--218.18 keV) and HET protons (7.045--89.46 MeV), which observed the first arriving particles. These channels were not affected by the enhanced levels of protons related to the arrival of the ICME to Solar Orbiter, as shown in Fig.~\ref{fig:Earth_solo_part_plasma}b (blue shaded area). The method used for fitting and the background window is detailed in Appendix \ref{app_VDA_Mario}. 

\begin{figure}[htb]
\centering
  \resizebox{1.0\hsize}{!}{\includegraphics{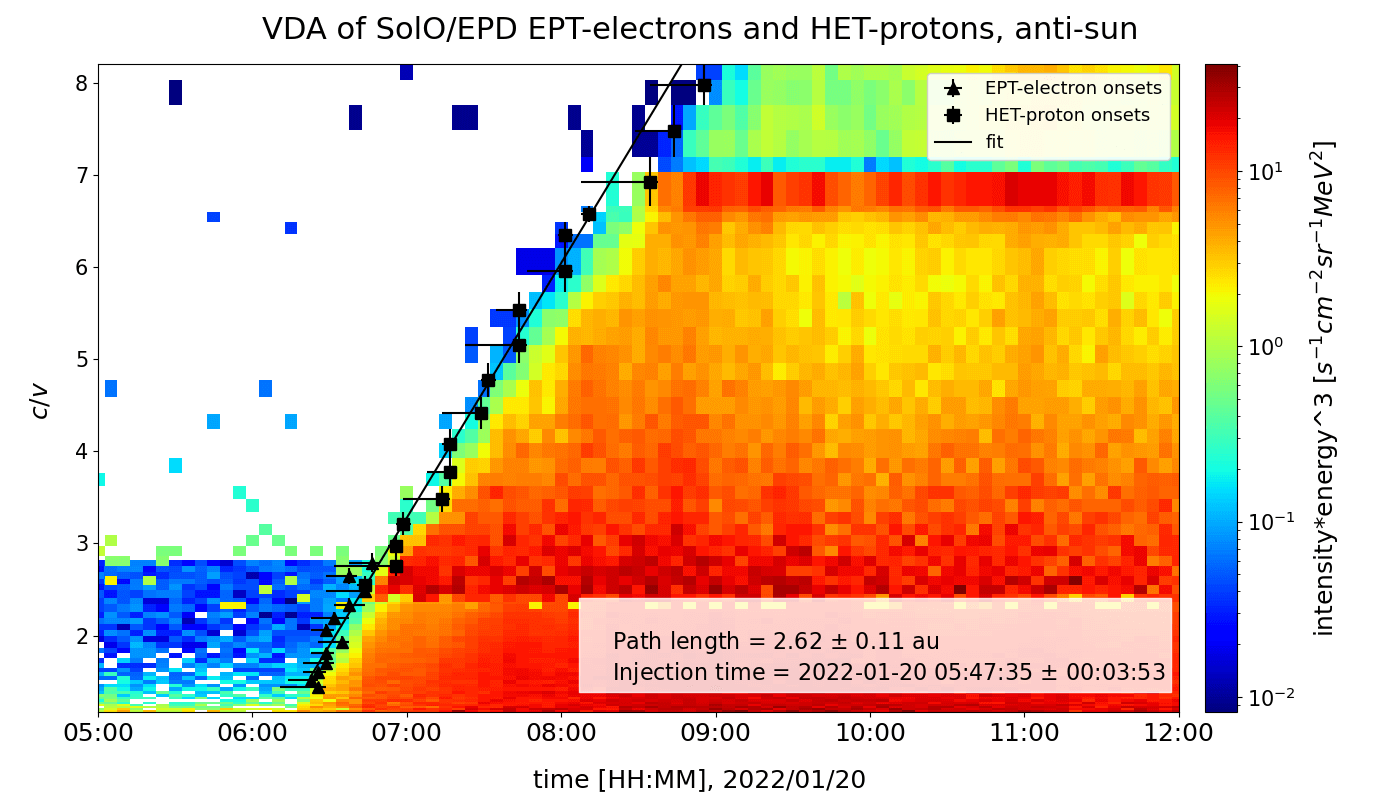}}
     \caption{Velocity dispersion analysis of the onset of the SEP event at Solar Orbiter. Electron and proton intensities (colour-coded) from EPT and HET sensors, respectively, as function of time and inverse speed ($c/v$ which is $1/\beta$ as used in Fig. \ref{fig:VDA_L1}). The colour-coded intensities are multiplied by the cubed energy to enhance the contrast. Over-plotted in black are the onsets of electrons (triangles) and protons (squares), and the velocity dispersion fitted line. The path length and injection time values shown in the legend are the result from bootstrapping (details given in the main text).} 
     \label{fig:VDA}
\end{figure}
Figure \ref{fig:VDA} shows the onsets for each energy channel indicated with a triangle (rectangle) for EPT electrons (HET protons), plotted on the particle spectrogram, with the corresponding uncertainties. We considered the bins of the channels as the uncertainties for the y-axis (c/v). In Appendix \ref{app_VDA_Mario} we detail how we estimated the uncertainties for the x-axis. Then we used an orthogonal distance regression (ODR) method to fit c/v against the onset times, to calculate the path length and the injection time. The linear fit is shown in Fig.~\ref{fig:VDA}.  

The final values of path length and injection time, using a bootstrapping method detailed in Appendix \ref{app_VDA_Mario} to estimate the uncertainties,  are given in the legend of Fig.~\ref{fig:VDA}. It shows an effective propagation path length of L=2.6 $\pm$ 0.1 au, much longer than the length of $\sim$0.99 au expected for a nominal Parker spiral field with the measured solar wind and scatter free propagation. It might indicate a relatively poor pitch-angle coverage or a non-standard interplanetary magnetic field topology. The injection time given is 05:48 UT $\pm$ 4 min on January 20 (time at the Sun). Using the light-travel time to Earth, the injection time is 05:56 UT $\pm$ 4 min. Within uncertainties, the injected times derived from Solar Orbiter and near Earth are in agreement.   

\begin{figure}[!t]
\centering
\includegraphics[width=0.44\textwidth]{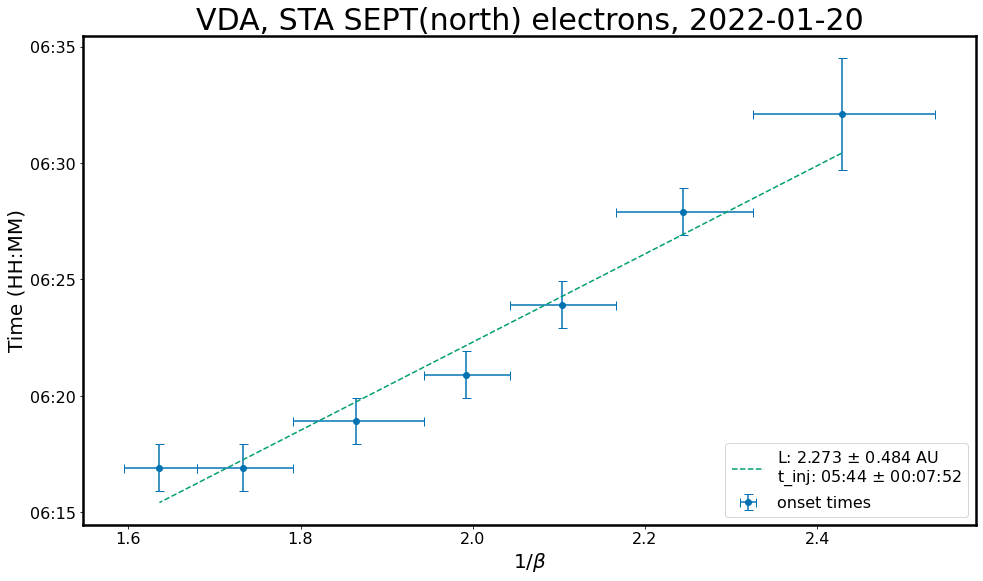}
\caption{Electron VDA of the SEP event at STEREO-A in the same format as Fig.~\ref{fig:VDA_L1}.  } 
     \label{fig:VDA_STA}
\end{figure}

\subsubsection{Solar energetic particle timing: STEREO-A}\label{sec:VDA_STEREO-A}

To estimate the path length and infer the injection time of the particles for STEREO-A spacecraft, we followed the same process as described in Sect.~\ref{sec:Particle_timing_Earth} for near-Earth spacecraft. We note that due to the presence of the MO at the time of the SEP onset at STEREO-A spacecraft (discussed in Sect.~\ref{subsec:Interplanetary_context_STEREO-A}), only electrons could be used for VDA since the elevated proton levels associated with the MO masked the proton onsets. For the electrons, we used energies from 45--145 keV (SEPT) measured by the north telescope, which registered the first arriving particles. The background time was chosen from to 02:00 to 05:20~UT, being short due to a previous event masking the background intensity.

The results of the VDA using SEPT electrons are shown in Fig.~\ref{fig:VDA_STA}. The results of the fitting show an effective path length travelled by the electrons of L=2.3$\pm$0.5~au, being much longer than the nominal Parker spiral length for STEREO-A ($\sim$1.15~au) using the measured solar wind speed. It might indicate a relatively poor pitch-angle coverage or a non-standard interplanetary magnetic field topology, although we note that the uncertainty is large. The injection time is 05:44~UT $\pm$ 8 min, or 05:52 UT $\pm$ 8 min shifted $\sim$8.2 min to compare with electromagnetic observations from 1 au. This timing is in agreement within uncertainties with the injection time derived from near-Earth and Solar Orbiter data.

\subsection{Solar energetic particle composition} \label{sec:particle_composition}

The elemental composition of this event was measured by EPD/SIS and EPD/HET on board Solar Orbiter, by SIT on board STEREO-A, and by ULEIS on board ACE. The differential energy spectral fluences measured by SIS and HET are shown in the top panel of Fig.~\ref{fig:SIS_obs}. The H and $^4$He spectra flatten at low energies, then steepen above a break at a few MeV/nucleon. The O and Fe spectra are similar but less certain due to the smaller energy range covered. These features are typical of large gradual SEP events \citep[e.g.][]{Desai2016a, Cohen2021}. For 1~MeV/nucleon the 20 January event fluence for O was $\mathrm{\sim4\times10^3}$ particles/(cm$^2$ sr MeV/nucleon), roughly a factor of 25 below the fluences in the large October--November (``Halloween'') 2003 events \citep{Cohen2005}, which are among the most intense events observed at 1~au in recent solar cycles. The top panel of Fig. \ref{fig:SIS_obs} shows fluence spectra for major elements. Dashed lines are Band functions fits \citep{Band1993} to H, $^4$He , O, and Fe. The resulting spectral fitting coefficients are listed in Table \ref{table:SIS}. They fall within the distribution
of results from the survey by \cite[][]{Desai2016}, which is  based on large gradual SEP events.

\begin{figure}[htb]
\centering
\includegraphics[width=0.44\textwidth]{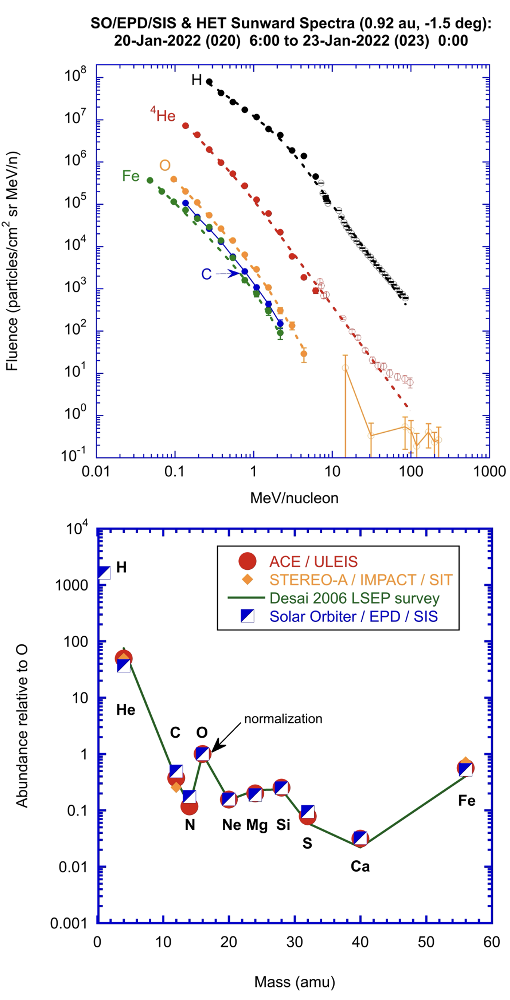}
\caption{Solar energetic particle fluences and relative abundances. Top panel: Fluence spectra from SIS (filled circles) and HET (circles) summed over the event, and fitted Band function spectra (dotted lines). Bottom panel: Abundances from 0.32--0.45 MeV/nucleon for the 2022 January 20 event compared with averages at the same energy from the survey of 64 large SEP events by \cite{Desai2006}. Blue half-filled squares are from Solar Orbiter SIS, filled red circles from ACE/ULEIS, and orange diamonds from STEREO-A/SIT. } 
     \label{fig:SIS_obs}
\end{figure}

The bottom panel of Fig.~\ref{fig:SIS_obs} shows the average elemental abundances measured between 0.32--0.45~MeV/nucleon for the 2022 January 20 SEP event measured at Solar Orbiter, ACE, and STEREO-A. The average abundances from the three spacecraft show a very similar pattern. Comparing this event with the average from the 64-event survey of \cite{Desai2016} measured at the same energy, it is clear that the composition of the 2022 January 20 event is typical for gradual SEP events. The measured $^3$He abundance was below 1\%. The maximum Fe/O abundance ratio at Solar Orbiter is around 0.64 at an energy of 0.19~MeV/nucleon, placing this event close to the average ratio found in the \cite{Desai2006} survey of gradual SEP events.

\subsection{Electron peak spectra}
\label{sec:particle_spectra}
\begin{figure*}[htb]
\centering
  \resizebox{0.8\hsize}{!}{\includegraphics{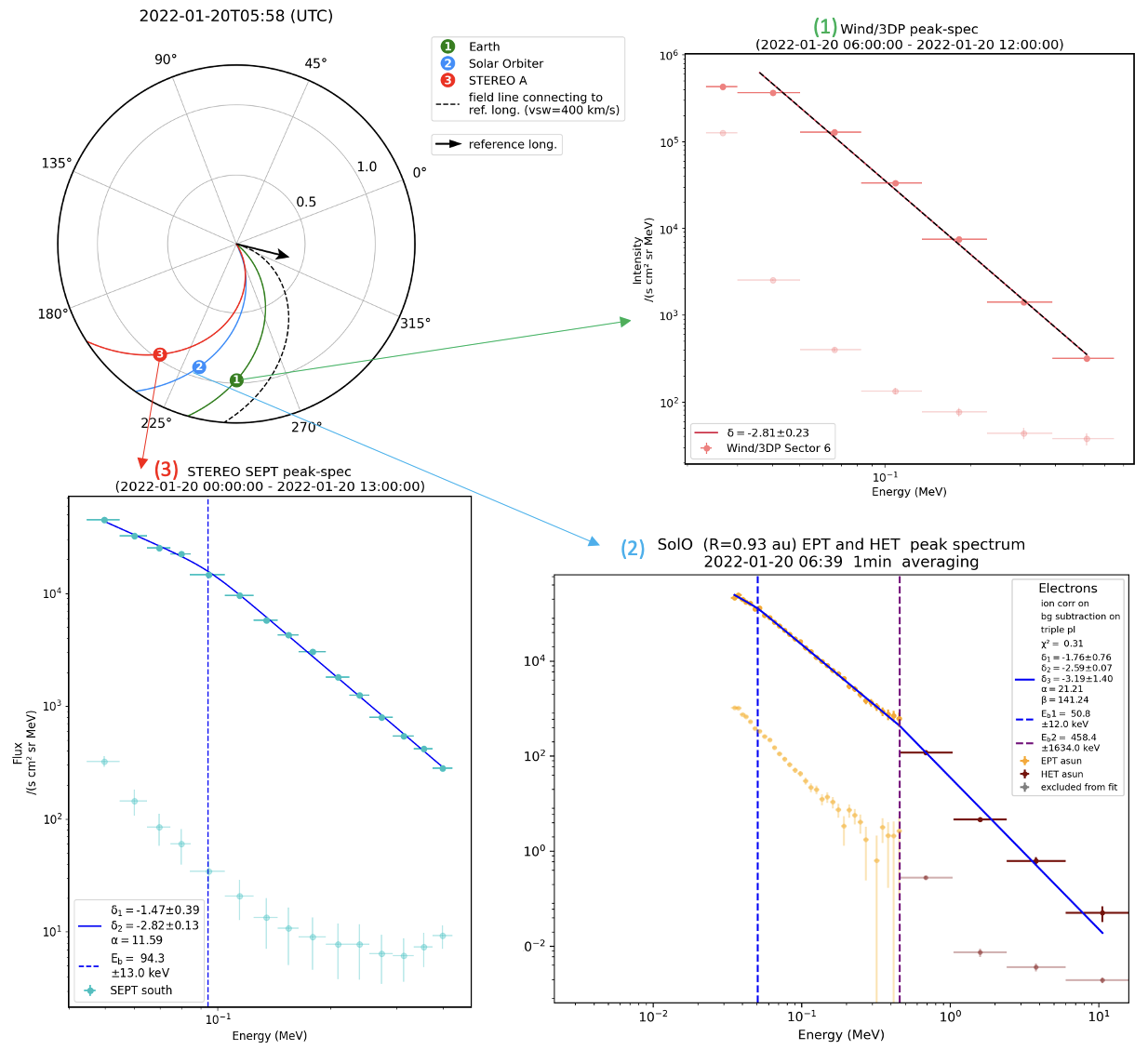}}
     \caption{Electron peak intensity spectra measured by Wind (1), Solar Orbiter (2), and STEREO-A (3). The legend shows the fit values: the spectral index ($\delta$\textsubscript{1}, $\delta$\textsubscript{2} $\delta$\textsubscript{3}) observed in between the spectral transitions: E\textsubscript{b}; and $\alpha$ ($\beta$), which determines the sharpness of the break(s) \citep{Strauss2020}. The lower and fainter set of points correspond to the pre-event background level. Details given in the main text.  }
     \label{fig:spectra_elec}
\end{figure*}
\begin{table*}
\caption{Electron peak spectra results. }  
\label{table:spectra}      
\centering          
\begin{tabular}{ccccccccc}      
\hline  Spacecraft & $\delta$\textsubscript{200}  & $\delta$\textsubscript{1}& $\delta$\textsubscript{2} & $\delta$\textsubscript{3}&E\textsubscript{b1} (keV)&E\textsubscript{b2} (keV)&$\alpha$&$\beta$\\    
  \hline
  
  Wind & -2.81$\pm$0.13 & -2.81$\pm$0.13 & -& -&-& -&-&- \\
  
  Solar Orbiter  & -2.59$\pm$0.07 & -1.76$\pm$0.76 & -2.59$\pm$0.07 & -3.19$\pm$1.40&51$\pm$12&458$\pm$1634&21.21&141.24  \\
  
  STEREO-A  & -2.82$\pm$0.13& -2.59$\pm$0.07 & -2.82$\pm$0.13 & -&94$\pm$13&-&11.59&- \\
 
\hline                    
\end{tabular}
\tablefoot{Parameters based on the method described by \cite{Dresing2020} and \cite{Strauss2020}. Details given in the main text.} 
\end{table*}
Following the method described by \cite{Dresing2020} and \cite{Strauss2020}, we determined the electron peak spectra, as observed by Wind, Solar Orbiter, and STEREO-A, with results shown in Fig.~\ref{fig:spectra_elec} and summarised in Table \ref{table:spectra}. In the case of Wind (panel 1 in Fig.~\ref{fig:spectra_elec}) and using sector 6 of the 3DP instrument, no spectral transition was found, representing a single power law shape according to
\begin{equation}
    I(E) = I_0 \left( \frac{E}{E_0}\right)^{\delta_1},
\end{equation} 
where $\delta_1$ represents the spectral index and $I_0$ is the intensity at $E_0=0.1$ MeV. We used sector 6 of the 3DP sensor instead of sector 7, which would cover a more field-aligned pitch-angle sector, because sector 7 was full of data gaps. 

For STEREO-A (panel 3 in Fig.~\ref{fig:spectra_elec}) we used the south telescope of SEPT, which presented the highest intensity peak. We found a broken power law to best describe the data represented by
\begin{equation}
    I(E) = I_0 \left( \frac{E}{E_0}\right)^{\delta_1} \left( \frac{E^\alpha+E^\alpha_b}{E^\alpha_0+E^\alpha_b}\right)^\frac{\delta_2-\delta_1}{\alpha}.
\end{equation} 
\noindent
This model yields a spectral transition at the energy $E_b$ and a second spectral index $\delta_2$ at energies above $E_b$. The parameter $\alpha$ describes the sharpness of the spectral transition. We note that there is a sudden drop around 07:30 UT in the intensity-time series caused by magnetic field changes (cf. Fig.~\ref{fig:anisotro_elec} 3). This drop potentially affects the peak intensities of the lower half of (or even all) the energy channels as they had not yet reached the peak. This means that especially the low energy channels, which usually reach their peak later, could in reality have higher peak intensities, which in turn could potentially lead to a steeper spectrum in that energy range. However, we are confident that the spectral break is not caused by this effect as the intensity drop rather affects also higher energies. However, we note that the break energy might be affected by this issue. 

In the case of Solar Orbiter (panel 2 in Fig.~\ref{fig:spectra_elec}), we used the EPT and HET anti-Sun telescopes, showing the highest intensity peak. Potentially due to the much higher energy resolution of the Solar Orbiter data we found a triple power law to best represent the observations, which is described by
\begin{equation}
    I(E) = I_0 \left( \frac{E}{E_0}\right)^{\delta_1} \left( \frac{E^\alpha+E^\alpha_{bl}}{E^\alpha_0+E^\alpha_{bl}}\right)^\frac{\delta_2-\delta_1}{\alpha} \left(\frac{E^\beta + E^\beta_{bh}}{E^\beta_0+E^\beta_{bh}}\right)^\frac{\delta_3-\delta_2}{\beta}.
\end{equation} 
This model yields two spectral transitions $E_bl$ at lower energies and $E_bh$ at higher energies and correspondingly three spectral indices $\delta_1$, $\delta_2$, and $\delta_3$. We note the high uncertainty of the second spectral transition. The parameters $\alpha$ and $\beta$ describe the sharpness of the two breaks, respectively. The spectral transition and indices below and above the spectral break are also summarised in Table \ref{table:spectra}.

For comparison, we selected the spectral index near 200 keV, namely $\delta$\textsubscript{200}. The spectral indices based respectively on Wind, Solar Orbiter, and STEREO-A data are similar within uncertainties and are summarised in the second column of Table \ref{table:spectra}.  The spectral indices observed in this event are clearly harder than a large sample of events (781 near-relativistic electron events measured by both STEREO) studied by \cite{Dresing2020}, who find $\langle\delta$\textsubscript{200}$\rangle$ = $-$3.5$\pm$1.4. Moreover, \cite{2022Dresing} analysed 33 energetic electron events that were related to coronal pressure waves. They derived a mean spectral index of $\langle\delta$\textsubscript{200}$\rangle$ = $-$2.5$\pm$0.3, similar to the indices found in this study ($\delta$\textsubscript{200} $\approx$ -2.6). 

\begin{figure}[htb]
\centering
  \resizebox{0.9\hsize}{!}{\includegraphics{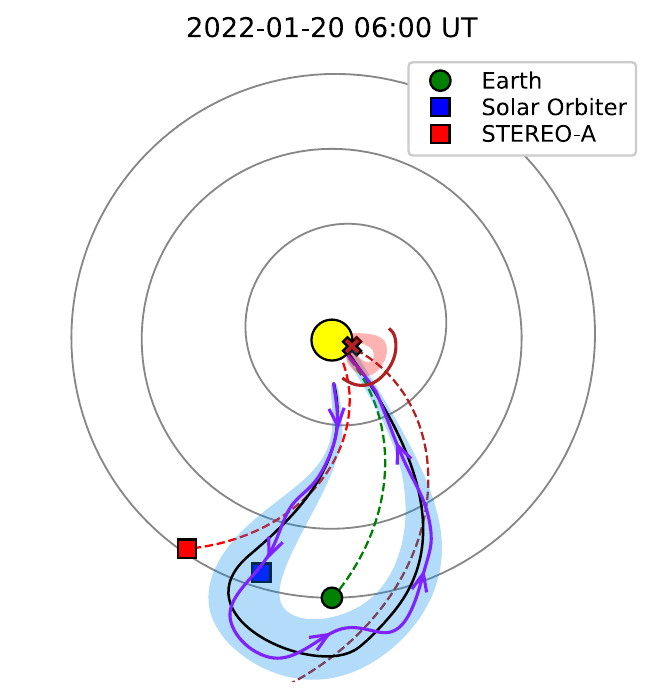}}
     \caption{Sketch showing interplanetary configuration of the 2022 January 20 SEP event. The Sun (not to scale) is shown at the centre indicated by the yellow circle. The grey circles represent, from the innermost outwards, the orbits of Mercury, Venus, and Earth. Earth, Solar Orbiter, and STEREO-A are shown by the green circle, blue, and red squares, respectively. The ICME corresponding to the CME erupting on 2022 January 16 is shown in blue. The CME and CME-driven shock associated with the SEP event on January 20 are indicated by the red shading and red curve, respectively. The dashed coloured lines indicate the nominal Parker spirals using measured solar wind speed. The rightmost dark-red dashed lines connects to the flare site using a nominal Parker spiral and 400~km~s\textsuperscript{-1}.    }
     \label{fig:scenario}
\end{figure}

 \begin{figure*}[htb]
\centering
  \resizebox{1.0\hsize}{!}{\includegraphics{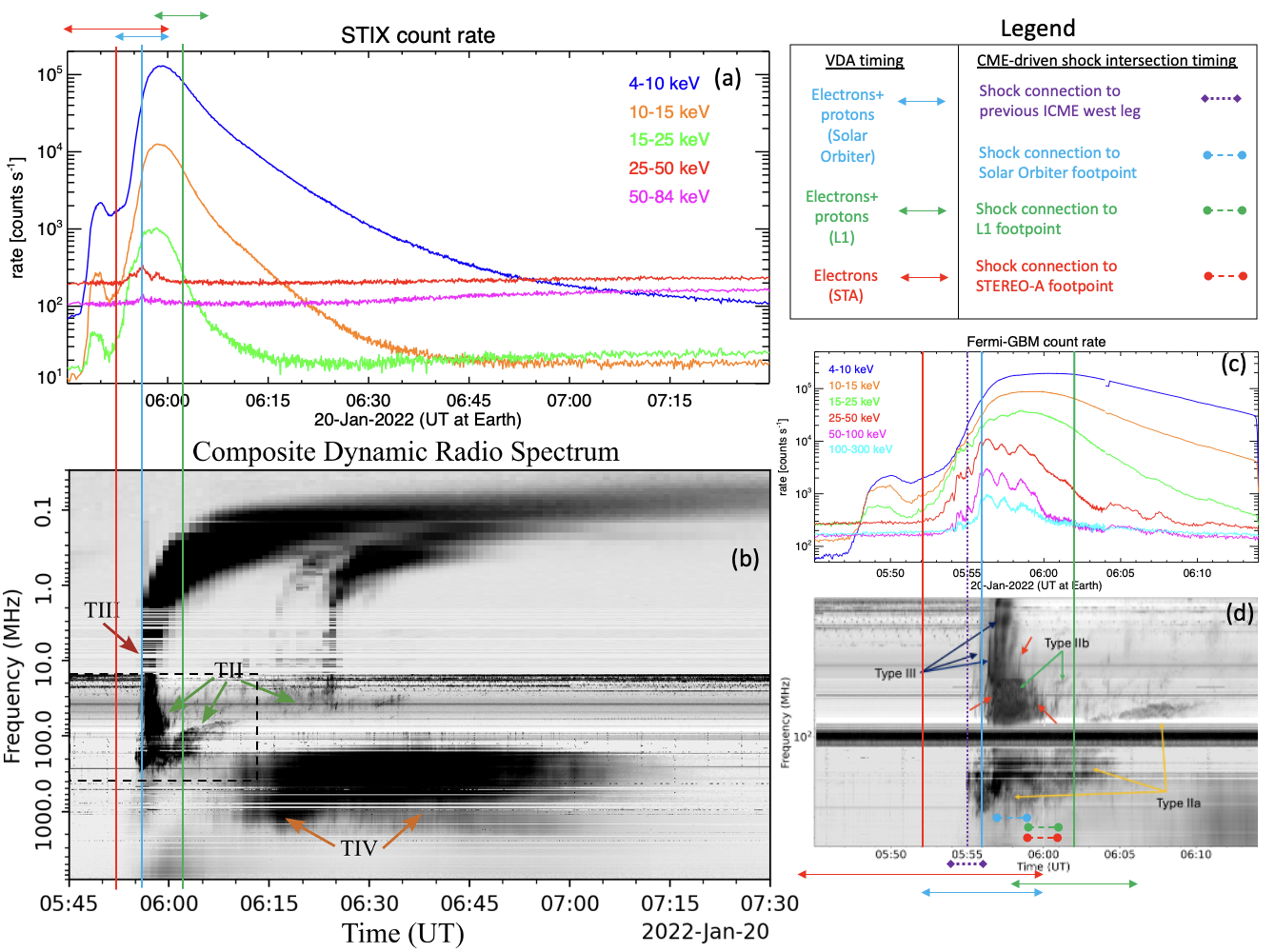}}
     \caption{Inferred SEP injection times shifted to 1 au (vertical lines with temporal error bars on top) overplotted on the radio spectrogram as observed from STEREO-A/WAVES and Earth (ASSA, and YAMAGAWA) and the X-ray count rates from Solar Orbiter/STIX. The zoom-in on the right corresponds to the dashed line square indicated on the left. It shows \textit{Fermi}-GBM X-ray count rates against the same radio spectrogram. The STIX times have been shifted by 30~s for comparison with electromagnetic observations from 1 au. Legend on the top right refers to lines in panels (a), (b), and (c). The observed radio structures are indicated in panels b and c.   Details given in the main text.} 
     \label{fig:radio_time}
\end{figure*}

\section{Solar energetic particle parent solar source: Remote-sensing observations and data analysis}
 \label{sec:solar_parent_activity}

\begin{figure}[htb]  
\centering
 \includegraphics[width=0.5\textwidth]{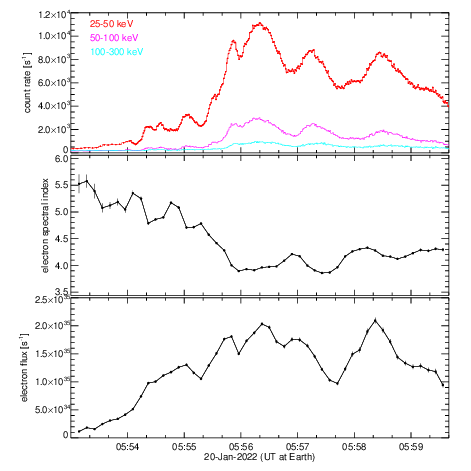} 
\caption{Results of the spectral analysis of the \textit{Fermi}-GBM data. \textit{Top}: GBM HXR count rates in three broad energy bands. \textit{Middle}: spectral index of injected electron flux. \textit{Bottom}: injected electron flux above the low-energy cutoff of 22~keV.} 
\label{fig:gbm_spectra}
\end{figure}

The National Oceanic and Atmospheric Administration (NOAA) AR number 12929 produced a series of eruptions around the time of the study.  A first detected CME was released from N08W30 (in Stonyhurst coordinates) at 20:48 UT on January 16, which arrived at Earth at 23:40 UT on January 18 and at Solar Orbiter at 17:10~UT on January 19, as discussed in Sect.~\ref{sec:in situ}.  This CME and its corresponding ICME are studied in more detail in Sect.~\ref{sec:Magnetic_cloud_16}, as it is affecting the particle propagation as observed by Solar Orbiter.  A second CME was launched at 17:00 UT on January 18 from N07W53 (in Stonyhurst coordinates), related to the SEP event on January 18. The particle increase related to this event is shown in Figs.~\ref{fig:Earth_solo_part_plasma} and \ref{fig:STA_particles_plasma}, whose background is affecting the onset of the SEP event on January 20.  A third eruption was observed to be released from N08W76 (in Stonyhurst coordinates; 325$^{\circ}$ in Carrington longitude) at 06:12 UT on January 20, related to the SEP event under study. This CME is also represented in the sketch of Fig. \ref{fig:scenario} as a red shading area. In the following, we present the remote-sensing observations and analysis of this third eruption. 

\subsection{Flare observation and analysis}
\label{sec:Flare}
An M5.5 flare was observed on 2022 January 20 at N08W76 (in Stonyhurst coordinates) by near-Earth assets such as SDO and by Solar Orbiter. Since we are mainly interested in energetic particles, we focus on hard X-ray (HXR) observations that constrain nonthermal electrons in solar flares. In Fig.~\ref{fig:radio_time}a, we plot X-ray count rates in five energy bands as recorded by STIX on Solar Orbiter. At lower energies (4--25~keV), the flare shows a smooth time profile, which is typical for thermal emission. It peaks at 05:58~UT (all STIX times have been shifted by 30~s to be consistent with observations from 1~au) and shows an extended gradual decay lasting more than 1.5 hours. Between 05:54 and 06:00~UT, three more impulsive peaks can be discerned at energies above 25~keV, which is consistent with nonthermal bremsstrahlung emission generated by accelerated electrons. However, the nonthermal emission is very weak. 

STIX provides Fourier-synthesis imaging capabilities \citep[cf.][]{Massa23}, so we reconstructed the HXR sources at thermal and nonthermal energies. We found a single coronal source above the solar limb, also at higher energies, where usually the emission is predominantly emitted by chromospheric footpoints. Since the flare is observed right at the solar limb as seen from Solar Orbiter, we conclude that the footpoints are actually occulted. This is corroborated by data from \textit{Fermi}-GBM, where the count rates show a much more pronounced nonthermal component above 25~keV (as shown in Fig.~\ref{fig:radio_time}c). While GBM has no imaging capability, we know from SDO/AIA that the flare was fully visible from Earth, and we are thus confident that GBM has complete coverage of the X-ray emission of this flare. The emission above 25~keV shows multiple impulsive peaks, including three major ones that extended to at least 300~keV.

Due to the full coverage of the flare from Earth's perspective, we used GBM data to get quantitative constraints on the electrons accelerated in the flare. We performed a series of spectral fits using the OSPEX (Object Spectral Executive) package\footnote{http://hesperia.gsfc.nasa.gov/ssw/packages/spex/doc/}, which is part of the SolarSoft IDL software library. We forward-fitted the background-subtracted GBM count spectra with a combination of an isothermal component and a nonthermal thick-target model assuming a power-law spectrum for the injected electrons \citep{1971Brown}. As GBM is not optimised for solar observations, the spectra suffer from pulse pileup, particularly during times of high count rates during solar flares. This mostly affects the thermal component and the transition to the nonthermal range. We therefore do not consider the thermal fits here. Concerning the transition to the high-energy power-law, we determined the effective low-energy cutoff in the early phase of the impulsive phase when pileup is still comparatively small. We found low-energy cutoffs around 22~keV, and then adopted this value as a constant parameter for all nonthermal fits. It should be stressed that this is the lowest cutoff energy that is consistent with the data, because the true cutoff is usually masked by the thermal component \citep[e.~g.][]{2020Warmuth}.

Figure~\ref{fig:gbm_spectra} shows the spectral fit results for the thick-target electron component together with the GBM count rates in the nonthermal energy range. We focus here on the impulsive phase of the flare. The top panel shows the GBM count rates in three broad energy bands that are dominated by nonthermal emission, as shown by the multiple impulsive peaks with typical duration of ${\approx}1$--2~min. It is thus clear that this flare was characterised by multiple discrete episodes of energy release and particle acceleration. The middle panel shows the power-law index $\delta$ of the electron flux spectrum. We note that when the count rates are high, the spectral index becomes lower, namely the spectrum hardens. This anti-correlation is known as the soft-hard-soft evolution \citep[e.g.][]{2004Grigis}. The hardest spectra are characterised by an index of $\delta \approx 4 $. Finally, the bottom panel of Fig.~\ref{fig:gbm_spectra} shows the total injected electron flux above the low-energy cutoff of 22~keV. Again, this is anti-correlated with the spectral index. During the impulsive phase, a total of $4.9 \pm 0.1 \times 10^{37}$ electrons were accelerated, which contained an energy of $2.5 \pm 0.1 \times 10^{30}$~erg. These values are typical for mid-M-class flares \citep{2020Warmuth}.

We note that the spectrum of the injected electrons deduced from the HXR observations is softer than the in-situ spectra discussed in Sect.~\ref{sec:particle_spectra}, namely $\delta_\mathrm{HXR} \geq 4$ as opposed to $\delta_\mathrm{in-situ} \approx 2.5$--2.8. This is consistent with what has been found by statistical studies of impulsive solar energetic electron events, which all show that the spectra of electrons precipitating on the Sun (assuming thick-target emission) are apparently softer than the spectra of the electrons injected into space \citep[cf.][]{2007Krucker,2021Dresing}. It is not yet clear whether this truly means that the injection spectrum is different for the downward- and upward-moving electrons, or whether this difference rather results from propagation effects, different acceleration mechanisms that might be involved, or modelling assumptions that are made for inferring the electron spectrum from the measured photon spectrum.

\subsection{Radio observations and analysis}
\label{sec:Radio}

In Fig.~\ref{fig:radio_time}b, we present a composite dynamic radio spectrum constructed using observations from several ground-based and space-based instruments. This provides an uninterrupted coverage of processes from the low corona to interplanetary space (8 GHz--30 kHz). The part of the spectrum from centimetric to metric wavelengths (in frequency, this corresponds to 8 GHz to 70 MHz) was constructed using data from the YAMAGAWA solar radio spectrograph, supplemented with data from the e-Callisto network of radio telescopes, and in particular with data from ASSA (80 MHz to 10 MHz). For the part from decametric to hectometric wavelengths (corresponding to 16 MHz to 30 kHz in frequency) we used data from SWAVES on board STEREO-A. Such a spectrum can be leveraged to distinguish between the nuances of particle acceleration and transport from the corona to the inner heliosphere \citep[e.g.][]{Ergun98,Vosh15,Badman22}.

The solar radio event presented here is rich with a number of different emission types such as type II (TII), type III (TIII), and type IV (TIV), marked in Fig.~\ref{fig:radio_time} b. In the low-decimetric to metric wavelengths (\(\ll\)1 GHz to 30 MHz) the observed radio emissions are mostly dominated by different types of plasma emission such as type II, III, and IV radio emission. These are associated with non-thermal electrons accelerated by propagating shock waves (TII), electron beams propagating along open and quasi-open magnetic field lines (TIII), and electrons trapped within rising post-flare loops or within CME flux ropes (TIV). The time evolution of the radio event, that is, the starting times of TII and TIII, are directly compared with the results from the VDA analysis (Sect. \ref{sec:particle_timing}) and discussed here.  

\subsubsection{Type II radio bursts}
\label{sec:Type IIs}

Two different TII lanes may be distinguished, Type IIa (TIIa) and Type IIb (TIIb) indicated in Fig. \ref{fig:radio_time}d, both with their own complexities. Multiple TII emissions from the same shock may have their sources at different regions of the shock, and therefore investigating them allows us to constrain regions where electrons are accelerated \citep[][]{Jebaraj2020,Jebaraj21}. TIIa starts from 250 MHz promptly at 05:55 UT suggesting shock formation early on during the event, as discussed in Sect.~\ref{sec:CME-driven shock_obse_anal}. TIIa drifts to lower frequencies at a rate of approximately \(df/dt \sim 10\) MHz per minute between 05:55 (1.09~$R_{\odot}$) and 06:11 UT (1.5~$R_{\odot}$). Using this drift rate, and the approximate coronal heights at which they are formed, which was obtained from a commonly used \cite{Newkirk61} coronal electron density model, we calculated the speed of the emitting source to be \(\sim 340\) km s\(^{-1}\). Previous research has often associated such slow propagation of the source with the shock's propagation within a streamer or at sector boundaries \citep{Kouloumvakos21,Morosan24}. It is noteworthy to mention the close correspondence between the speed of the EUV wave, discussed in Sect. \ref{sec:CME-driven shock_obse_anal}, and the speed deduced from the TIIa drift. Moreover, the derived height at which TIIa was formed (1.09~$R_{\odot}$) is low in the corona where the EUV wave propagates \citep{warmuth2015large}. TIIa continues up to the hectometer wavelengths, 3 MHz in the frequency spectra where it stops at 06:40 UT.

\begin{figure}[htb]
\centering
  \resizebox{1.0\hsize}{!}{\includegraphics{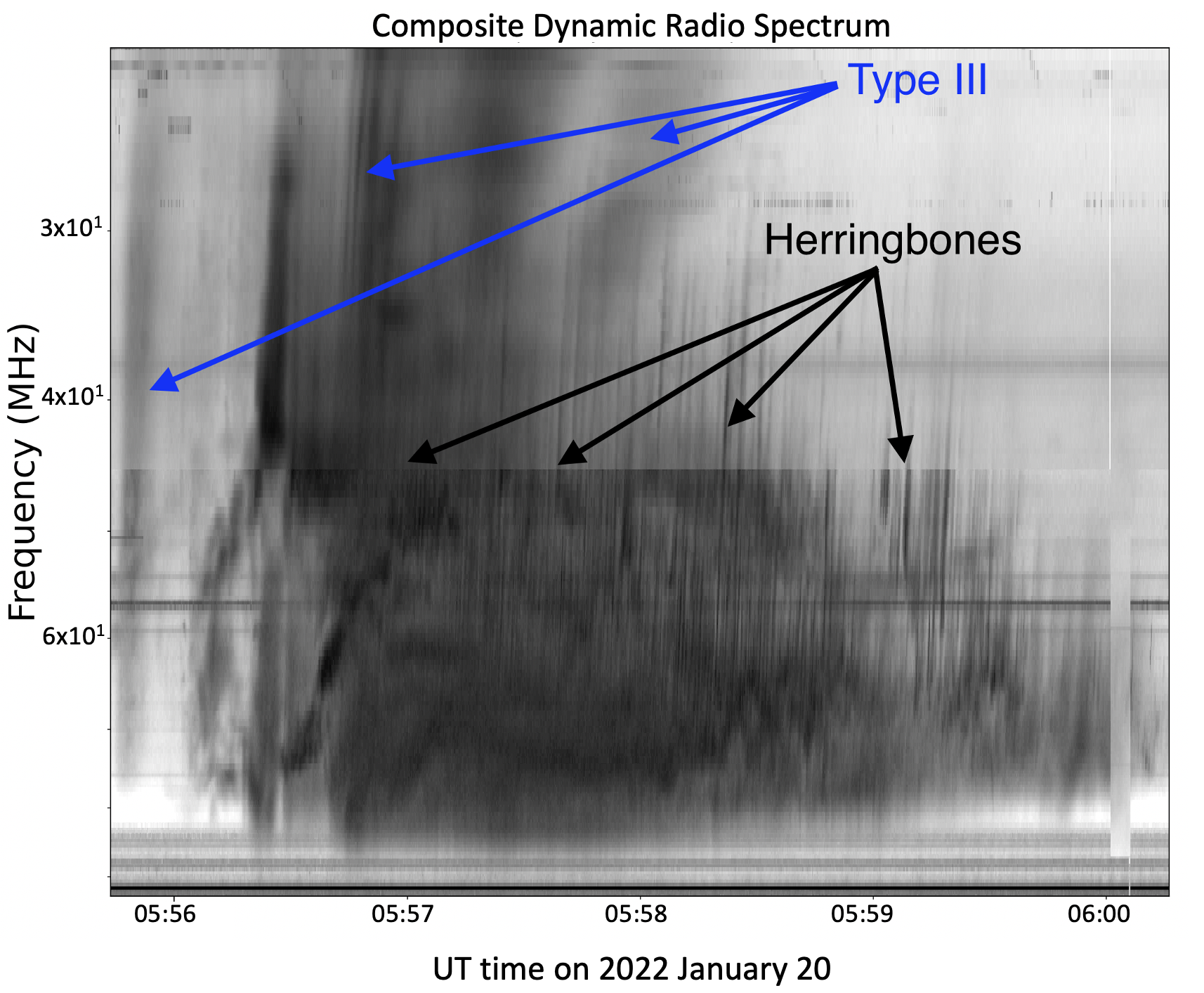}}
     \caption{Zoom-in of Fig. \ref{fig:radio_time}b. It shows in detail the Type IIIs and HBs radio structures, indicated with the blue and black arrows, respectively.  }
     \label{fig:HBs}
\end{figure}

TIIb exhibits a far more complex structure. It is first characterised by a spectral kink-like morphology, previously linked to source propagation through regions of varying density \citep{Kouloumvakos21,koval2023morphology}. It features distinct herringbone (HB) structures which are observed for a brief period between 05:56:30 and 06:00:00 UT within the 80 to 50 MHz range (corresponding to \(\sim 1.5 R_\odot\)). They are indicated with red arrows in Fig. \ref{fig:radio_time}d and with the black arrows in the zoomed-in plot in Fig. \ref{fig:HBs}. HBs are known to be electron beams accelerated by a nearly perpendicular shock wave, emanating from a backbone that represents the width of the shock's nearly perpendicular region \citep{Mann18,morosan2022shock}. TIIb is observed for about ten minutes and drifts to 16 MHz (decameter wavelengths) by 06:05 UT. By applying a Newkirk density model to estimate the speed of the source linked with these emissions, we find it to be approximately \(1400\) km s\(^{-1}\). This estimated shock speed is in agreement with that derived from the spheroid 3D reconstruction in Sect. \ref{sec:CME-driven shock_obse_anal}, which was 1433 km s\(^{-1}\).   

\begin{figure*}[htb]
\centering
  \resizebox{1.0\hsize}{!}{\includegraphics{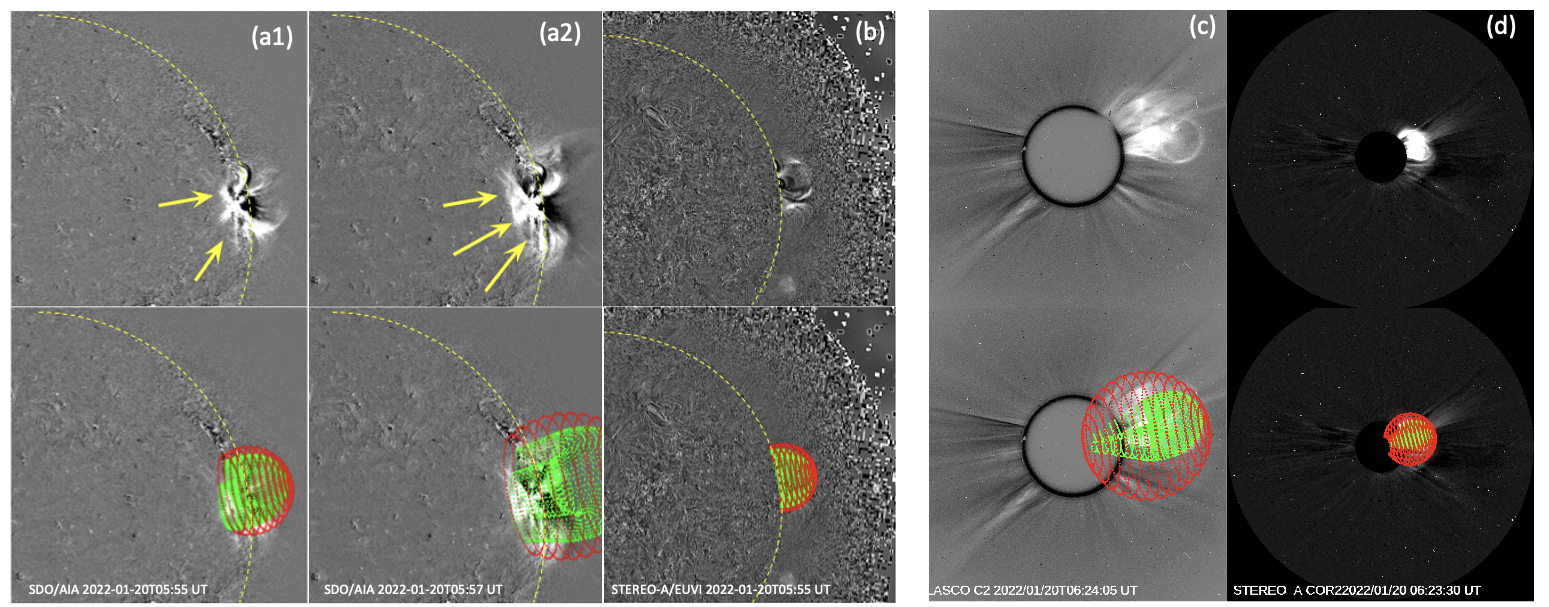}}
     \caption{EUV and coronagraph images and GCS 3D reconstruction of the CME (green mesh) and associated driven shock (red mesh) as seen by two different points of view: STEREO/EUVI (b) and STEREO/COR2-A (d); SDO/AIA (a) and SOHO/LASCO-C2 (c), at different times. Details given in the main text.    } 
     \label{fig:CME}
\end{figure*}
\subsubsection{Type III radio bursts}
\label{sec:Type IIIs}
In the meter to kilometre wavelengths (corresponding to 80 MHz to 30 kHz), we identified one group of TIII emissions close to the flare time, as indicated with the blue arrows in Figs. \ref{fig:radio_time} and \ref{fig:HBs}. This indicates electrons streaming away from the corona during this time period. They are observed to start around 80 MHz (\(\sim 1.45 R_{\odot}\) based on the \cite{Newkirk61} density model) at 05:55:40 UT and are seen across the deca-hecto-kilometric wavelengths as observed from space-based receivers. STEREO-A and L1 spacecraft did not measure any Langmuir waves at the time of the energetic particle event and the type III radio burst. The Radio and Plasma Waves \citep[RPW;][]{Maksimovic20b,Maksimovic2021,Vecchio2021} on Solar Orbiter was not observing during this time period and therefore we cannot be conclusive about the lack of Langmuir waves at Solar Orbiter. However, given that Solar Orbiter was located between L1 and STEREO during the time of the SEP event, it is highly unlikely that it would have observed local wave, which suggests that the TIII emitting electron beams never traverse the vicinity of any spacecraft. This group of TIII are observed until 05:58 UT at metric wavelengths ($\sim$60-30 MHz) where most individual TIII within the group seem to originate.

\subsubsection{Co-temporal Type III bursts and HBs}
\label{sec:Cotemporal Type III+HBs}

The co-occurrence of TIII and TIIB suggests that some of the electron beams accelerated by the shock (manifesting as HB) may contribute to the group of TIII bursts. Given the morphological similarities between HBs and TIII, it may be speculated that some TIII observed in low frequency spacecraft observation (\(<15\)~MHz) may have also been continuations of the HBs. Correlation between near-relativistic electrons observed in situ and HBs emitted by the coronal shock have been qualitatively discussed in prior studies, such as \cite{Jebaraj23}.  Following their conclusions, we may suggest that the shock strongly interacted with the field lines where the flare-accelerated electrons propagated. Since a near-perpendicular shock geometry (with respect to the upstream magnetic field) is required to generate the HB structures, the lateral regions of the shock are the most-likely hotspots for such interactions. It is also worth noting that the HB and TIII bursts occur co-temporally with the HXR peak observed by \textit{Fermi}-GBM (Fig. \ref{fig:radio_time}c). While, qualitatively this lends credibility to a shock re-acceleration phenomena, it is impossible to quantify these correlations due to the lack of precise X-ray and radio imaging. 

We provide below a scenario where the above qualitative result is self-consistent. The acceleration mechanism which is invoked is the fast-Fermi process, particularly in its relativistic form \citep{Leroy84,Kirk1994,Jebaraj23b}. If the electrons accelerated during reconnection at the flaring site interact with a near-perpendicular shock (\(\theta_\mathrm{Bn} > 85^{\circ}\), $\theta$\textsubscript{Bn} being the angle between the shock normal vector and the upstream magnetic field line), they may be re-accelerated resulting in beams. The mechanism is such that only particles occupying a limited portion of pitch-angle space—the range of angles between a particle’s velocity and the magnetic field—may undergo the adiabatic reflection process. Since this occurs in a frame where the shock is at rest and the magnetic field is aligned with the shock normal, the incoming electrons are reflected upstream along the magnetic field lines, resulting in field-aligned beams. These beams can subsequently become two-stream unstable and simultaneously emit plasma radiation, which manifests in the radio spectrogram as HB or TIII bursts \citep{holman1983,Krasnoselskikh85,Mann18,mann2022}. The process is highly efficient and would result in a significantly changed electron spectra than the ones accelerated by flares. This is corroborated by the fact that the electron spectra discussed in Sect. \ref{sec:particle_spectra} deviates from the photon spectra, likely due to shock modification.

\subsubsection{Solar phenomena--SEP timing comparison}
\label{sec:RAdio_timing}

In Fig. \ref{fig:radio_time} we include the  VDA timing results from Sect. \ref{sec:particle_timing} to compare with the HXR and radio signatures. The red, blue, and green vertical lines represent the injection times derived for STEREO-A, Solar Orbiter, and near-Earth spacecraft, respectively. The uncertainties of these onset times are represented by the arrows in the top panel a and bottom panel d. We also present in Table \ref{table:timing} a summary of this inferred SEP injection times and the timing of the solar phenomena discussed above (HXR peaks, TIIs, HBs, and TIIIs). For the spacecraft with less uncertainty in the VDA analysis (Solar Orbiter, t$_{\mbox{inj}}$=05:56$\pm4$ min; Near-Earth, t$_{\mbox{inj}}$=06:02$\pm4$ min), the injection times are co-temporal with the emission of TIIa, TIIb, HBs, TIIIs (partly co-temporal with the HBs), and the nonthermal HXR peaks. In the case of STEREO-A, the inferred SEP injection time shows a larger uncertainty (t$_{\mbox{inj}}$=05:52$\pm8$ min), however this time it is still co-temporal with the solar phenomena mentioned above.     
\subsection{Extreme ultraviolet and coronagraph observations and analysis}
\label{sec:EUV_observations}

The extreme ultraviolet (EUV) observations of the solar eruption associated with the SEP event under study  have been examined in detail by \cite{2022Zhang}. We include here a summary of the most relevant information from that study and further observations and analysis using the EUV and coronagraph imagery as presented below.  

\subsubsection{Observation and analysis of the coronal mass ejection }
\label{sec:CME_obser_analysis}
The early phase of the eruption started before 05:51:30 UT on 2022 January 20, as shown in Fig.~1 of \cite{2022Zhang} in the hot channels of AIA at 131{\AA} and 94{\AA}. The overlying loop is tardy during the slow rise of the flux rope observed at the hot channels. It is pushed upward to form the leading front of a CME as the hot flux rope accelerates \citep{2013Cheng}. The final speed of the flux rope and the overlying loops are close to each other ($\sim$830 km~s$^{-1}$). At 05:55:04 UT the eruption is clearly observed by SDO/AIA at the west solar limb, as shown in the left part of Fig.~\ref{fig:CME} (top panel a1). About 30 minutes later,  at 06:24:05 UT, the flux rope has evolved high enough in the corona to be fully observed by both SOHO/LASCO (panel c) and STEREO/COR2-A (panel d) coronagraphs.

To characterise the CME associated with the SEP event, mainly in terms of final coronal CME speed, width, and location, we took advantage of the multi-view spacecraft observations and reconstructed the 3D CME to minimise projection effects using the graduated cylindrical shell \citep[GCS;][]{Thernisien2006GCS,Thernisien2011} model. The GCS model uses the geometry of what looks like a hollow croissant to fit a flux-rope structure using coronagraph images from multiple viewpoints. The sensitivity (deviations) in the parameters of the GCS analysis is given in Table 2 of \cite{Thernisien2009}. The COR1/2-A and C2 and C3 quasi-simultaneous images were used to fit the flux-rope shape of CME at different times. The routine used for the reconstruction is \textit{rtcloudwidget.pro}, available as part of the \textit{scraytrace} package in the SolarSoft IDL library\footnote{\url{http://www.lmsal.com/solarsoft/}}. The main CME reconstruction period, using two vantage points of view, covered from 05:50 UT to 07:54 UT on 2022 January 20. 

The lower panels of Fig.~\ref{fig:CME} show on top of the coronagraph images, the GCS fit analysis for the CME (green mesh) and the spheroid model fit for the  CME-driven shock (red mesh), discussed below. The 3D reconstruction shows that the CME follows a radial path with a Stonyhurst (Carrington) latitude and longitude of $10^{\circ}$ and 74$^{\circ}$ (323$^{\circ}$), respectively. The tilt angle ($\gamma$), namely the inclination of the flux rope with respect to the ecliptic plane, does not show deviations, staying at a fixed value of $40^\circ$. The CME speed at the leading edge estimated from the linear fit to the height--time measurements is 1410~km~s$^{-1}$. The uncertainty of the CME speed is considered to be 7\% of the value based on \cite{Kwon2014}. The width or total angular extent of the CME is 51$^{\circ}$, based on \citet{Dumbovic2019}, where the semi-angular extent in the equatorial plane is expressed by ${R\textsubscript{maj}-{(R\textsubscript{maj}-R\textsubscript{min})} \times |\gamma|/90}$. The value of $R\textsubscript{maj}$ (face-on CME half-width) is calculated by adding $R\textsubscript{min}$ (edge-on CME half-width) to the half-angle, and $R\textsubscript{min}$ was calculated as the $\arcsin(aspect~ratio)$. The CME width deviation was derived from the mean half-angle error, estimated by \cite{Thernisien2009} as +13$^{\circ}$/$-7^{\circ}$.
Thus, at the latest time of the 3D reconstruction at 07:54 UT,  corresponding to a CME height of 16.10~$R_{\odot}$, the narrow CME ($\sim$51$^{\circ}$) is propagating in the direction W74N10 with a relatively high speed ($\sim$1410~km~s\textsuperscript{-1}). Figure \ref{fig:scenario}, which depicts a sketch showing the interplanetary configuration of the 2022 January 20 SEP event, shows this CME represented by the red shading. 

\begin{figure*}[htb]
\centering
  \resizebox{1.0\hsize}{!}{\includegraphics{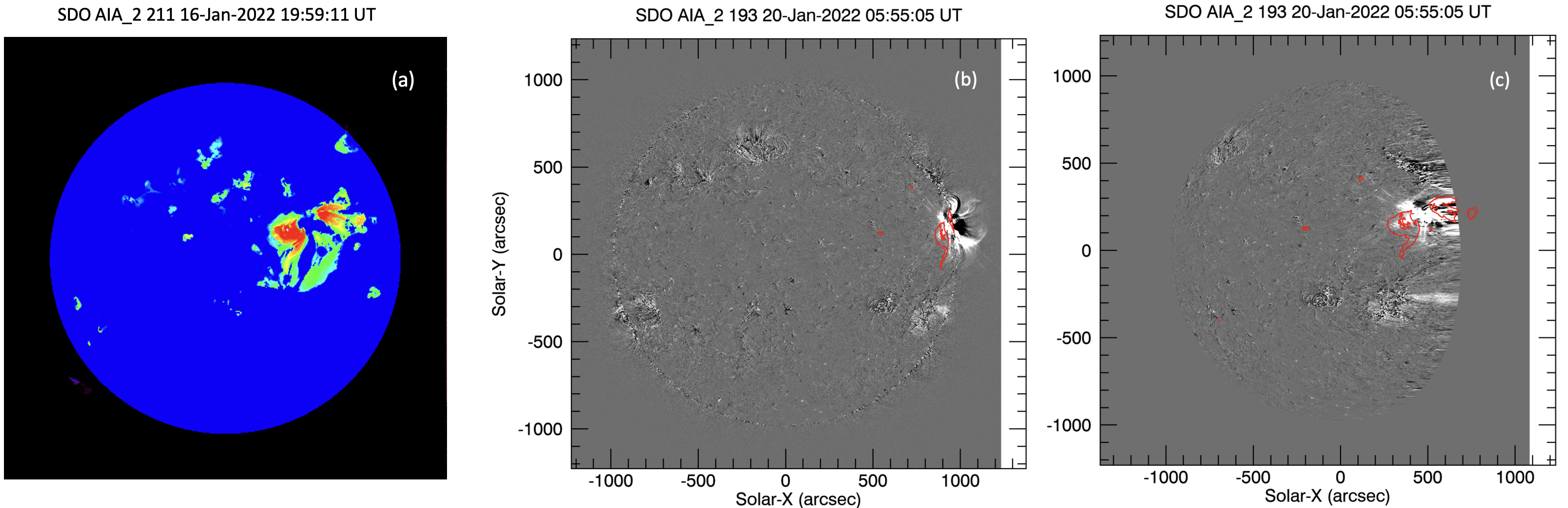}}
     \caption{Evolution of AR 12929 from 16 to 20 of January 2022. \textit{(a)} Image taken by SDO/AIA\_2 211{\AA} on 2022 January 16 at 19:59:11 UT, showing the accumulated pixels of the dimming areas in the period of time of observation. \textit{(b)} It shows the CME-driven shock and the EUV wave on January 20 along with the dimming lobes shown in panel a overplotted in red in the position they would have been four days later. \textit{(c)} the  image shown in panel b is de-rotated to the time of panel a using the solar differential rotation formula by \cite{1990Howard}. Details given in the main text. }
     \label{fig:CME_ICME}
\end{figure*}
\subsubsection{CME-driven shock observation and analysis}
\label{sec:CME-driven shock_obse_anal}

The CME eruption leads to the formation and propagation of an EUV wave \citep[shown in figure~1 of][]{2022Zhang}. The signatures of the EUV wave propagating on the solar surface,  are  clearly visible from 05:52 to 06:09 UT on 2022 January 20 in AIA images, as indicated with the yellow arrows in panels a1--a2 of Fig. \ref{fig:CME}. The EUV wave on the solar disk extends to about 365 Mm from the source region with a speed of 373 km s\textsuperscript{-1}. This value is in agreement with the TIIa drift of 340~km~s\textsuperscript{-1} derived from radio observations in Sect. \ref{sec:Radio}. The successful eruptions of the flux rope and the overlying system of loops evolve higher in the corona into a fast and wide CME, which drives a shock wave \citep[shown in figure~4 of][]{2022Zhang}. This shock wave is observed as a well-formed bubble over the west solar limb by SDO/AIA at 05:55:04 UT, shown in the top panel a1 of Fig.~\ref{fig:CME}. 

In order to gain a detailed understanding of the magnetic connectivity to the CME-driven shock associated with the SEP event, the coronal shock 3D reconstruction, shown as the red mesh in lower panels of Fig. \ref{fig:CME}, was performed using the model developed by \cite{Olmedo2013}. The model uses a spheroid shape to fit the CME-driven shock using quasi-simultaneous images from COR1 and COR2, and from C2 and C3. The images underwent a basic process for calibration, and base-difference or running-difference procedure was used to highlight the front of the shock better. The process of the fitting are explained in detail by \cite{2021Rodriguez-Garcia}. The main shock reconstruction period, using two vantage points of view, covered from 05:55 UT to 07:54 UT on 2022 January 20, when the shock height changed from $\sim$1.27 R\textsubscript{$\odot$} to $\sim$16.10 R\textsubscript{$\odot$}.

The parameters of the 3D reconstructed shock (red mesh in Fig.~\ref{fig:CME}) are consistent during the main reconstruction period. The resultant spheroid is oblate ($e$=0.28) and the self-similarity coefficient ($\kappa$) is $\sim$0.52. The longitude and latitude values show that the origin at the Sun of the coronal shock is located at W73N10. Lastly, the coronal shock speed, estimated as the linear fit of the evolution of the shock height, is 1433 km s\textsuperscript{-1}. The uncertainty of the CME-driven shock speed is considered to be 8\% of the value, following \cite{Kwon2014}. We note that the shock speed deduced from the spheroid reconstruction is in good agreement with the estimated type II drift ($\sim$1400 km s\textsuperscript{-1}) deduced in Sect.~\ref{sec:Radio}.

We used the 3D reconstruction of the CME-driven shock to estimate the first time the shock wave intersects the magnetic field lines connecting to the near-Earth, Solar Orbiter, and STEREO-A spacecraft. For this purpose and to have a reliable value of the uncertainty of the crossing time, we utilised the magnetic field lines given by the Magnetic Connectivity Tool\footnote{\url{http://connect-tool.irap.omp.eu/}}, which uses the measured solar wind and both a fixed value of high (800 km s\textsuperscript{-1}) and low (300 km s\textsuperscript{-1}) speed  of the solar wind to estimate a set of magnetic field lines connecting the spacecraft to the solar surface (2.5~R\textsubscript{$\odot$}). These sets of lines are modelled back to the solar surface using the PFSS model, with a bundle of 100 magnetic field lines for each solar wind speed value. We note that the assumption of nominal IP magnetic field lines is likely not valid for Solar Orbiter.  

The first time the shock intersects more than 50\% of the number of lines in the bundle is 05:58 UT $\pm$ 1 min for Solar Orbiter and 06:00 UT $\pm$ 1 min for both near-Earth spacecraft and STEREO-A. This timing is summarised in Table \ref{table:timing} and indicated in the right panel of Fig.~\ref{fig:radio_time} with the horizontal dashed segments on the bottom axis, following the colour code of the spacecraft, namely green, blue, and red for near-Earth probes, Solar Orbiter, and STEREO-A, respectively. These shock-lines-intersection times can be compared with the results given by the VDA analysis from Sect. \ref{sec:particle_timing}, already indicated in the same Fig. \ref{fig:radio_time} as vertical lines  discussed above. In the case of near-Earth spacecraft, the injection of the particles is estimated to happen two minutes after the first connection to the shock. For Solar Orbiter and STEREO-A, the injection time is respectively estimated three and eight minutes earlier than the first time of connection to the shock. However, the timing is co-temporal for the three spacecraft if we consider the time uncertainties. 

We also estimated for the three spacecraft the angle between the 3D shock normal and the magnetic field lines at the intersection \(\theta_{\mathrm{Bn}}\),   being \( \theta_{\mathrm{Bn}} = 82^\circ \), 71$^{\circ}$, and 84$^{\circ}$, for near-Earth, STEREO-A, and Solar Orbiter, respectively. The heights at the time of first connection between the shock wave and the magnetic field lines linking to the near-Earth, Solar Orbiter, and STEREO-A spacecraft are \(1.38 \pm 0.18 R_{\odot}\), \(1.47 \pm 0.15 R_{\odot}\), and \(1.39 \pm 0.20 R_{\odot}\), respectively. These heights are consistent with our estimated formation heights of HBs, which is approximately \(1.5 R_{\odot}\). In the presence of open magnetic fields in the laterally expanding shock regions, this configuration meets the steep geometry requirement for generating HB radio bursts. This alignment further supports the electron acceleration scenario proposed in Sect. \ref{sec:Radio} and the early connection to the escaping electron beams from the shock.

\begin{figure*}[htb]
\centering
  \resizebox{0.9\hsize}{!}{\includegraphics{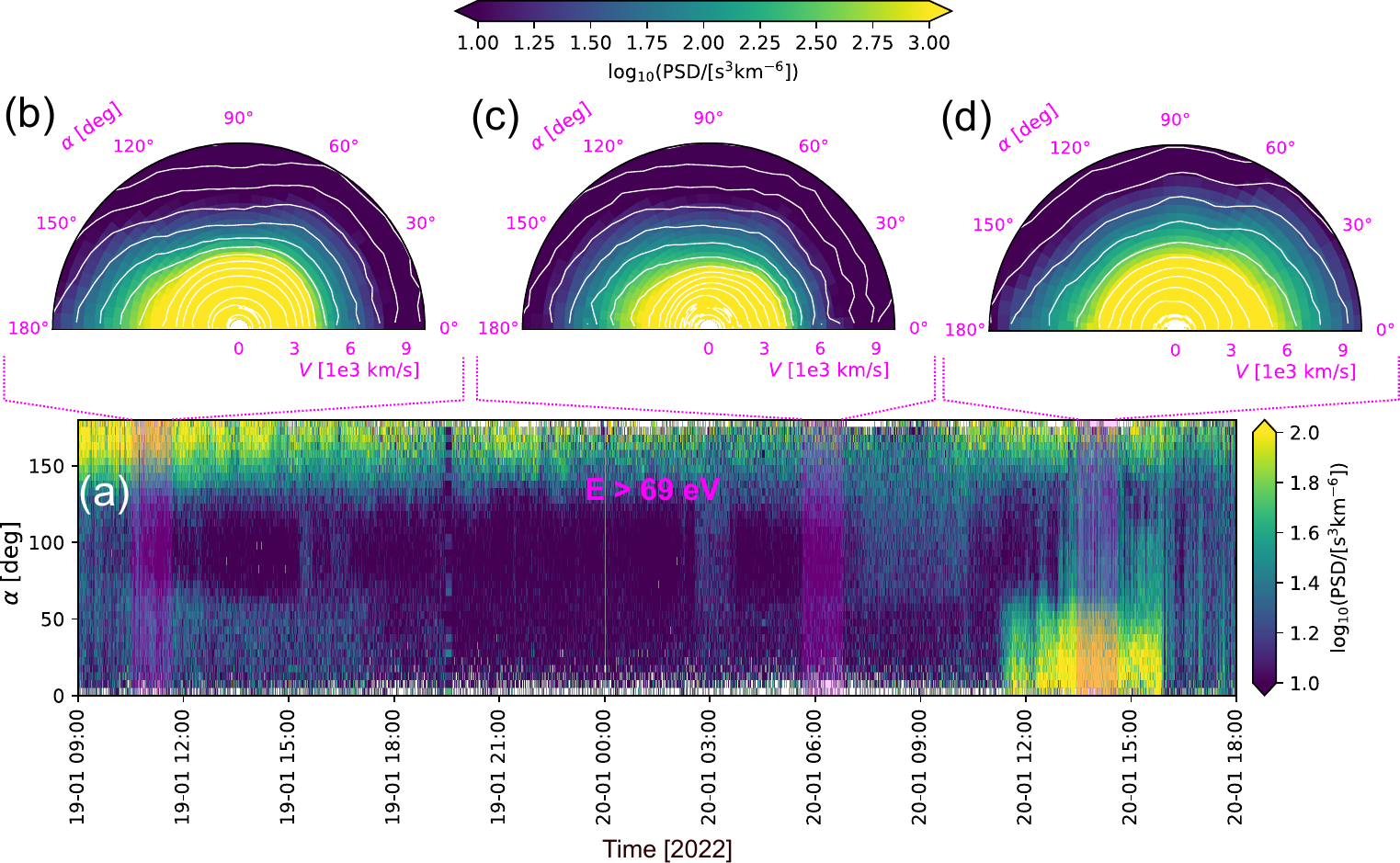}}
     \caption{Pitch-angle distribution function of solar wind electrons. (a) Pitch angle distribution of solar wind electrons with energies between ~69 eV and ~5 keV, for the time interval from 09:00 on 2022 January 19 to 18:00 on January 20. (b), (c), and (d) are 2D speed, pitch-angle distributions averaged over the three selected ~70 minutes intervals, marked by the magenta shadowed regions in panel a.}
     \label{fig:SWA_EAS}
\end{figure*}
\section{Observations and analysis of the CME/ICME on 2022 January 16 }

\label{sec:Magnetic_cloud_16}

At the time of the onset of the SEP event under study, Solar Orbiter was embedded in an ICME, namely from January 19 at 08:02 UT to January 20 at 17:52 UT. This ICME passed near Earth on January 18 at 22:57 UT and left the near-Earth environment on January 20 at 00:11 UT, a few hours before the SEP onset, as discussed in Sect. \ref{sec:in situ}. An inspection of the STEREO and SOHO coronagraph images, and the CDAW SOHO LASCO CME catalogue\footnote{\url{https://cdaw.gsfc.nasa.gov/CME_list/}} \citep{Yashiro2004} together with the near-Earth ICME list provided by I. Richardson and H. Cane\footnote{\url{http://www.srl.caltech.edu/ACE/ASC/DATA/level3/icmetable2.htm}\label{footnote Richardson list}} \citep{Richardson2010}, revealed that this ICME is most likely associated with a CME that appeared in LASCO C2 field of view at 20:48 UT on 2022 January 16. As this ICME might influence the transport of solar energetic particles to the Solar Orbiter location, we present below the CME evolution in the corona and in the heliosphere with some detail. 

\subsection{Observation and analysis of the January 16 CME }
\label{sec:CME_January_16}

The CME detected by SOHO/LASCO C2 at 20:48 UT on 2022 January 16 is associated with a flare erupting at 17:42 UT from AR 12929, the same region as for the event on January 20. Based on GOES observations, the flare being classified as C1.1 level, peaked at 17:48 UT and was located on W27N07. We used the GCS model described in Sect. \ref{sec:CME_obser_analysis} to derive the 3D morphology and average speed of the CME close to the Sun. 

The 3D fitting of the CME  shows a tilted flux-rope ($\gamma = -45^{\circ}$) with a speed of $\sim$773 km s\textsuperscript{-1}. The ecliptic CME width based on \cite{Dumbovic2019} is estimated as 38$^{\circ}$.  Based on the reconstructed CME nose longitude and latitude, the CME left the near-Sun environment propagating towards the Stonyhurst direction W25N17 (Carrington 315) at 15~R\textsubscript{$\odot$}. However, to fit the in situ observations of the ICME, based on arrival time at the locations being encountered, namely Earth and  Solar Orbiter as presented in Sects.~\ref{subsec:Interplanetary_context_Earth} and \ref{subsec:Interplanetary context_Solar Orbiter} respectively, the CME should have being oriented towards the south and east, specifically towards Stonyhurst W08N04 (Carrington 298). This is consistent with previous studies that fast CMEs turn to be blocked by the background solar wind ahead and deflected to the east \citep{wang2004deflection}. 
The additional observations from the STEREO-A/HI cannot confirm if the CME rotated in the interplanetary space after leaving the solar corona or if the difference in the CME nose position is due to inherent uncertainties associated with the GCS fitting \citep[e.g.][]{Verbeke2023,Kay2024}. 

Figure~\ref{fig:CME_ICME} shows the evolution of AR 12929 during 2022 January 16--20. Panel a presents an image taken by SDO/AIA/211{\AA} on January 16 at 19:59 UT, which shows the accumulated pixels of the dimming areas within the period of observation of the CME erupting on January 16. Panel b presents a base-difference image  
taken by SDO/AIA/193{\AA} on 2022 January 20 at 05:55:05 UT, showing the EUV wave, as discussed in Sect.~\ref{sec:CME-driven shock_obse_anal}. We overplotted in red the position of the dimming lobes shown in panel a as they have would rotated in time from January 16 to January 20. To better visualise the location of the footpoints of the CME on January 16 relative to the centre of the source region of the CME on January 20, the  image is now de-rotated to four days earlier using the solar differential rotation formula of \cite{1990Howard}, as shown in panel c.  

Using the EUV wave velocity of 373 km s\textsuperscript{-1}, as discussed in Sect.~\ref{sec:CME-driven shock_obse_anal}, we estimated that the EUV wave reaches the footpoints of the CME that erupted on 2022 January 16, within 3$\pm$1 (western footpoint) and 5$\pm$1 (eastern footpoint) minutes respectively after being firstly observed at 05:52 UT. Thus, the first time that the EUV wave related to the SEP event on January 20 intersects the centroid of the west dimming lobe is at 05:55 UT $\pm$ 1 min. This time is indicated in Table \ref{table:timing} and in Fig.~\ref{fig:radio_time} as a purple vertical dashed line in panels (c--d). We note that the inferred injection time for the particles observed by Solar Orbiter is co-temporal within uncertainties with the intersection time of the EUV wave associated with the January 20 SEP event with the western leg of the January 16 ICME, represented by the centroid of the dimming lobes.

\subsection{Observation and analysis of the January 16 ICME }
\label{sec:ICME_January_16}

The details about the solar wind and plasma data related to the ICME arriving at Earth and Solar Orbiter near the SEP event onset were presented in Fig.~\ref{fig:Earth_solo_part_plasma} and discussed in Sect.~\ref{sec:in situ}. It is unusual that an ICME directed towards the Earth and Solar Orbiter locations arrived $\sim$9 hours earlier at 1 au, as Solar Orbiter was located near 0.96 au. This might be related to the ICME propagating along a high-speed stream observed at near-Earth spacecraft before the arrival of the ICME, as shown in panel 7 of Fig.~\ref{fig:Earth_solo_part_plasma}a. A coronal hole (not shown) was located to the south-west of AR 12929, affecting the plasma conditions in which the ICME propagates,  which may have caused distortion in the ICME shape, as discussed by \cite{Rodriguez-Garcia2022CME}. 

The parameters of the CME in terms of speed (773 km s\textsuperscript{-1}) and orientation (W08N04) inserted in the ENLIL model fit well the observed arrival time at near-Earth and Solar Orbiter locations, presented in Figs.~\ref{fig:Earth_solo_part_plasma} and \ref{fig:STA_particles_plasma}. Figure~\ref{fig:enlil} shows how ENLIL simulates an earlier arrival of the ICME at Earth than at Solar Orbiter, being a flank arrival for both locations, as indicated with the yellow circle in the figure. This relative configuration of the ICME is represented in Fig. \ref{fig:scenario} by the blue shading. The ICME passed left the near-Earth (green circle) environment a few hours before arriving to Solar Orbiter (blue square) located at 0.96 au.

Figure \ref{fig:SWA_EAS} shows the pitch-angle distribution function of solar wind electrons with energies between ~69 eV to ~5 keV, built from Solar Orbiter SWA-EAS and MAG observations between 09:00 on 2022 January 19 and 18:00 on January 20. From the start of the shown interval, at 09:00 on January 19, to 18:00 on the same day, there is a clear beam flowing anti-parallel to the magnetic field (peak at pitch angle 180) and a faint beam flowing parallel to the magnetic field (secondary peak at pitch-angle~0). From~18:00 on January 19 until~11:00 on January 20, we observe only the anti-parallel beam signature, and then, we have a signature of a bi-directional beam, lasting for~4 hours. After 15:00 on January 20, the pitch-angle distributions do not exhibit clear beam signatures. In panels b, c, and d, we show three 2D speed, pitch-angle distribution functions, averaged over~70 minutes intervals within the three time periods characterised by different electron distribution function properties. The distribution in Fig. \ref{fig:SWA_EAS}b, is stretched towards the antiparallel direction, as there is a clear antiparallel beam, dominating over a much more “faint”, and broad parallel beam. The distribution in \ref{fig:SWA_EAS}c, which is co-temporal with the SEP onset, shows only the antiparallel beam. The distribution in \ref{fig:SWA_EAS}d has a signature of a bi-directional beam, resulting in a highly anisotropic distribution function. 

Therefore, the PAD function of the solar wind electrons is in agreement with the following. (1) A flank arrival of an ICME to Solar Orbiter from the beginning of the shown interval ($\sim$9 UT on January 19) until 15 UT on January 20, with the presence of a flux-rope structure with both legs still connected to the Sun between 11:50 and 16:00 UT on 2022 January 20; (2) At the time of the SEP onset the solar wind particles were propagating towards the Sun, as the pitch-angle is 180 and the local IMF vector was pointing outwards (Fig. \ref{fig:Earth_solo_part_plasma}b). This is congruent with the anti-sunward flux observed in energetic particles measured by Solar Orbiter; (3) At the time of the SEP onset, the eastern leg of the ICME passing through Solar Orbiter is disconnected from the Sun, as we observe only the anti-parallel beam signature.

The ICME reconstruction using the EC analytical model \citep{2018aNievesChinchilla}  is shown in Appendix \ref{appendix_ICME_reconstruc}. 
The results are not coherent, as the central magnetic fields are pointing to opposite directions at Solar Orbiter and near Earth. This result could be related to the flank nature of the encounter at both locations, and/or the potential deformation of the shape of the ICME in the heliosphere during propagation.

\section{Tracing the interplanetary propagation of the energetic particles}
\label{sec:interplanetary_propagation}

Assuming that the energetic particles observed by Solar Orbiter are injected inside the western leg of the ICME on 2022 January 16, as discussed below, we could estimate the field line twisting taking into account the energetic particle timing. The level of magnetic fluctuations inside MCs is generally lower than in the solar wind \citep{2005Dasso} and therefore energetic particle propagating inside MCs tend to have long mean-free paths. Based on \cite{2011Kahlertracing}, using a simple cylindrical flux-rope approximation, the number of field rotations N from the Sun to 1 au is given by
   
\begin{equation}
N=\frac{1}{2\pi} \frac{X}{R}\sqrt{\frac{L^2}{X^2}-1}\,, 
\label{eq1}
\end{equation} 
where L is the total field line length, R is the radius of the flux rope and X is the axial field line length. The VDA in Sect. \ref{sec:Particle_timing_Earth} provides an estimation of the field line length of L= 2.6$\pm$0.1 au. Using the GCS analysis (Sect.~\ref{sec:CME_January_16}) extrapolated to Solar Orbiter's location and taking into account that the CME did not centrally sweep over the spacecraft, the axial field line length of the western (longer) leg when it reaches Solar Orbiter can be estimated as 2.07 au, with an estimation of the total loop length at 0.92 au of 3.21 au. Assuming R/X $\sim$ 0.05--0.3 \citep{2011Kahlertracing}, Eq.~\ref{eq1} gives an estimation of N $\sim$ 0.4--2.74 turns along the longer leg of the ICME. Hereafter we use `longer leg' to denote the most distant ICME leg relative to Solar Orbiter, which connects the spacecraft to the Sun from the anti-sunward direction. This results is in agreement with the values found by \cite{2011Kahlertracing}, namely\ 1--10 turns along the full MC length. These low number of field line rotations is represented in Fig. \ref{fig:scenario} by the purple line winding around the main axis depicted in black. The orientation of the magnetic field line is indicated in agreement with the in situ observations (local IMF vector pointing outwards at Solar Orbiter location), as presented in Table \ref{table:ADAPT-WSA connectivity}, and discussed in Sect. \ref{sec:solar_orbiter_aniso}.   

\section{Summary and discussion}
\label{sec:summary_discussion}
On 2022 January 20, Solar Orbiter observed a SEP event showing strong sunward-directed beams for the first arriving particles, as presented in Figs. \ref{fig:anisotro_elec} 2 and \ref{fig:anistrop_protons} 2. The presence of velocity dispersion evidenced a solar origin, confirmed by radio and remote-sensing observations. Solar Orbiter was located at 0.92 au and 18$^{\circ}$ 
eastwards of near-Earth spacecraft, which measured usual antisunward-directed particles. At the time of the SEP onset, based on solar wind and magnetic field signatures discussed in Sect.~\ref{sec:in situ}, Solar Orbiter was crossing the eastern flank of an ICME present in the heliosphere that erupted from the Sun four days earlier on 16 January from the same AR as the one related to the SEP event. This ICME is well simulated by ENLIL model, as shown in Fig. \ref{fig:enlil}. An IP shock is impacting the spacecraft at 08:02~UT on January 19 and an MC arrives at 03:28~UT on 20 January, just before the particle onset, being observed until 17:52~UT. This ICME had passed the near-Earth environment a few hours before the SEP onset (shown in Fig. \ref{fig:scenario}) and did not appear to affect the particle propagation.  However, this ICME could still have played a role in forming the overall SEP pitch-angle distributions, which showed distinct discontinuities between the two pitch-angle hemispheres as discussed in Sect. \ref{sec:anistropies} and Appendix \ref{app:pitch-angle space}. The SEP event was widespread in the heliosphere, as it was observed by Solar Orbiter, near-Earth spacecraft, STEREO-A, and MAVEN, namely spanning at least a longitude of $\sim$160$^{\circ}$, as presented in Fig. \ref{fig:solar_mach}.   
The solar source related to the SEP event  was located at AR 12929, close to the west limb as observed by Earth at the time of the particle onset.  

An M5.5 flare was observed  erupting from AR 12929 located at Stonyhurst N08W76 and  peaking at 05:58 UT on 2022 January 20. The flare was characterised by multiple discrete episodes of energy release and particle acceleration. Several nonthermal HXR peaks were observed being co-temporal with TIIs and TIIIs radio bursts, the latter starting at 80 MHz, as discussed in Sect. \ref{sec:Radio}. The spectrum of the injected electrons deduced from the HXR observations in Sect. \ref{sec:Flare} is softer than the in-situ spectra discussed in Sect.~\ref{sec:particle_spectra}, namely $\delta \geq 4$ as opposed to $\delta \approx 2.5-2.8$, summarised in the second column of Table \ref{table:spectra}. We note that in itself, this mismatch does not rule out a flare-related origin of the interplanetary electrons, as similar relations were found to be typical for impulsive electron events where no CME nor CME-driven shock was present.

The overlying loop of the eruption is pushed upward to form the leading front of a CME at 05:50 UT on 2022 January 20. The CME eruption leads to the formation and propagation of an EUV wave on the solar surface \citep[shown in figure~1 of][]{2022Zhang},  clearly visible from 05:52 to 06:09 UT on 2022 January 20 in AIA images with a speed of 373 km s$^{-1}$. The EUV wave intersects at 05:55 UT $\pm$ 1 min for the first time the centroid of the west dimming lobe from the CME erupting on January 16, passing Solar Orbiter at the time of the SEP event. This  time is co-temporal with the inferred injection time of the particles (05:56 UT $\pm$ 4 min) observed by Solar Orbiter based on the VDA analysis presented in Sect. \ref{sec:particle_timing}. A CME-driven shock was observed early at 05:55 UT. As discussed in Sect. \ref{sec:CME-driven shock_obse_anal}, the first time the 3D shock intersects the magnetic field lines based on PFSS model connecting to both near-Earth spacecraft and STEREO-A is 06:00 UT $\pm$ 1 min. 
This timing is also in agreement with the injections times derived for STEREO-A (05:52 UT $\pm$ 8 min) and near-Earth spacecraft (06:02 UT $\pm$ 4 min).  Regarding connectivity, there is a good agreement between the PFSS model and observations except for Solar Orbiter. The polarity of the source region from where the particles were ejected is negative based on near-Earth and STEREO-A in situ data, but positive based on Solar Orbiter in situ observations. This is in agreement with the particles observed by Solar Orbiter propagated inside the previous ICME erupting on January 16.  

Radio observations of the SEP event shown in Sect.~\ref{sec:CME-driven shock_obse_anal} suggest that a shock wave formed in the low corona since TIIa started at 05:55 UT. The shock speed estimated from TIIa (\(\sim\)340 km s$^{-1}$) is in agreement with the speed of the EUV wave (373 km s$^{-1}$) observed starting at 05:52 UT, discussed above. A second Type II, TIIb was also identified and featured HB structures which are observed for a brief period of time between 05:57 and 06:00 UT. The co-temporal occurrence of HB and TIII radio bursts seem to suggest that they may be physically related such that some TIII may emanate from TIIb. That is, the electron beams generating HB may also generate TIII radio bursts. The TIIb is observed for about ten minutes and approached the decameter wavelengths ($\sim$16 MHz) at 06:05 UT. The shock speed estimated from TIIb (\(1400\) km s\(^{-1}\)) is also in agreement with the 3D CME-driven shock speed (1433 km s\(^{-1}\)) estimated from coronagraph data (Sect. \ref{sec:CME-driven shock_obse_anal}). The coronal TII radio bursts provide evidence that the shock was a significant particle accelerator, further supported by the co-temporal occurrence of TIII bursts and the most intense part of TIIb, known as HB bursts, which also align with the solar release time of the energetic particles.

The PAD in panel 2 in Figs.~\ref{fig:anisotro_elec} and \ref{fig:anistrop_protons} shows that particles arriving to Solar Orbiter propagated mostly anti-parallel to the magnetic field direction (distribution peaking at pitch-angle 180$^{\circ}$). The local magnetic field vector at the time of the SEP onset is pointing outwards at Solar Orbiter, indicating that the energetic particles propagated towards the Sun. This is in agreement with the PAD of the solar wind electrons. Figure \ref{fig:SWA_EAS} shows that  at the time of the SEP onset only the anti-parallel beam is observed. This agrees with the solar wind electrons propagating towards the Sun inside the western (longer) leg of an ICME connected to the Sun and the eastern (shorter) leg being disconnected. In the case of STEREO-A and near-Earth spacecraft, the analysis of the anisotropies in Sect. \ref{sec:anistropies} correspond to particles propagating from the Sun. 

We also used the VDA analysis to estimate an effective particle propagation length of L = 1.4$\pm$ 0.1~au for near-Earth observers, which is close to the nominal Parker spiral length for near Earth ($\sim$1.08~au) using the measured solar wind speed. However, in case of Solar Orbiter the effective length travelled by the particles is estimated to be L = 2.6$\pm$0.1~au, much longer than the length of $\sim$0.99 au expected for a nominal Parker spiral field with the measured solar wind and scatter free propagation. It might indicate a non-standard interplanetary magnetic field topology. This long effective path agrees with the particles propagating inside an ICME to arrive to the Solar Orbiter location. We note that STEREO-A also presents a long effective path length of L = 2.3$\pm$0.5~au.  We speculate that the presence of the previous ICME could distort the field line leading to STEREO-A out of the ecliptic making it longer. However, we note the relatively poor pitch-angle coverage and the high uncertainty of the path length.  Further analysis might be needed to discuss the apparently long path followed by the particles to reach STEREO-A, which is out of the goal of this study.  

From this VDA analysis we derived also the estimated injection time of the particles, as discussed above, being similar within uncertainties for the three spacecraft near Earth, Solar Orbiter, and STEREO-A. This timing is summarised in Fig. \ref{fig:radio_time} and Table \ref{table:timing} to compare with radio and HXR signatures presented above. For the three spacecraft, the estimated injected times are co-temporal with the presence of nonthermal HXR peaks, Type IIs, HBs, and Type IIIs starting at 80 MHz.  

We determined the electron peak spectra, as observed by Wind, Solar Orbiter, and STEREO-A, as summarised in Table \ref{table:spectra}.  For comparison, we selected the spectral index near 200 keV, namely $\delta$\textsubscript{200}. The $\delta$\textsubscript{200} indices found in this study $-$2.8 <~$\delta$\textsubscript{200} <$-$2.5 are similar to the study by \cite{2022Dresing}.   They derived a mean spectral index of $\langle\delta$\textsubscript{200}$\rangle$ = $-$2.5$\pm$0.3 analysing 33 large gradual SEP events that were related to coronal pressure waves. Moreover, the elemental composition measured by EPD/SIS and EPD/HET are typical of large gradual SEP events \citep{Desai2016a, Cohen2021}. 

At Solar Orbiter, based on several indicators, namely (1) the energetic and solar wind particles propagating towards the Sun, (2) the solar origin of the energetic particles based on VDA, (3) the anisotropy pattern, (4) the long effective path,  (5) the presence of the ICME that erupted on January 16 at the location of Solar Orbiter, (6) and the early connection of the EUV wave with the west lobe of the ICME, we argue that the energetic particles of the SEP event on 2022 January 20 propagated inside the ICME that erupted on January 16 and arrived to Solar Orbiter, travelling along the longer (western) leg of the ICME. 

This configuration is shown in Fig. \ref{fig:scenario} that shows the sketch of the interplanetary configuration of the 2022 January SEP event. STEREO-A and Earth are connected to the solar source through the nominal Parker spirals indicated with the dashed coloured lines using the measured solar wind speeds. We note that STEREO-A, being located to the east of Solar Orbiter, is estimated to be magnetically connected to the same region as near-Earth spacecraft, based on the PFSS model (not shown in Fig. \ref{fig:scenario}). The injection times of both spacecraft are in agreement with the intersection times of the respective magnetic field lines, based on the PFSS model, and the reconstructed CME-driven shock, as discussed above.   

Solar Orbiter is embedded in an ICME, shown with a blue shading, with the axial magnetic field line in black. The longer leg of the ICME was still anchored to the solar surface, based on the solar wind electron PAD. The solar source identification in Sect.~\ref{sec:solar_parent_activity} indicates that such leg is connecting to AR12929, and connected to the EUV wave and CME-driven shock related to the particle event, indicated as a red curve. The winding of the magnetic field lines of the magnetic flux rope is found to be of moderate size. The number of magnetic field turns in the MC structure inferred using the particle timing is below 6. The calculated particle path is around 30\% longer than the modelled lengths of the loop legs. This result is in agreement with previous observations of energetic particles inside ICMEs \citep[e.g.][]{2011Kahler_a, 2016DresingINJE, Palmerio2021}. 
We note that the ICME shows evidence of deformation of the front, since the ICME arrived at 1 au (Earth) before arriving at Solar Orbiter, located at 0.92 au. These observations support the importance of considering ejecta as irregular or deformable structures rather than "rigid" bodies and their propagation direction can be significantly influenced by the ambient solar wind \citep{wang2004deflection, Rodriguez-Garcia2022CME}. 

The injection of energetic particles into both inside the western loopleg --Solar Orbiter--, and outside --near-Earth and STEREO-A-- of the magnetic cloud requires an extended acceleration region that is most likely provided by the associated coronal shock, which is indicated by the associated TII radio burst, shown in Fig. \ref{fig:radio_time}. However, diverging magnetic field lines in the low corona could also provide this extent. A further analysis of the event, showing (1) connection to the shock in agreement with the particle injection time; (2) harder electron spectra of the measured in situ energetic particles as compared to those reconstructed from flare HXR spectra; (3) hard electron in situ spectra similar to events related to coronal pressure waves; (4) particle composition typical of large gradual events; and (5) the presence of TIII bursts starting at 80 MHz and being co-temporal with the TII, indicates that the main accelerator of the particles might be the CME-driven shock. 

\section{Conclusions}
\label{sec:Conclusions}
This work illustrates the importance of  the preconditioning of the heliosphere and the interplanetary magnetic field in the transport and spread of SEPs. Our main conclusions can be summarised as follows:
\begin{description}
 
 \item[$\bullet$ Solar source:] The solar source associated with the widespread SEP event on 2022 January 20 is likely the shock driven by the CME eruption observed near the west side from Earth's perspective. 
 
 \item[$\bullet$ Particle injection:] The energetic particles are injected over a wide angular region into and outside of a previous MC ejected on 2022 January 16 present in the heliosphere at the time of the particle onset on January 20. The sunward-propagated particles measured by Solar Orbiter are produced by the injection of particles in the longer (western) leg of the MC, which is still anchored to the Sun. 
\end{description}

\begin{acknowledgements}
LRG\ acknowledges support through the European Space Agency (ESA) research fellowship programme.
The UAH team acknowledges the financial support by the Spanish Ministerio de Ciencia, Innovación y Universidades FEDER/MCIU/AEI Projects ESP2017-88436-R and PID2019-104863RB-I00/AEI/10.13039/501100011033 and by the European Union’s Horizon 2020 research and innovation program under grant agreement No. 101004159 (SERPENTINE).
ND is grateful for support by the Research Council of Finland (SHOCKSEE, grant No.\ 346902). ND, CP, AW, and RV acknowledge funding by the European Union’s Horizon Europe research and innovation program under grant agreement No.\ 101134999 (SOLER).
LAB acknowledges the support from the NASA program NNH17ZDA001N-LWS (Awards Nr. 80NSSC19K0069 and 80NSSC19K1235).
EP acknowledges support from NASA's LWS (grant no.\ 80NSSC19K0067) and LWS-SC (grant no.\ 80NSSC22K0893) programmes.
AK acknowledges financial support from NASA NNN06AA01C (SO-SIS Phase-E) contract.
ICJ acknowledges the support of Academy of Finland (SHOCKSEE, grant 346902).
AW acknowledges support by the German Space Agency
(DLR), grant numbers 50 OT 2304.
JG thanks the support from National Natural Science Foundation of China (Grant Nos. 42188101, 42130204, 42474221.
TL acknowledges support from the UK Science and Technology Facilities Council (STFC) through grants ST/V000934/1 and ST/Y002725/1.
COL acknowledges support from the NASA LWS program (grant no. 80NSSC21K1325)  and the MAVEN project
funded through the NASA Mars Exploration Program.
RV also acknowledges funding by the Research Council of Finland (FORESAIL, grant No. 352847).
The authors acknowledge the different SOHO, STEREO instrument teams, and the STEREO and ACE science centers for providing the data used in this paper. Solar Orbiter is a space mission of international collaboration between ESA and NASA, operated by ESA. This research has used PyThea v0.7.3, an open-source and free Python package to reconstruct the 3D structure of CMEs and shock waves (GCS and ellipsoid model). ENLIL simulation results have been provided by the CCMC at NASA Goddard Space Flight Center (GSFC) through their public Runs on Request system (\url{http://ccmc.gsfc.nasa.gov}; run ID Laura\_Rodriguez-Garcia\_121523\_SH\_1). The WSA
model was developed by N. Arge, currently at GSFC, and the ENLIL Model was developed by D. Odstrcil, currently at George Mason University.

\end{acknowledgements}

\begin{flushleft}

\textbf{ORCID iDs} 
\vspace{2mm}

Laura Rodríguez-García \orcid{https://orcid.org/0000-0003-2361-5510}

Raúl Gómez-Herrero \orcid{https://orcid.org/0000-0002-5705-9236}

Nina Dresing \orcid{https://orcid.org/0000-0003-3903-4649}

Laura A. Balmaceda \orcid{https://orcid.org/0000-0003-1162-5498}

Erika Palmerio \orcid{https://orcid.org/0000-0001-6590-3479}

Athanasios Kouloumvakos \orcid{https://orcid.org/0000-0001-6589-4509}

Francisco Espinosa Lara \orcid{https://orcid.org/0000-0001-9039-8822}

Christian Palmroos \orcid{https://orcid.org/0000-0002-7778-5454}

Immanuel C. Jebaraj \orcid{https://orcid.org/0000-0002-0606-7172}

Alexander Warmuth \orcid{https://orcid.org/0000-0003-1439-3610}

Georgios Nicolaou \orcid{https://orcid.org/0000-0003-3623-4928}

Ignacio Cernuda \orcid{https://orcid.org/0000-0001-8432-5379}

Teresa Nieves-Chinchilla \orcid{https://orcid.org/0000-0003-0565-4890}

Annamaria Fedeli \orcid{https://orcid.org/0000-0001-9449-4782}

Christina O. Lee \orcid{https://orcid.org/0000-0002-1604-3326}

Christina M. S. Cohen \orcid{https://orcid.org/0000-0002-0978-8127}

Jingnan Guo \orcid{https://orcid.org/0000-0002-8707-076X}

Timo Laitinen \orcid{https://orcid.org/0000-0002-7719-7783}

Glenn M. Mason \orcid{https://orcid.org/0000-0003-2169-9618}

George C. Ho \orcid{https://orcid.org/0000-0003-1093-2066}

Olga Malandraki \orcid{https://orcid.org/0000-0002-4751-6835}

Rami Vainio \orcid{https://orcid.org/0000-0002-3298-2067}

Javier Rodr{\'i}guez-Pacheco \orcid{https://orcid.org/0000-0002-4240-1115}

\end{flushleft}
%
%
\bibliographystyle{bibtex/aa}
\bibliography{bibtex/biblio.bib}

\clearpage
\onecolumn

 \begin{appendix}
 \section{ENLIL model}
 \label{append:ENLIL_model}

ENLIL is a global 3D MHD model that provides a time-dependent background characterisation of the heliosphere outside 21.5~R\textsubscript{$\odot$}. ENLIL uses time-dependent magnetograms as a background, into which spheroidal-shaped high-pressure structures without any internal magnetic field can be inserted to mimic observed CME-associated solar wind disturbances. ENLIL-modeled CMEs have an artificially higher thermal pressure to compensate for the lack of a strong magnetic field \citep[][and references therein]{Odstrcil2004}. To improve the characterisation of the heliosphere, multipoint coronagraph observations are used to infer CME parameters, using the GCS model described in Sect. \ref{sec:CME_obser_analysis}. The inner boundary condition is given by the Wang-Sheeley-Arge (WSA) V5.2 model, using inputs from the standard quick-reduce zero-point corrected magnetograms from GONG (GONGZ), available on the National Solar Observatory website\footnote{\url{ftp://gong2.nso.edu/QR/zqs/}}.  The reliability of the CME arrival predictions depends strongly on the initial CME input parameters, such as speed, direction, and width \citep[e.g.][]{Mays2015,Kay2020,Palmerio2022}, but also on the errors that can arise in the ambient model parameters and on the accuracy of the solar wind background derived from magnetograms and coronal field modelling assumptions \citep[e.g.][]{Lee2013,Jin2022,2023Ledvina}. Based on \cite{Wold2018}, the mean absolute arrival-time prediction error in ENLIL is expected to lie around 10.4 $\pm$ 0.9 hours, with a tendency to an early prediction of $-4.0$ hours.

The preconditioning of the heliosphere and the interaction of the IP structures that might be present at the onset time can actively influence the magnetic connectivity of the different spacecraft. Therefore, the ENLIL simulation time ranges from 
January 15 to January 25 (i.e.\ from five days before to five days after the SEP event). This interval encompasses the possible previous CME that may influence the particle propagation at the onset time, and the ICME evolution through the IP medium up to 2.1 au. For this purpose, the GCS 3D reconstruction process presented in Sect.~\ref{sec:CME_obser_analysis} was also used for the ten relevant prior CMEs erupting in the time range of January 15 to January 20. We used the CME LE parameters (position and speed) rather than the bulk (bright core, if present) as they often capture the overall and great impact of the high-pressure structures better. The CME and model set-up parameters, and the results of the simulations are available on the Community Coordinated Modeling Center (CCMC) website.\footnote{\url{https://ccmc.gsfc.nasa.gov/database_SH/Laura_Rodriguez-Garcia_121523_SH_1.php}}

 \section{Additional SEP pitch-angle distributions}
 \label{app:pitch-angle space}
\begin{figure*}[htb]
\centering
  \resizebox{0.9\hsize}{!}{\includegraphics{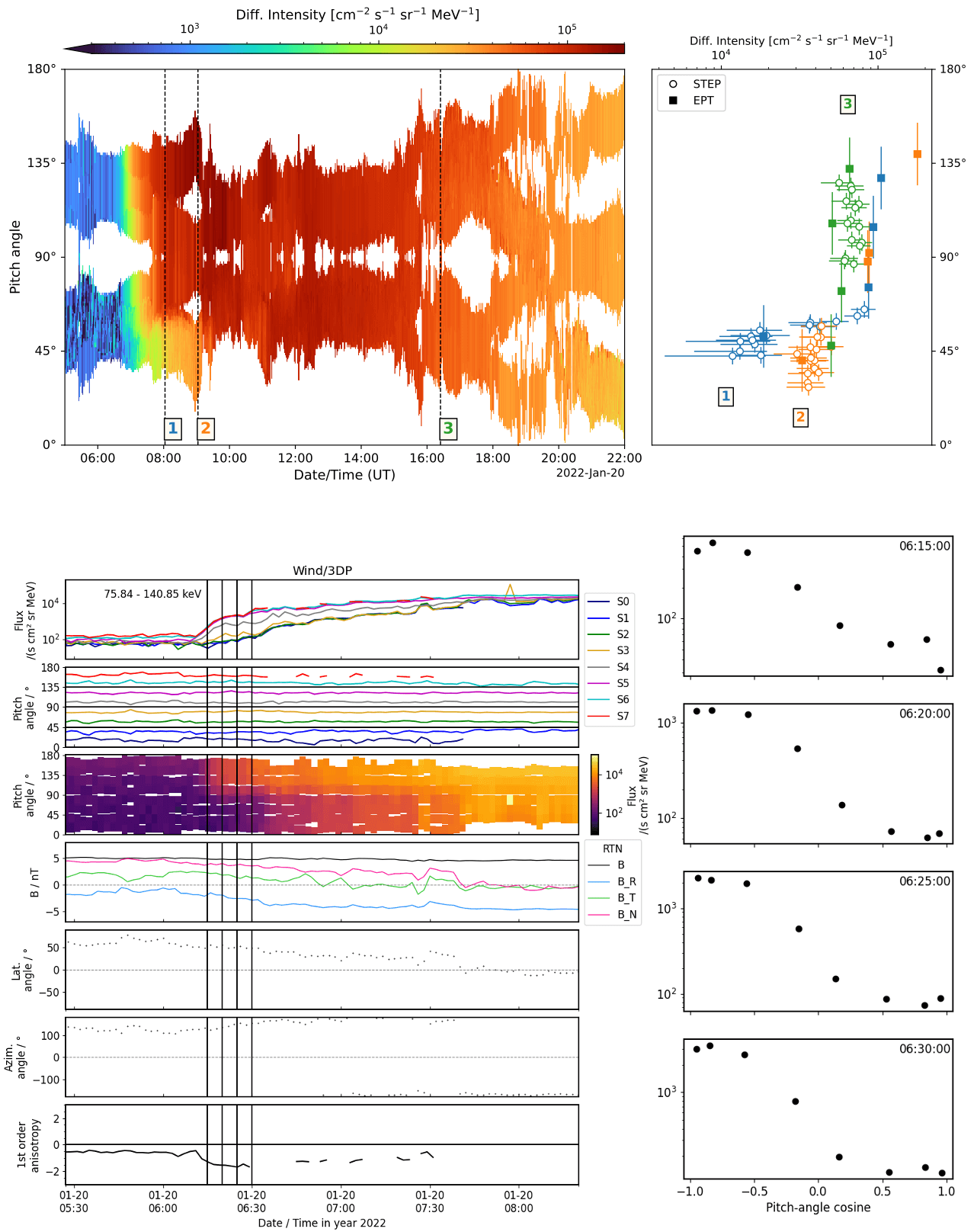}}
     \caption{Pitch-angle space. \emph{Top}: Reconstruction of the electron pitch-angle distribution as observed by Solar Orbiter in the energy range of 30 to 50~keV, using data from EPT and STEP. The top-right panel shows three slices through the pitch-angle distribution at different times during the event, indicated by labels 1, 2 and 3 on the top-left panel. EPT and STEP measurements are represented by filled squares and empty circles, respectively.  \textit{Bottom}: Left from top to bottom: electron intensity in the eight sectors of Wind/3DP, corresponding pitch angles of the bin centres, combined pitch-angle distribution with electron intensities marked by colour-coding, magnetic field magnitude and RTN-components, magnetic field latitudinal and azimuthal angles, and first-order anisotropy. Right: 2-dimensional pitch-angle distributions at the times marked by vertical lines in the plot on the left.}
     \label{fig:Pa_space_solo}
\end{figure*}
 Figure \ref{fig:Pa_space_solo} shows the electron PAD as measured by Solar Orbiter (top) and Wind (bottom). The top panel shows the reconstruction of the electron pitch-angle distribution as observed by Solar Orbiter in the energy range of 30 to 50~keV, using data from EPT and STEP. EPT provides wider pitch-angle coverage while STEP, with its segmented detector, offers finer resolution. The top-right panel shows three slices through the pitch-angle distribution at different times during the event, indicated by labels 1, 2 and 3 on the top-left panel. During the prompt phase of the event (label 1), the pitch-angle distribution exhibits a discontinuity around 60$^\circ$. As the event progresses (labels 2 and 3), the distribution becomes more isotropic. This might be related to the particle population streaming from the anti-Sun direction into the backward pitch-angle hemisphere uniformly, not filling the whole pitch-angle range evenly with particles.   
 
The lower panel of Fig. \ref{fig:Pa_space_solo} shows respectively from top to bottom the electron intensity in the eight sectors of Wind/3DP, corresponding pitch angles of the bin centres, combined pitch-angle distribution with electron intensities marked by colour-coding, magnetic field magnitude and RTN-components, magnetic field latitudinal and azimuthal angles, and first-order anisotropy. On the right we show the 2-dimensional pitch-angle distributions at the times marked by vertical lines in the plot on the left.  These show that the particle beam forms a plateau over a $\mu$-range from -1 to -0.5, but not until $\mu=0$. Therefore, there is a high difference between the two hemispheres in pitch angle space. The presence of the ICME that was ejected on January 16 could be a reason of this discontinuity in the pitch-angle hemispheres, as it might block part of the backward streaming SEP distribution. 

\section{VDA analysis: definition and methods used}
\label{sect:app_VDA}

Under the assumptions of being injected simultaneously and propagate scatter-free and without adiabatic cooling, the onset times of the energetic particles follow a velocity dispersion pattern $t_{onset}(v)=t_{inj}+L/(c*\beta(v)$), where $t_{inj}$ and $t_{onset}$ are the SEP injection time at the Sun and observation time at the spacecraft, respectively, $L$ the effective path length, and $\beta$= $v/c$, where $v$ represents the particle velocity \citep[e.g.][]{Vainio2013}. Thus, when the onset times, determined at a number of energies, and plotted as a function of the reciprocal of the particle velocities at respective energies, the slope of a curve fitted to the data indicates the effective path length $L$ and the intercept with the y-axis gives the release time $t_{inj}$.
\subsection{Poisson-CUSUM-bootstrap hybrid method}
\label{app:CUSUM_method}
The Poisson-CUSUM-bootstrap hybrid method finds distributions of particle onset times by taking random samples from the pre-event background and mappping the CUSUM parameters (mean and standard deviation of the pre-event background) of the samples to the onset times. The modified hybrid method is explained in detailed in Palmroos et al.\ 2024 (under review in A\&A). The method also applies this bootstrapping on the data while varying the integration time, in order to find the most probable onset time regardless of time resolution used, accompanied by the respective 95\% confidence intervals.

The background window for Wind data from which the parameters for the hybrid method were calculated was set to 01:30--05:40 UT on 2022 January 20. This window starts after the previous ICME has left near-Earth spacecraft and the elevated electron levels due to ion contamination decreased, as shown in Fig.~\ref{fig:Earth_solo_part_plasma} left (blue shaded area and horizontal line in panel 1). For ERNE protons we used a background window from 16:00 UT on 2022 January 19 to 05:00 UT on January 20. We note that ERNE protons were not affected by the previous ICME that arrived to Earth, as discussed above. We used the proton channels between 13 and 50 MeV, where velocity dispersion was observed in the onset times and the peak-to-background intensity ratios were 860--3890. 

\subsection{Sigma-threshold-bootstrap method}
\label{app_VDA_Mario}
We used the following procedure to estimate the onsets: (1) We integrated the intensities of pairs of consecutive energy channels to enhance the statistics; (2) For each pair of channels and using the time series of the particle intensities, we defined a sliding window of 9 minutes width, in which we averaged the intensity (mean) and calculated uncertainty using the error propagation (sigma); (3) We defined a threshold value calculated as the mean value plus 4 sigma above the background level; (4) If the intensities of the five following time stamps after the window were above the threshold value, we considered the first one as an onset candidate; (5) The sampling window advances one time step and the onset condition is tested again. This is repeated until the end of the time series; (6) To avoid choosing a candidate within the background level, we added the restriction in which for each onset candidate, the following consecutive two time stamps should be also onset candidates. Thus, we created a series of new onset candidates that fulfills the aforementioned restriction; (7) Choosing the first one as the final onset. 

We define the lower and higher uncertainties for the x-axis by looking at the time series and taking into account the different scenarios depending on the background level, statistics and rising phase. In case of channels with previous background almost nonexistent (it usually happens at higher energies), we sometimes find by eye a few counts before the onset which are very likely onset candidates but they are not found by the method. We consider the earliest of these counts as the lower uncertainty, while the upper uncertainty is the time resolution of the time series. Other cases are high background level and/or slow rising phase. In one of these two cases or combination of both, we consider the lower uncertainty as the earliest time when we see by eye that the SEP event start to increase but still not detected by the method, while the upper uncertainty is considered as the point where the increase is very clear to have started due to its steepness.

To determine more reliable mean values and uncertainties for the path length and injection time,  we used bootstrapping. In this process, for each pair of channels, we modify randomly the value of the onset as one of these three:  the value calculated by the method described above, and the lower and upper limits of the value based on the uncertainties. Moreover, we deleted a random number of onsets between zero and four. Then we did the fit with ODR and repeated this process 10.000 times to obtain the Gaussian distributions for the path length and injection time. We considered the mean of these Gaussians as the final values for path length and injection time. For the uncertainty we multiplied the standard deviation by the Student's t for a confidence level of 95\%. 

\section{SIS spectra and spectrograms}
\label{sec:appendix_SIS_spectra}

We fitted the spectra in Fig. \ref{fig:SIS_obs} with the 2-slope Band functional form which has been used in surveys of large SEP events \citep{Band1993, Desai2016, Mewaldt2012}. The fit coefficients are listed in Table~\ref{table:SIS}, following the notation used by \cite{Desai2016}. In the table, column 2 is a normalisation constant C, columns 3 and 4 are $\mathrm{\gamma_\alpha}$  and $\mathrm{\gamma_\beta}$, the low and high energy power law indices, and column 5 is the spectral break energy in MeV/nucleon. The values from the spectral fittings are similar to the survey results of \cite{Desai2016}, for example their mean values for O measured in 36 events was $\mathrm{\gamma_\alpha = 1.21\pm0.10}$, and $\mathrm{\gamma_\beta = 3.74\pm0.17}$. The Oxygen $\mathrm{\gamma_\alpha}$ in Table~\ref{table:SIS} is higher than the mean shown by \cite{Desai2016}, but lies with the distribution of results from their survey \citep[][see their Fig. 4(b)]{Desai2016}. The Oxygen $\mathrm{\gamma_\beta}$ in the Table~\ref{table:SIS} is close to the mean by \cite{Desai2006} survey. The spectral break energies in Table~\ref{table:SIS} decreases with increasing particle mass, as observed in previous studies \citep{Cohen2021, Desai2006, Mewaldt2005}.  The fits to H, 4He, O, and Fe are shown as dotted lines in Fig. \ref{fig:SIS_obs}. \cite{Li2009} modelled the energy dependence of spectral breaks, finding that a dependence on the break energy can be ordered by (Q/A)$\mathrm{^\alpha}$ where Q is the ion charge state and A is the atomic number, and $\alpha$ depends on the shock geometry. The partially ionised state of elements O and above leads to a decrease in the break energy. 

Figure~\ref{fig:SIS_spectrograms} shows spectrograms for H and $^4$He for the sunward- and anti-sunward pointing SIS telescopes. At energies above a few MeV/nucleon this event showed highly unusual intensities wherein the anti-sunward telescope intensities exceeded those of the sunward looking telescope. The implications of this are discussed in Sect. \ref{sec:summary_discussion}. At energies below $\sim$1 MeV/nucleon the intensity variations were typical for large SEP events with an initial large (factor of 10 or more) sunward/anti-sunward anisotropy that decayed after the initial rise phase of the event.  

\begin{table}[h!]
\caption{Band spectral fit parameters} \label{table:SIS}
\centering
\begin{tabular}{cccccc}
\toprule
Element & $C$ & $\gamma_\alpha$ & $\gamma_\beta$ & $E_B$ \\
\midrule
H & $1.66 \times 10^7$ & 1.17 & 2.53 & 3.33 \\
$^4$He & $2.13 \times 10^5$ & 1.83 & 2.50 & 1.86 \\
O & $5.27 \times 10^3$ & 1.88 & 3.24 & 1.74 \\
Fe & $2.03 \times 10^3$ & 1.75 & 4.00 & 1.20 \\
\bottomrule
\end{tabular}
\end{table}

\begin{figure}[htb]
\centering
\includegraphics[width=0.47\textwidth]{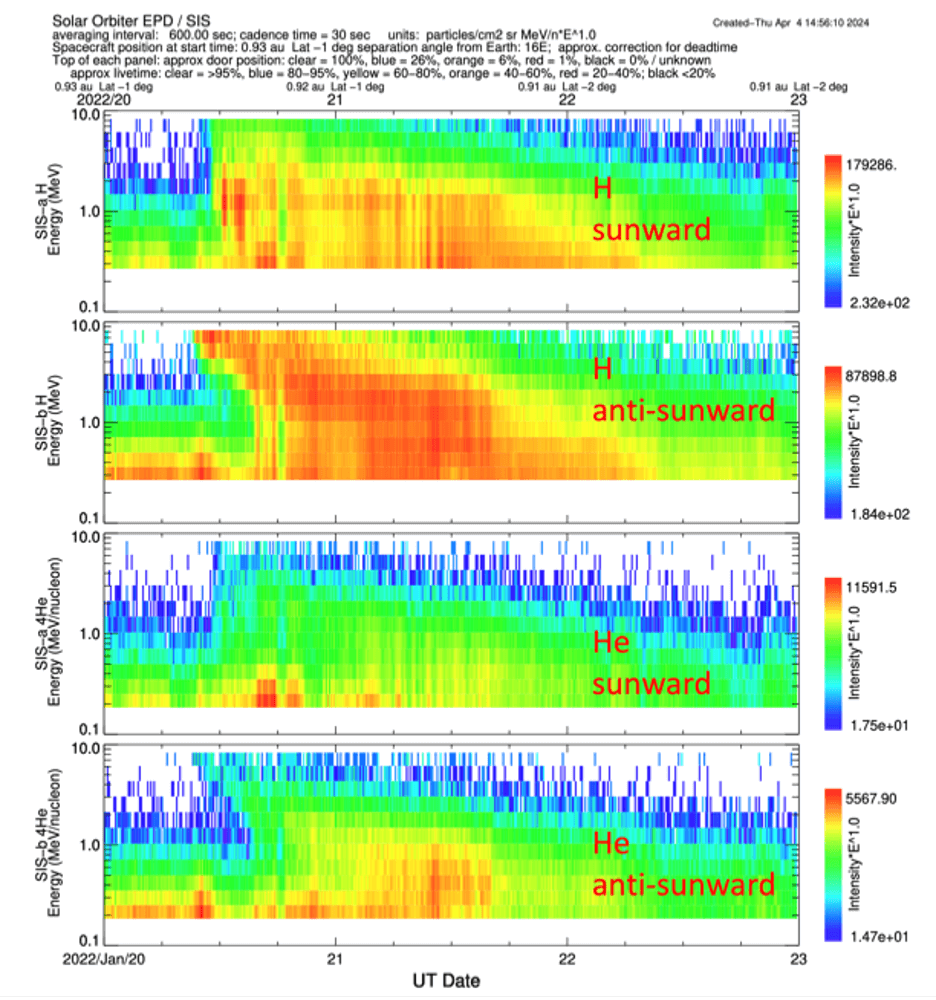}
\caption{Intensity spectrograms for H and He from the sunward and anti-sunward pointing telescopes, with the energy/nucleon scale multiplied by energy to increase the clarity of the higher energies. We note the anti-sunward telescope saw higher intensities than the sunward pointing telescope during the early portion of the event, as observed by the electrons discussed in Sect. \ref{sec:anistropies}.} 
     \label{fig:SIS_spectrograms}
\end{figure}

\section{January 16 ICME reconstruction}
\label{appendix_ICME_reconstruc}

 \begin{table*}[htbp]
\caption{ EC model fit parameters in RTN coordinates}
\label{table:ECmodel}
\begin{tabularx}{1\textwidth}{cccccccccc} 
\hline
\hline
s/c\ & Longitude  & Tilt & Rotation &Ellipse ratio  &Cross-section &Distance & $\chi\textsuperscript{2}$& Chirality&Solar wind\\
&$\phi$ (deg)&$\theta$ (deg)&$\xi$ (deg)&$\delta$ (-)&R (au)&$Y_0$ (au)& &&(km s$^{-1}$)\\
\hline
(1)&(2)&(3)&(4)&(5)&(6)&(7)&(8)&(9)&(10)\\

\hline
Wind MC&  157 & 17 &  7 & 0.70 & 0.10 &  0.075& 0.15&Positive&593\\
SolO MC&214& -10& 70& 0.79& 0.051&-0.026&0.27&Negative&460\\
 \hline
\\
\end{tabularx}
\footnotesize{\textbf{Notes.} Column 1: Spacecraft. Column 2: MFR axis longitude ($\phi$=[0...360]$^{\circ}$). Column 3: inclination of the flux rope with respect to the equatorial plane ($\theta$=[-90...+90]$^{\circ}$). Column 4: MFR rotation about its central axis  ($\xi$=[{0...180}]$^{\circ}$). Column 5: MFR distortion (ratio between major and minor ellipse axis, $\delta$=[0...1]). Column 6: MFR size. Column 7: distance from the spacecraft trajectory to the MFR axis (negative value means that the spacecraft is crossing the upper part of the structure). Column 8: goodness of the fitting ($\chi\textsuperscript{2}$=[0...1]). Column 9: MFR handedness. Column 10: average solar wind speed used for the fitting.}
\end{table*}
\begin{figure*}[htb]
\centering
  \resizebox{1.0\hsize}{!}{\includegraphics{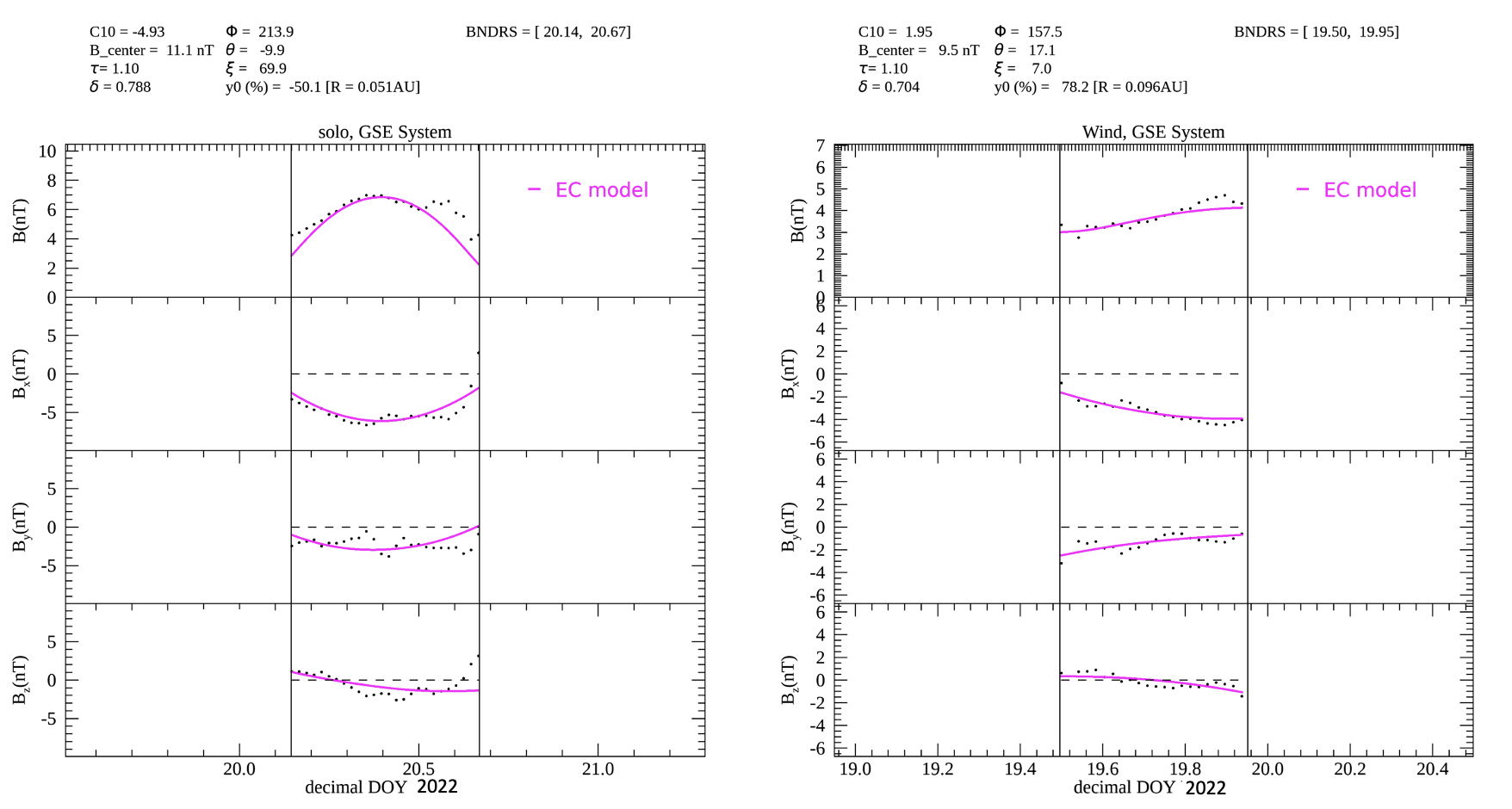}}
     \caption{Comparison of the EC model  fitting results (pink) with Wind (\textit{left}) and Solar Orbiter (\textit{right}) magnetic field observations spanning the MC. \textit{From the top}, the panels display the magnetic field strength and the three magnetic field B\textsubscript{RTN} components, respectively. }
     \label{fig:ICME_recons}
\end{figure*}
Several models for reconstructing MCs from in-situ observations have been established, such as the concept of a flux rope in a force-free configuration \citep{Burlaga1988,Lepping1990} or models that relax the force-free conditions  \citep[e.g.][]{Owens2006}. \cite{2018aNievesChinchilla} developed the Elliptical-Cylindrical analytical MFR model for MCs (hereafter the EC model) as an approach to considering the distorted cross-section of the magnetic field topology as a possible effect of the MFR interaction with the solar wind. However, all the models describe a limited subset of the properties of an MC as they are based on one-dimensional measurements along a line cutting through the structure, and it is not uncommon for different reconstruction techniques to display discrepant results \citep[e.g.][]{AlHaddad2013, Lynch2022}.

The analytical MFR model or EC model was applied to reconstruct the MC present within the ICME at the locations of Wind and Solar Orbiter. The MC reconstructions are local, based on the magnetic field measured in situ at each location. The EC model assumes an MFR magnetic topology, that is, an axially symmetric magnetic field cylinder with twisted magnetic field lines of elliptical cross-section. Therefore, the EC model allows us to consider cross-section distortion as a consequence of the interaction of the flux rope with the solar wind. The MC time intervals chosen for the EC model analysis correspond to the blue shadings in Fig. \ref{fig:Earth_solo_part_plasma} left (Wind) and Fig. \ref{fig:STA_particles_plasma} left  (Solar Orbiter). Column (10) in Table \ref{table:ECmodel} shows the average solar wind speed used for the fitting. The trajectory of the spacecraft through the MC is inferred by using the minimisation of the $\chi\textsuperscript{2}$ function to obtain a set of parameters that best fit the measured data \citep{2018bNieves-Chinchilla}. Table \ref{table:ECmodel} lists the obtained $\chi\textsuperscript{2}$ function and the EC model fit parameters in RTN coordinates. The MFR orientation in space is given by three angles: the central magnetic field longitude, $\phi$ (equal to 0$^{\circ}$ in the spacecraft-Sun direction), the tilt angle, $\theta$ (where positive values represent north of the equatorial plane), and the MFR rotation about its central axis, $\xi$. The geometry of the flux rope is given by the ratio between the major and minor ellipse axis, $\delta$, and the size by the cross-section major radius, R. $Y_0$ is the impact parameter, which  represents the closest approach to the MFR axis, where a positive value means that the spacecraft is crossing the lower part of the structure. Finally, the chirality or handedness of the flux rope is shown in Col. 9. In Fig. \ref{fig:ICME_recons}, the magnetic field data from Solar Orbiter (left) and Wind (right) are shown, along with the EC model fitting (smooth pink lines). The changes in the magnetic field  components are not well captured, especially at the rear part of the MC.

According to Table \ref{table:ECmodel}, Wind observes the MFR axis approximately between the perpendicular and the radial direction, based on the magnetic field longitude value ($\phi$=214$^{\circ}$), close to 270$^{\circ}$, while Solar Orbiter, with a longitude angle closer to 180$^{\circ}$ ($\phi$=157$^{\circ}$) might observe the flux rope closer to a flank. We note that the central magnetic field are pointing to opposite directions. The tilt angle ($\theta$) shows a difference between the observatories, with a northwards tilt in Wind ($\theta$=17$^{\circ}$) and a southwards tilt in Solar Orbiter ($\theta$=-10$^{\circ}$).  The disagreement in the MFR rotation about its central axis, $\xi$, for Wind and Solar Orbiter means that the orientation of the ellipse’s major axis is dissimilar in space. In the context of the other two angles, the respective $\xi$ value of 7$^{\circ}$ and 70$^{\circ}$ for Wind and Solar Orbiter, means that the distorted structure is parallel and perpendicular to the spacecraft trajectory.

The radius of the MFR cross-section, R is higher at Wind location than at Solar Orbiter, which is expected due to the expansion of the structure. The closest distance to the MFR axis, Y\textsubscript{0}, is positive (negative) for both Wind (Solar Orbiter), so that Wind (Solar Orbiter) spacecraft would be crossing the lower (upper) part of the structure. Wind is crossing further to the MFR axis than Solar Orbiter. The chirality for Wind (Solar Orbiter) is positive (negative) corresponding to right-handed (left-handed) flux ropes.  The fitting results based on $\chi\textsuperscript{2}$ (Col. 8 in Table \ref{table:ECmodel}) give satisfactory results. However, visual inspection of Fig. \ref{fig:ICME_recons}, which shows the comparison between the fitting in pink and the magnetic field observations by Wind (left) and Solar Orbiter (right), and the final interpretation of the position of the clouds lead to non-physical results. This is probably related to the boundaries selection for the fitting, the flank arrival of the cloud to both locations of Wind and Solar Orbiter, and the potential deformation of the shape of the ICME in the heliosphere during propagation in the heliosphere. 

 \end{appendix}

\end{document}